\documentclass[12pt]{article}

\usepackage{scicite}
  
\usepackage{times}

\usepackage{amsmath}
\usepackage{amsfonts}
\usepackage{amssymb}
\usepackage{graphicx}
\usepackage{pgfplots}

\newenvironment{sciabstract}{%
\begin{quote} \bf}
{\end{quote}}

\newcounter{lastnote}

\title{Synthetically Non-Hermitian Nonlinear Wave-like Behavior in a Topological Mechanical Metamaterial}

\author
{Haning Xiu$^{1\dagger}$, Ian Frankel$^{2\dagger}$, Harry Liu$^{3\dagger}$, Kai Qian$^{2}$, Siddhartha Sarkar$^{3}$, \\ Brianna C. Macnider$^{2}$, Zi Chen$^{1\ast}$, Nicholas Boechler$^{2\ast}$, Xiaoming Mao$^{3\ast}$\\
\\
\normalsize{$^{1}$Department of Surgery, Brigham and Women's Hospital/}\\\normalsize{Harvard Medical School, Boston, MA 02115,}\\
\normalsize{$^{2}$Department of Mechanical and Aerospace Engineering, University of California San Diego,}\\\normalsize{La Jolla, CA 92093,}\\
\normalsize{$^{3}$Department of Physics, University of Michigan, Ann Arbor, MI 48109,}\\
\\
\normalsize{$^\ast$Corresponding author. E-mail: maox@umich.edu,}
\\
\normalsize{nboechler@eng.ucsd.edu, zchen33@bwh.harvard.edu}
\\
\normalsize{$^\dagger$: These authors contributed equally to this work.}
}

\date{}

\begin{document} 

% Double-space the manuscript.

\baselineskip16pt

% Make the title.

\maketitle

% Place your abstract within the special {sciabstract} environment.

\begin{sciabstract}
Topological mechanical metamaterials have enabled new ways to control stress and deformation propagation. Exemplified by Maxwell lattices, they have been studied extensively using a linearized formalism. Herein, we study a two-dimensional topological Maxwell lattice by exploring its large deformation quasi-static response using geometric numerical simulations and experiments. We observe spatial nonlinear wave-like phenomena such as harmonic generation, localized domain switching, amplification-enhanced frequency conversion, and solitary waves. We further map our linearized, homogenized system to a non-Hermitian, non-reciprocal, one-dimensional wave equation, revealing an equivalence between the deformation fields of two-dimensional topological Maxwell lattices and nonlinear dynamical phenomena in one-dimensional active systems. Our study opens a new regime for topological mechanical metamaterials and expands their application potential in areas including adaptive and smart materials, and mechanical logic, wherein concepts from nonlinear dynamics may be used to create intricate, tailored spatial deformation and stress fields greatly exceeding conventional elasticity.

\end{sciabstract}

\section*{Teaser}
Found map of 2D topological material static deformation to 1D nonlinear dynamics, enabling new control of stress and strain.

%This one-sentence summary (125 characters with period) provides a snapshot of your research for non-specialist readers and should complement rather than repeat the title.

\section*{Introduction}

\noindent The study of topological band theory in condensed matter physics has led to novel classes of materials termed topological insulators \cite{hasan2010colloquium,zangeneh2020topological} and topological superconductors \cite{qi2011topological}, which support localized modes at the materials' edges that are highly robust to defects and perturbation \cite{asboth2016short}. The stability of these modes stems from topological protection conferred by the material's bulk properties. Topologically nontrivial materials have been shown to support unidirectional, backscattering-immune mode propagation, thus, facilitating the development of new superconducting devices \cite{beenakker2016road} with applications in areas such as quantum computation \cite{aasen2016milestones,PhysRevB.102.125407}, as well as magnetoelectronic \cite{gilbert2021topological} and optoelectronic devices \cite{chorsi2022topological}. Recently, topological band theory has also been applied to the mechanical domain, which has enabled the creation of topological mechanical metamaterials (TMMs) that support phenomena such as energy localization and immunity to backscattering at finite frequencies, and a new ability to design and control quasi-static and spatiotemporally-varying stress and deformation fields in materials \cite{yang2015topological,xin2020topological,huber2016topological,bertoldi2017flexible,singh2021design}. 

\begin{figure}[t!]
\centering
  \includegraphics[width=6.3in]{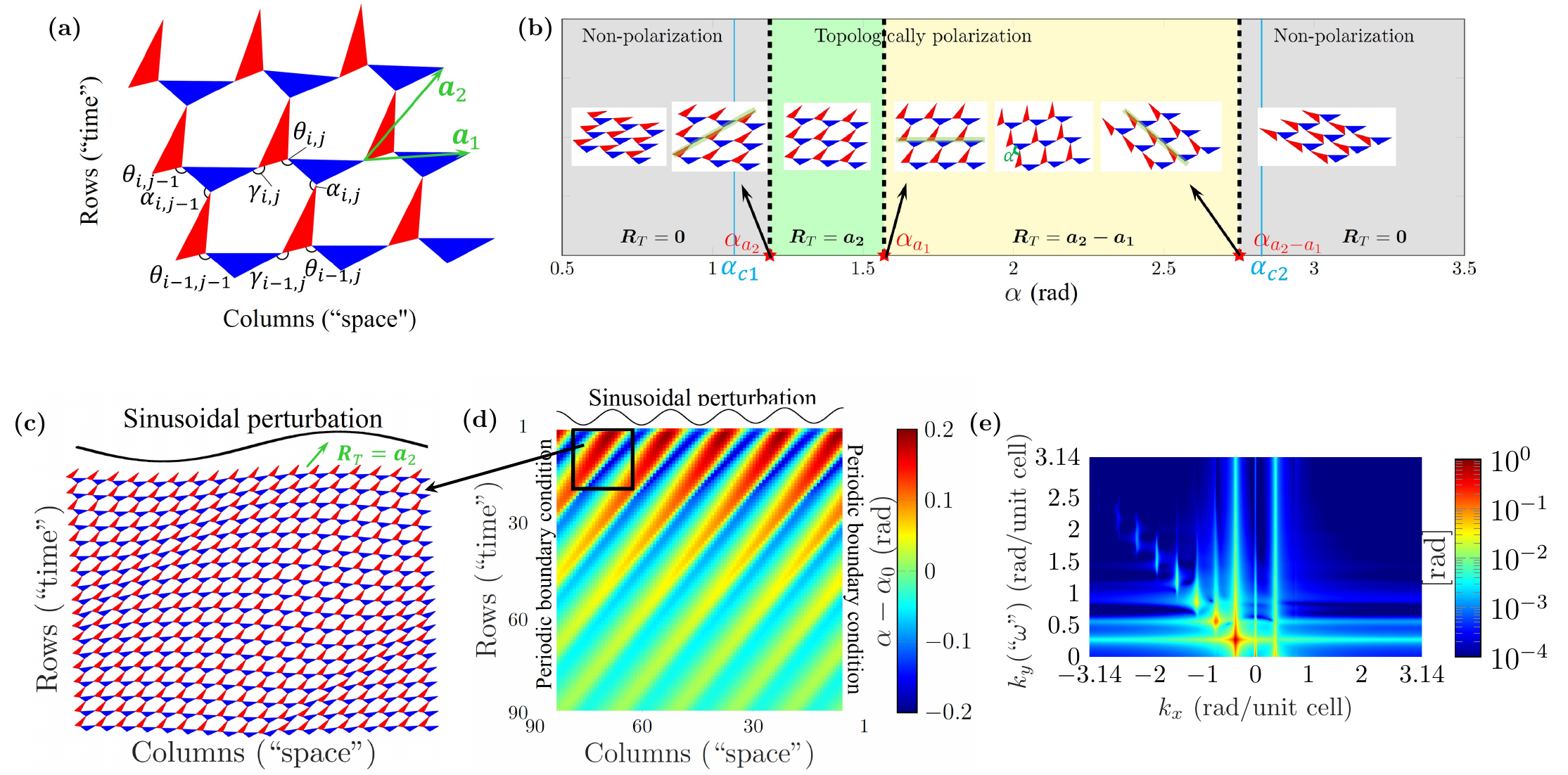}%
\caption{\textbf{Zero energy configuration, polarization diagram, and subsequent nonlinear wave-like behavior in the zero energy deformation field of a deformed kagome Maxwell lattice. The analogy between the 2D static deformation fields and a 1D, nonlinear, non-Hermitian wave equation is denoted in the axes labels}. (a) Geometry of the deformed kagome lattice we study here, where $\theta$, $\alpha$, and $\gamma$ are internal angles between the red and blue triangles in the unit cells (with normalized dimensionless side lengths of (0.4, 0.8, 1) and (0.5, 0.7, 1) for red and blue triangles, respectively). (b) Topological transitions of the lattice shown in (a) by twisting $\alpha$. The black arrows point to configurations at the boundary between polarization domains. The blue vertical lines denote angles between which $\det \epsilon_0<0$ and the linearized ZM deformation of the homogeneous lattice obeys hyperbolic PDEs, outside of which $\det \epsilon_0>0$ and the lattice obeys elliptic PDEs. (c) A zoomed-in view of a calculated section of an initially homogeneous lattice with $\mathbf{a_2}$ polarization, $\alpha_0=1.3344$, and periodic boundary conditions on left and right, perturbed by a sinusoidal static signal with $k_x=0.349$ rad/unit cell and $\varepsilon=20$ mrad. (d) The full lattice corresponding to the section shown in (c). (e) The 2D Fourier transform of the deformation field shown in (d), where the nonlinear phenomena of harmonic generation can be seen. Subscripts $x$ and $y$ denote the ``columns'' and ``rows'' axes, respectively. } 
\label{geometry_topology_lattice}
\end{figure}

Topological mechanical metamaterials at the Maxwell point (as shown in Fig.~\ref{geometry_topology_lattice}(a)), where the number of degrees of freedom (DOFs) balances with the number of constraints in the bulk, are a subclass of TMMs (referred to as ``Maxwell lattices'') in which modes having zero energy, referred to as ``zero'' or ``floppy'' modes (ZMs) \cite{kane2014topological,mao2018maxwell}, arise. These ZMs have a topological nature described by a polarization vector that is analogous to the topological invariant seen in the Su-Schrieffer-Heeger model~\cite{Su1979} and they localize such that, in the linkage-limit, the edges the polarization vector points towards have zero stiffness and the opposite are rigid \cite{kane2014topological}. The direction of the polarization vector is controlled by the lattice's geometry and tunable
through a soft strain \cite{rocklin2017transformable}. In the presence of interfaces or topological defects, this polarization can result in internal localized states of self-stress (SSSs) and ZMs \cite{kane2014topological,paulose2015topological}. Further, due to the balanced numbers of DOFs and constraints in the bulk, such lattices are holographic and the state of the zero-energy configuration of a $d-$dimensional material can be fully prescribed from its $(d-1)$ dimensional boundary. In special cases such as twisted kagome lattices, the mechanisms can be written as conformal transformations ~\cite{sun2012surface,lubensky2015phonons}. Such holography adds additional levels of deformation control since the bulk state can be controlled via the boundary. Due to their intrinsic scalability and high degree of control over deformation and stress fields through the tuning of the topological polarization vector, Maxwell lattices have been suggested for future use related to robotics, impact and energy absorption, tear resistance, nanoscale manufacturing via origami, and acoustic and phonon logic and computation devices (see, e.g., logic strategies via multistable metamaterials \cite{zhang2021hierarchical})  \cite{zhou2020switchable,silverberg2015origami,zunker2021soft,mcinerney2020origami,bossart2021oligomodal,bilal2017intrinsically}.

% None of these references were relevant: \cite{wang2021stiffness,brandenbourger2019non,zhalmuratova2020reinforced}
% \cite{alamri2018dynamic,davami2019dynamic,habib2018fabrication}
% \cite{PhysRevApplied.14.054035}
% ,li2021computation,liu2020topological

Despite this revealed wealth of applications, the study of Maxwell lattices has been confined, by and large, to the linear, small deformation limit \cite{mao2018maxwell,lubensky2015phonons,kane2014topological,rocklin2017transformable,rocklin2017directional}. Intriguing nonlinear effects such as topological solitons have been revealed in one-dimensional (1D) Maxwell chains~\cite{chen2014nonlinear,lo2021topology}. In two-dimensional (2D) topological Maxwell lattices, the study of nonlinear effects have been so far limited to perturbation theories~\cite{zhou2020switchable}. This is an important gap, as nonlinear systems do not obey superposition and, as such, support an ability to control the spatiotemporal allocation of energy in materials that vastly exceeds their linear counterparts \cite{patil2021review,lapine2014colloquium,scott2006encyclopedia} through phenomena such as self-localization \cite{chen2012optical,dauxois2006physics}, frequency conversion and dynamic tunability \cite{boyd2020nonlinear,nayfeh2008nonlinear}, and chaos \cite{strogatz2018nonlinear}, as well as rich interplay with finite-frequency topological states \cite{zhou2020switchable,manda2021nonlinear,zangeneh2020topological,smirnova2020nonlinear,chaunsali2021stability,tempelman2021topological,pernet2022gap,zhou2022topological}. As already suggested for Maxwell lattices in the linear regime, we envision that combining nonlinear responses with the strong localization, non-reciprocity, and the robust nature of topological protection will lead to an important expansion of the ability to tailor spatiotemporal stress, deformation, and energy fields, with application areas demonstrated for nonlinear dynamical systems ranging from impact mitigation \cite{nesterenko2013dynamics} to neuromorphic \cite{markovic2020physics} and ultrafast mechanoacoustic computation \cite{raja2021ultrafast,kippenberg2018dissipative}. 

In this work, we show that ZMs of 2D topological Maxwell lattices map to waves in 1D non-Hermitian (active or damped \cite{el2018non,ashida2020non}) and non-reciprocal dynamical systems, and a rich set of nonlinear phenomena arise, offering precise remote control of complex zero-energy spatial deformation patterns. Historically, space-time mappings have brought critical insight in many fields in science, from polymer physics to quantum criticality, and time crystals \cite{gennes1968soluble,Sondhi1997,kolner1994space,wilczek2012quantum}. Using exact geometric calculations and subsequent experimental validation of nonlinear ZMs in deformed kagome TMMs, we observe spatial nonlinear wave-like phenomena including harmonic generation, localized topological domain switching, amplification enhanced frequency conversion, and solitary waves. Results presented here are scale-free, material independent, and add a new dimension to mechanical metamaterials engineering, wherein deformation fields can be predicted and intricately designed using insights derived from the analysis of nonlinear waves in non-Hermitian systems \cite{el2018non,chen2012optical,patil2021review,chong2018coherent,xia2021nonlinear,cross1993pattern,akhmediev2008dissipative}.

%%%%%%%%%%%%%%%%%%%%%%%%%%%%%%%%%%%%%%%%%%%%%%%%%%%%%%%%%%%%%%%%
%%%%%%%%%%%%%%%%%%%%%%%%%%%%%%%%%%%%%%%%%%%%%%%%%%%%%%%%%%%%%%%%

\section*{Results}

\subsection*{Topological polarization and analogy to 1D dynamical systems}

\noindent In Maxwell lattices, the number of DOFs and the number of constraints are identical in the bulk. By the Maxwell-Calladine index theorem \cite{calladine1978buckminster,kane2014topological,sun2012surface,mao2018maxwell}, this equality indicates that the difference between the number of ZMs and SSSs is proportional to the size of the open boundary. By manipulating the unit cell geometry, the ZMs can be localized at the boundaries of the lattice at which the topological polarization vector $\mathbf{R}_T$ points. Considering a finite 2D deformed kagome lattice consisting of $N_x$ (number of columns) by $N_y$ (number of rows) unit cells (with two triangles per unit cell), the total number of nodes and bonds (edges of triangles) under open boundary conditions are $N = 3 N_x N_y + N_x + N_y$ and $N_b = 6 N_x N_y$, respectively. Consequently, the number of ZMs $N_0$, is given by $N_0 = 2N-N_b+N_s=2 N_x + 2 N_y+N_s$ ($N_s=0$ for a open boundary conditioned lattice). Removing the number of rigid body DOFs of the whole lattice, the remaining number of nontrivial ZMs is $2N_x+2N_y-3$. For our lattice, shown in Fig.~\ref{geometry_topology_lattice}(a), its configuration is described by a set of angles $\{\alpha_{i,j},\theta_{i,j},\gamma_{i,j}\}$ defined for each unit cell at $i$-th row and $j$-th column. For a homogeneous lattice, all $\{\alpha_{i,j},\theta_{i,j},\gamma_{i,j}\}$ are set to be the same in each unit cell, leaving only one free angle (i.e., the Guest-Hutchinson  mode, labeled as $\alpha$ here) to determine the geometry of the homogeneous configuration as is shown in Fig.~\ref{geometry_topology_lattice}(b). This angle also determines the topological polarization $\mathbf{R}_T$  of the lattice~\cite{kane2014topological}, which is defined via the phase winding $\phi(k)$ of the determinant of the equilibrium matrix $\mathbf{Q}$ that maps tension on the bonds to the total force on the sites in momentum space $k$, where [$\mathbf{Q}(k)=|\mathbf{Q}(k)|^{i\phi(k)}$], and lattice vectors $\mathbf{a}_i$, such that $\mathbf{R}_T=\sum_{i} a_i \frac{1}{2\pi} \oint dk_i\cdot \nabla_{k_i} \phi(k)$. When $\mathbf{R}_T=0$ (with a proper gauge), all edges of the lattice have ZMs, while when $\mathbf{R}_T \neq 0$, the polarization vector points towards the ``soft'' edges that the ZMs are localized to, such that the edges opposite to the direction of $\mathbf{R}_T$ becomes the ``hard'' edges. As shown in Fig.~\ref{geometry_topology_lattice}(b), the deformed kagome lattice experiences topological transitions at three critical angles $\alpha_{\mathbf{a_2}}$, $\alpha_{\mathbf{a_1}}$, and $\alpha_{\mathbf{a_2-a_1}}$. When $\alpha<\alpha_{\mathbf{a_2}}$ or $\alpha>\alpha_{\mathbf{a_2-a_1}}$, the lattice has $\mathbf{R}_T=0$. Between these two critical angles, the lattice is topologically polarized, and $\mathbf{R}_T$ has two distinct directions, $\mathbf{a_2}$ or $\mathbf{a_2-a_1}$, separated by $\alpha_{a_1}$. 

We first consider the deformation of our TMMs in the continuum limit, where ZMs are determined by partial differential equations (PDEs), such that to third derivatives at linear order of the $x$ component of the displacement vector, $u_x$: 
\begin{equation}\label{EQ:PDE}
   ( [\epsilon_{0yy}\partial^2_x - 2 \epsilon_{0xy}\partial_x\partial_y +  \epsilon_{0xx}\partial^2_y]+[C_1\partial^3_x + C_2\partial^2_x\partial_y + C_3\partial_x\partial^2_y] ) u_x=0,
\end{equation}
with $\epsilon_0$ being the Guest-Hutchinson mode, a soft strain always present in Maxwell lattices~\cite{guest2003determinacy} (see SI for the derivation of this PDE). As discussed in Ref. \cite{rocklin2017transformable}, this type of soft, spatially varying modes $u$ generally arise in all materials in which a homogeneous strain $\epsilon_0$ is soft. In Maxwell lattices, this soft strain $\epsilon_0$ is guaranteed to exist and cost exactly zero energy \cite{guest2003determinacy}, and the spatially varying soft modes $u$ take the form of \emph{exact zero energy} modes, protected by the Maxwell-Calladine index theorem \cite{rocklin2017transformable,lubensky2015phonons}. Such soft strain $\epsilon_0$ can also accidentally arise due to geometric singularities in over-constrained lattices, such as planar quadrilateral kirigami, where these soft modes $u$ cost a small amount of elastic energy even when the hinges are considered perfect \cite{zheng2022continuum,PhysRevE.99.013002,czajkowski2022conformal,zheng2022modeling,czajkowski2022duality}. Importantly, in Maxwell lattices, the fact that these ZMs are exact zero energy makes them scale free and materials independent.

Solutions to this PDE to the quadratic order (first square brackets) can be obtained by considering the case with prescribed $k_x$ (wave number along $x$), where the ZM is given by $k_y=\frac{\epsilon_{0xy}\pm \sqrt{-\det \epsilon_0}}{\epsilon_{0xx}} k_x$. When $\det \epsilon_0>0$, corresponding to the Guest-Hutchinson mode being a dilation dominant (auxetic) mechanism and the PDE being elliptic, $k_y$ is complex with an imaginary part $k_y''\propto \pm k_x$, describing a pair of ZMs localized on the top and bottom edges, respectively. With proper coordinate transformations, these ZMs are mapped to conformal transformations~\cite{sun2012surface,rocklin2017transformable}.  Adding terms with higher order derivatives only quantitatively changes these ZMs. In the opposite case, $\det \epsilon_0<0$, corresponding to the Guest-Hutchinson mode being a shear dominant (non-auxetic) mechanism and the PDE being hyperbolic, $k_y$ is real, describing a pair of bulk ZMs. Unless fine-tuned, when terms of higher order derivatives (the second square brackets in Eq.~(\ref{EQ:PDE})) are introduced, the solution of $k_y$ becomes complex, with the imaginary part $k_y''\propto k_x^2$ being higher order, indicating slower decay.  Importantly, the sign of these decay rates is determined by the topological polarization $\mathbf{R}_T$, in all cases. The same conclusion can be reached by starting with given $k_y$. We note that most known cases of topologically polarized 2D Maxwell lattices belong to the hyperbolic case ($\det \epsilon_0<0$)~\cite{rocklin2017transformable}.

For the non-auxetic case, the mapping to a hyperbolic equation suggests that an analogy can be made between the \emph{2D spatial} PDE of Eq. (\ref{EQ:PDE}) and a \emph{1D space-time} PDE. As shown in Fig.~\ref{geometry_topology_lattice}, the specific analogy between 2D spatial deformation and 1D spatiotemporal deformation we propose herein has progression across columns, or in the $x$ direction, correlating with space, and progression across rows, or in the $y$ direction, correlating with time evolution in the 1D analog dynamical system ($y \leftrightarrow t$). In Eq.~\ref{EQ:PDE}, the 2nd order cross derivative term thus provides a conservative non-reciprocity along the $x$ direction as the ``waves'' propagate in ``time'' ($y$). Interestingly, the 3rd order terms become \emph{non-conservative} in the analog system, making the wave equation, ``synthetically'' non-Hermitian \cite{ashida2020non}. The $y$-component of the topological polarization in the 2D spatial lattice thus translates to a spatially uniform activity/damping in the 1D space-time lattice. Similarly, the $\partial_x\partial^2_y u_x$ term provides a non-reciprocal activity in the effective 1D lattice. 

Given aforementioned analogy to a 1D non-reciprocal, non-Hermitian, spatiotemporal system, we aim to study ``wave propagation'' in our 2D spatial TMM. With the expected polarization-dependent spatial amplification/decay, we expect large amplitude deformations outside the confines of a linear, small deformation approximation, leading to the proliferation of rich nonlinear phenomena. To this end, we numerically calculate the exact nonlinear ZM configuration for chosen homogeneous configurations, with periodic boundary conditions on its left and right edges, and then an applied perturbation to the soft or hard edge of the lattice such that the $\theta$ angles $\theta_{1,j}=\theta_0+f(j)$, where $\theta_0$ is the initial homogeneous $\theta$ value. Given three angles and fixed edge lengths of the triangles, a hexagon is fully determined to within a choice of a single convex or concave angle (Fig.~\ref{geometry_topology_lattice} and see the SI for details). We choose the convexity where the complementary angle (angle on the opposite side across the hinge) to $\theta$ is always less than $\pi$, which allows us to span the entire topological polarization range. By solving iteratively through each row starting with the edge where the perturbation is applied, the entire lattice can be determined geometrically, without approximation. Periodic boundary conditions are implemented by using Newton's method and numerically solving for a compatible periodic solution at each row. Within the context of our analogy to a 1D spatiotemporal system, this is as if we are applying an initial condition across the entire lattice, and then letting the system evolve in time. 

%%%%%%%%%%%%%%%%%%%%%%%%%%%%%%%%%%%%%%%%%%%%%
\subsection*{Linear and weakly nonlinear response}
\begin{figure}[t!]
\centering
  \includegraphics[width=6.8in]{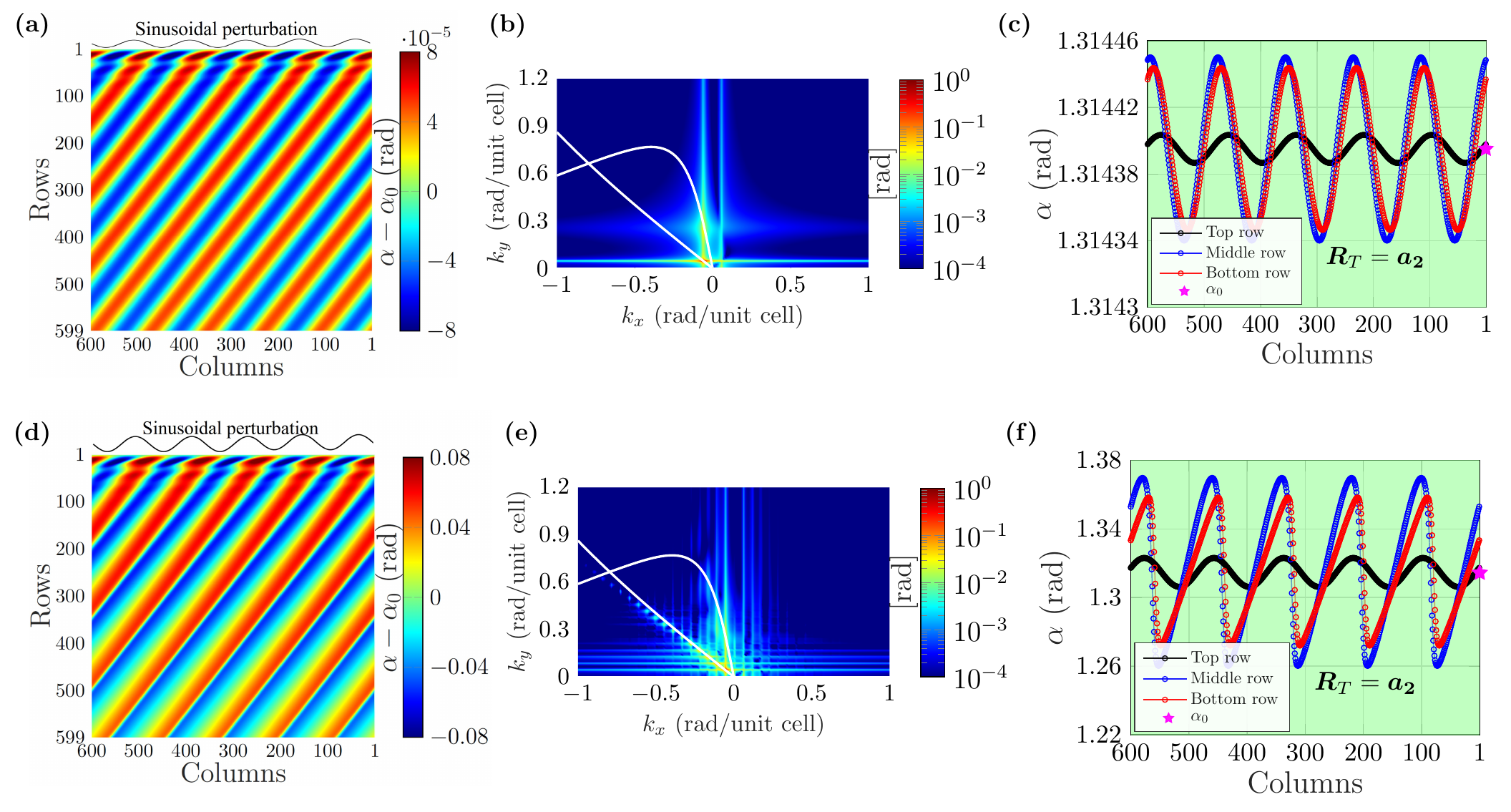}%
\caption{\textbf{Linear and weakly nonlinear response due to a sinusoidal perturbation applied to the soft (top) edge of a kagome Maxwell lattice deep in the $\mathbf{a_2}$ polarized region}. The lattice has $\alpha_0=1.3144$ rad and $k_x=0.0524$ rad/unit cell. (a-c) Linear response at $\varepsilon=1$ $\mu$rad. (d-f) Weakly nonlinear response and harmonic generation at $\varepsilon=1$ mrad. (a,d) Deformation field. (b,e) 2D Fourier transform of (a,d). White lines denote the real part of the ZM modes predicted by linear theory. (c,f) Select rows of (a,d). The pink star in (c,f) denotes the initial homogeneous angle, and the background shading denotes the topological phase (always $a_2$ polarization in this case).}\label{a2_polarization_wave}
\end{figure}

\noindent We start by verifying wave characteristics of ZMs in 2D Maxwell lattices in the linear and weakly nonlinear regimes using our exact geometrical numerical method. Because the lattice satisfies the Maxwell criterion, this ZM configuration is exactly geometrically
determined, independent of materials and length scales, in contrast to low
energy modes studied in Refs. \cite{zheng2022continuum,czajkowski2022conformal,zheng2022modeling,czajkowski2022duality}. We start with a homogeneous lattice deep in the $\mathbf{a_2}$ topologically polarized phase. The critical angles between which the lattice is hyperbolic, $\alpha_{c1}, \alpha_{c2}$, are shown in 
Fig.~\ref{geometry_topology_lattice}(b). We then apply a low-amplitude sinusoidal perturbation to the soft edge (the top row), such that $f(j)=\varepsilon \sin{(k_x j)}$. The resulting deformation field is shown in Fig. \ref{a2_polarization_wave}(a-c), and is described by a superposition of two ZMs that decay into the bulk that closely match expectations from the linear theory (further described in the SI). A 2D Fourier transform of the deformation field can be seen in Fig.~\ref{a2_polarization_wave}(b), overlaid with two white lines denoting the real part of the wave number $k_y$ of the two ZMs predicted by linear theory. Of the two ZMs, one has a shorter y-direction wavelength (higher ``frequency'' in the effective 1D spatiotemporal system) with faster decay, which is part of a highly dispersive branch, and the other a longer y-direction wavelength with slower decay, which is part of a weakly dispersive branch. The initial increase in amplitude of the deformation field with distance from the perturbation (decreasing row number) that can be seen in Fig.~\ref{a2_polarization_wave}(a,c) is due to the input phase of the two ZMs and coherent interference. Fig.~\ref{a2_polarization_wave}(d-f) details the same system shown in Fig.~\ref{a2_polarization_wave}(a-c), but with a larger initial perturbation, inducing the nonlinear phenomena of harmonic generation \cite{nayfeh2008nonlinear}, similar to Fig.~\ref{geometry_topology_lattice}.

%%%%%%%%%%%%%%%%%%%%%%%%%%%%%%%%%%%%%%%%%%%%%
\subsection*{Strongly nonlinear phenomena}
\begin{figure}[h!]
\centering
  \includegraphics[width=6.8in]{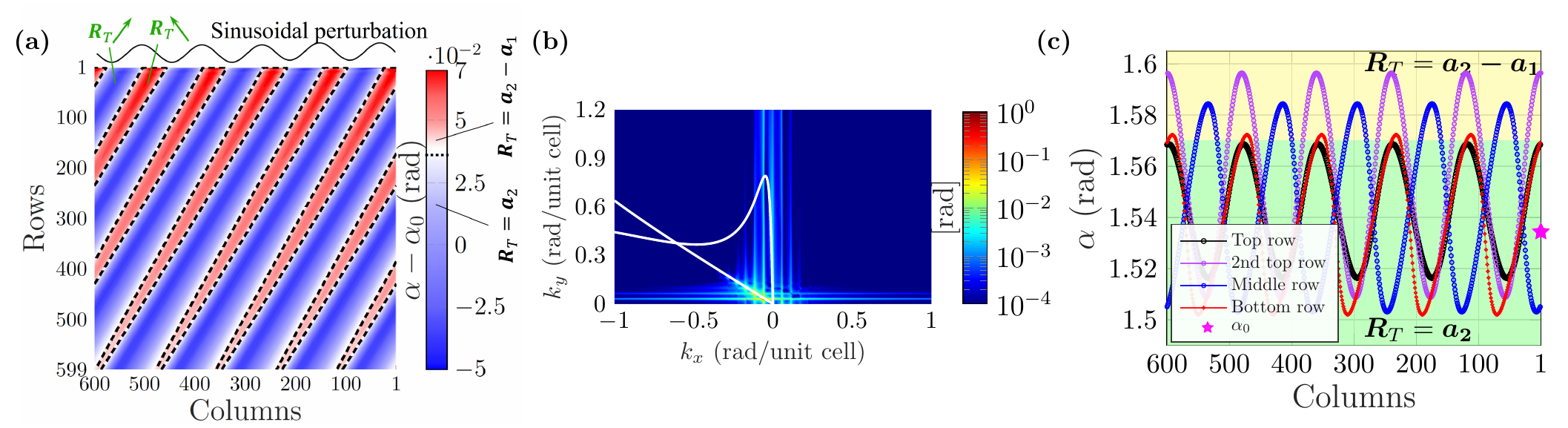}%
\caption{\textbf{Intrinsic localized topological polarization switching and domain formation due to a sinusoidal polarization applied to the soft edge (top) of the deformed kagome Maxwell lattice near the border of the $\mathbf{a_2}$ polarized phase}. The lattice has $\alpha_0=1.5344$ rad, $k_x=0.0524$ rad/unit cell, and $\varepsilon=1$ mrad. (a) Deformation field, (b) 2D Fourier transform of (a) with white lines representing the real part of the ZMs solved from the linear theory, and (c) Select rows of (a). The pink star in (c) denotes the initial homogeneous angle, and the background shading denotes the topological phases. The dashed black lines in (a) denote boundaries between regions of different polarizations. 
}\label{a2a1_switch_wave}
\end{figure}

\noindent We now proceed to explore more strongly nonlinear phenomena arising in these lattices. In particular, we show three examples, namely, automatic and localized topological polarization switching, amplification enhanced frequency conversion, and solitary wave formation. The first example, the switching of topological polarization as a result of nonlinear waves, occurs when the lattice is in the $\mathbf{a_2}$ phase close to the boundary with the $\mathbf{a_2-a_1}$ phase. As shown in Fig.~\ref{a2a1_switch_wave}, for this case, a sinusoidal perturbation causes regions of deformation significant enough to cause local transitions to $\mathbf{a_2-a_1}$ polarization. The boundaries between regions of different polarization are known to support internal SSSs \cite{paulose2015topological}, which has been shown to have implications for lattice fracture \cite{zhang2018fracturing}. We highlight that this domain switching is a strictly nonlinear effect, as it requires finite deformation. Such finite deformation effects may lead to boundary-defined (holographic) programmable topological domain walls.

\begin{figure}[h!]
\centering
\includegraphics[width=6.8in]{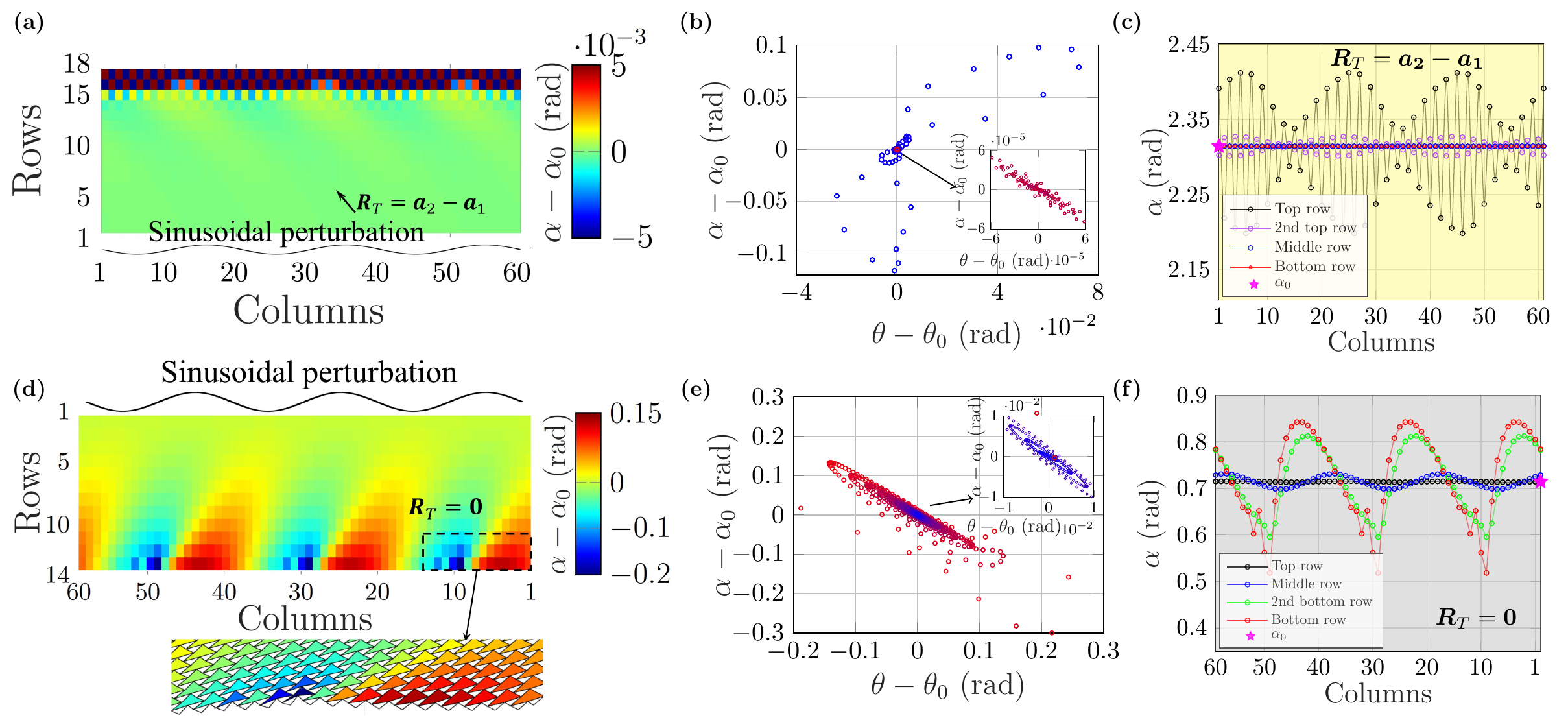}%
\caption{\textbf{Nonlinear wave amplification and frequency conversion in Maxwell lattices in response to sinusoidal perturbation}. (a-c) Hard edge perturbation (from the bottom) of an $\mathbf{a_2-a_1}$ polarized lattice ($\alpha_0=2.3144$ rad), where $k_x=0.314$ rad/unit cell and $\varepsilon=1$ $\mu$rad. (d-f) Perturbation (from the top) of an unpolarized lattice with $\alpha_0=0.7144$ rad, with $k_x=0.314$ rad/unit cell and $\varepsilon=1$ mrad. (a,d) Deformation field. The configuration of the lattice in the dashed box is shown below the plot. (b,e) Phase space of (a,d), where blue to red color gradient denotes a progression from top to bottom rows, respectively. The insets provide a zoomed-in view near the perturbation. (c,f) Select rows of (a,d). The pink stars in (c,f) denote the initial homogeneous angles, and the background shadings denote the polarization regions.} 
\label{Rt0_a2a1_polarization_wave}
\end{figure}

The second example, amplification enhanced frequency conversion, occurs when the lattice is excited from the hard edge of a polarized lattice, or either edge of a non-polarized lattice. In the case of perturbing from the hard edge, following the linear theory, we expect the perturbation to project to two ZMs which both grow exponentially into the bulk. In the context of our analogy with the 1D, non-reciprocal, non-Hermitian system, this would map to either an active system evolving forwards in time, or a damped system evolving backwards in time. In Fig.~\ref{Rt0_a2a1_polarization_wave}(a-c), we show the calculation for the sinusoidal perturbation of a lattice deep in the $\mathbf{a_2-a_1}$ domain. As shown in Fig.~\ref{Rt0_a2a1_polarization_wave}(a), the deformation field amplifies into the bulk. We stop the calculation at 18 rows, after which a zero strain solution cannot be found, due to overlap or non-connection of the triangles, such that the lattice is ``broken''. To better illustrate the  "time evolution" of the mode, we show the trajectory of the lattice in the phase space of the angles.  In Fig.~\ref{Rt0_a2a1_polarization_wave}(a,c) the growth in amplitude can be seen to be accompanied by the generation of higher frequency wave components, which is connected to the blue loops in the angular phase space plot of Fig.~\ref{Rt0_a2a1_polarization_wave}(b). 

The other case where the ZM amplifies is when an non-polarized lattice is perturbed from either edge. In this case, in the linear theory, one ZM grows and the other decays. A generic perturbation projects to both ZMs and the growing one is observed far from the edge. Here we study a sinusoidal perturbation on a lattice deep in the $\mathbf{R}_T=0$ domain. As mentioned above, linear ZMs in this lattice are described by conformal transformations. In Fig.~\ref{Rt0_a2a1_polarization_wave}(d-f), using the same conventions as Fig.~\ref{Rt0_a2a1_polarization_wave}(a-c), we show the deformation of the perturbed lattice. In contrast to the hyperbolic case of hard-edge perturbation, in Fig.~\ref{Rt0_a2a1_polarization_wave}(d,f), we see the formation of ``kinks''. At linear order, these kinks can be understood as a signature of conformal transformations, which have a one-to-one correspondence with complex analytic functions. All analytic functions periodic in $x$ can be expanded in the basis of $e^{ikz}$, which features these kinks. Higher order terms, both in $u$ and in derivatives, lead to further complex features of these kinks. Such kink formation may find future use in applications that can take advantage of deformation amplification or stress concentration. In the SI, we show further examples for sinusoidal perturbation of $\mathbf{R}_T=0$ and $\mathbf{a_2-a_1}$ lattices that are closer to the polarization boundaries, wherein domain switching can be observed. 

The third---and perhaps the most intriguing---example, solitary waves, occur when the lattice is subject to localized perturbations. Typically described as localized waves that maintain their shape as they propagate with constant, often amplitude-dependent, speed and shape, solitary waves are one of the most canonical phenomena that emerge from nonlinear systems \cite{chen2012optical,dauxois2006physics,scott2006encyclopedia}. Herein, we distinguish solitary waves from the more restrictive localized type of wave referred to as ``solitons'', which  ``reappear virtually unaffected in size or shape'' following collisions \cite{zabusky1965interaction}. While solitary waves are most commonly considered in conservative systems, they have also been studied in a wide range of non-conservative (i.e. non-Hermitian) systems \cite{el2018non,chen2012optical,patil2021review,chong2018coherent,scott2006encyclopedia,cross1993pattern,akhmediev2008dissipative}. 

\begin{figure}[h!]
\centering
  \includegraphics[width=6.8in]{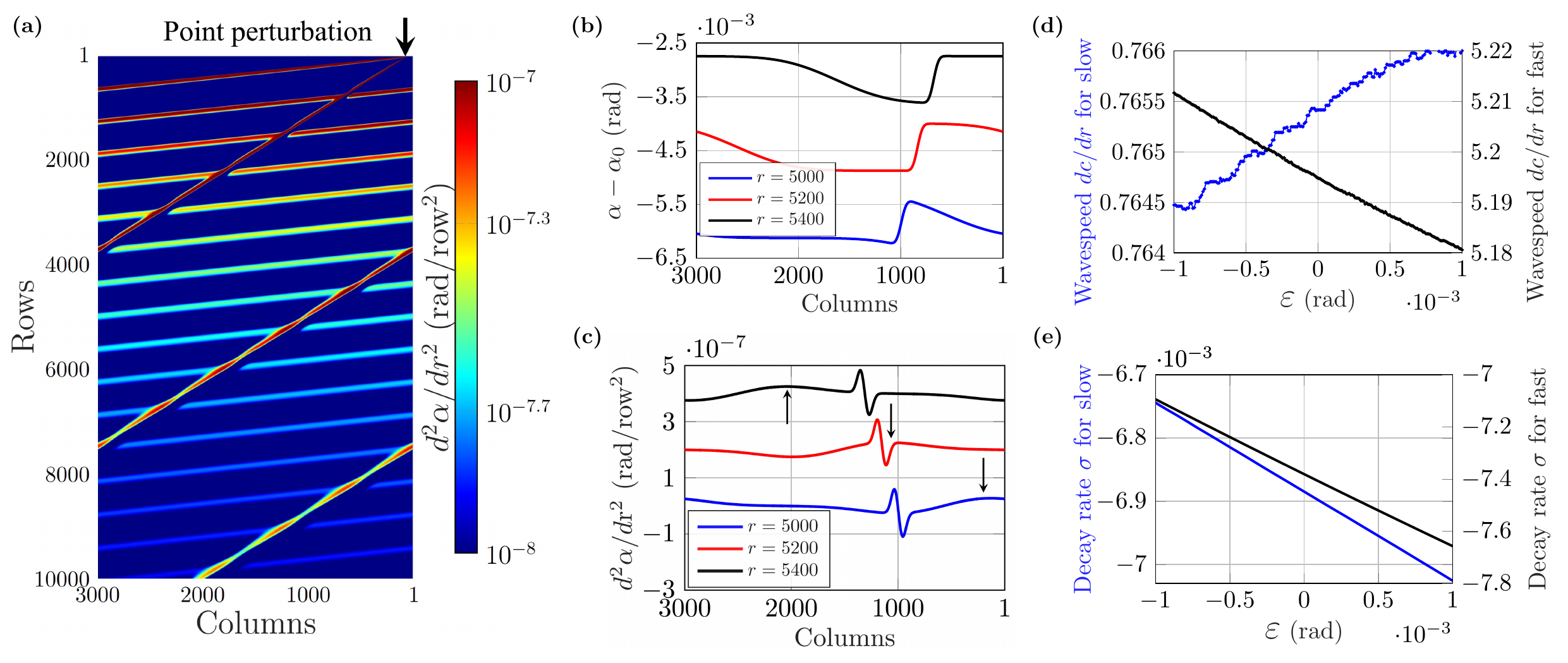}%
\caption{\textbf{Formation of two solitary waves resulting from a point perturbation at the edge of an $a_2$ polarized Maxwell lattice and their collisions}. The lattice has $\alpha_0=1.3144$ rad, and the point perturbation is applied to the top row at column 50. (a-c) Perturbation amplitude $\varepsilon=0.23$ mrad. (a) $d^2\alpha/dr^2$ as a function of space. (b) $\alpha-\alpha_0$ and (c) $d^2\alpha/dr^2$ for rows 5000 (blue), 5200 (red), and 5400 (black). Rows 5200 and 5400 are sequentially offset by 1.25 mrad in (b) and 0.2 $\mu$rad/(unit cell)$^2$ for visualization purposes. The black arrows point to the fast moving, spatially wider, solitary wave. (d) ``Speed'' ${dc}/{dr}$ of the solitary waves as a function of $\varepsilon$. (e) Decay rate of the peak-to-peak magnitude of $d^2\alpha/dr^2$ of the solitary waves $\sigma$, defined  $\frac{d^2\alpha}{dr^2}_{max}-\frac{d^2\alpha}{dr^2}_{min}=Ae^{\sigma r}$.
} \label{a2_soliton_wave}
\end{figure}

\begin{figure}[h!]
\centering
  \includegraphics[width=2.5in]{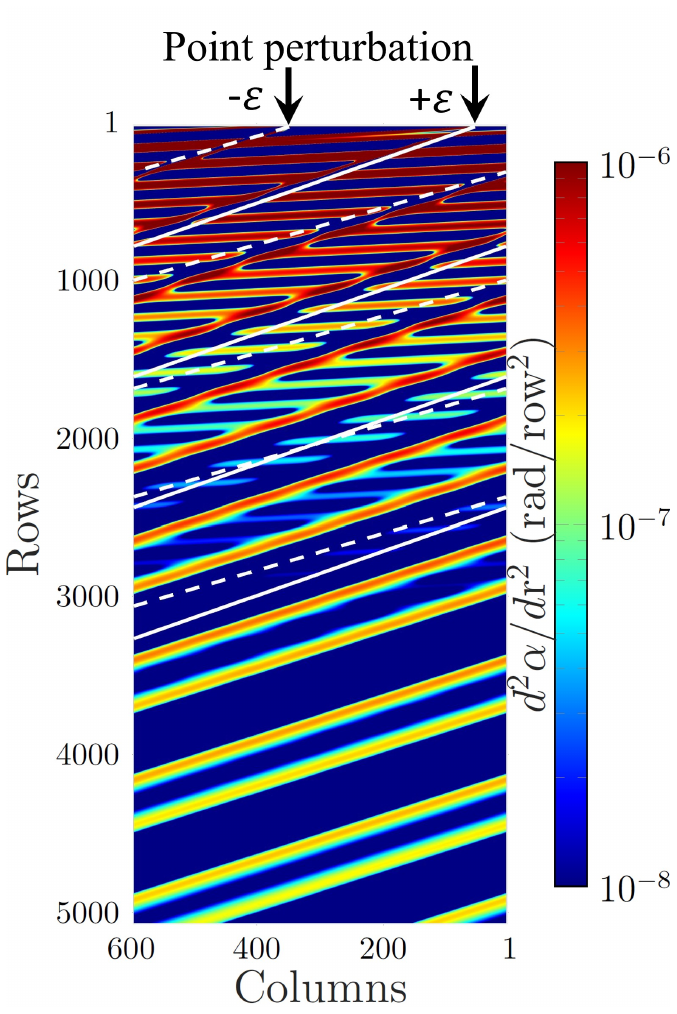}%
\caption{\textbf{Repulsion between two solitary waves, visualized via $\mathbf{d^2\alpha/dr^2}$, for two point perturbations applied to the soft edge (top) of a lattice with the same homogeneous angle as in Fig.~\ref{a2_soliton_wave}}. Perturbations of $\pm \varepsilon=0.6$ mrad are applied at columns 50 and 350, respectively. The solid and dashed lines demonstrate the predicted behavior of their perspective perturbations, respectively, based on the speeds calculated in the SI.
} \label{a2_two_soliton_wave}
\end{figure}

To explore the possibility of such localized traveling structures in our system, we induce a point perturbation at the top of a lattice with the same homogeneous configuration as studied in Fig.~\ref{a2_polarization_wave}, and sweep a range of perturbation amplitudes, where $f(j)=\varepsilon \delta{(j-j_p)}$ and $j_p$ is column to which the perturbation is applied. As can be seen in Fig.~\ref{a2_soliton_wave}(a-c), using the language of the 1D spatiotemporal system analogy, we see two main solitary waves emerge with nearly constant speeds and interact with each other: one with a fast wave-speed and decay rate, and the other with a slow wave-speed and decay rate. In Fig.~\ref{a2_soliton_wave} and below, we use $r$ and $c$ to denote rows and columns, respectively, in the derivative terms. Figure~\ref{a2_soliton_wave}(d,e) shows the dependence of speed and decay rate of the two solitary waves on perturbation amplitude. At first glance, it appears that there is a minimal interaction between the two solitary waves as they intersect. However, additional calculations for the same conditions as described in Fig.~\ref{a2_soliton_wave}, but with five times fewer columns, and thus more collisions between the waves (see the SI), shows a significantly greater variation in the speeds and decay rates of the solitary waves. This suggests that the two solitary waves do interact upon their collisions. Additional data for the peak-to-peak decay rate in terms of $\alpha$ and the evolution of average $\alpha$ with increasing row number is included in the SI. Interestingly, both the narrow and wide lattices have perturbation amplitudes for which the decay rate of $\alpha$ is zero across the sampled rows (in contrast to $d^2\alpha/dr^2$), which is reminiscent of solitary waves in non-conservative systems, where nonlinearity, dispersion, and loss/gain balance to form a traveling wave packet of constant shape \cite{el2018non,chen2012optical,scott2006encyclopedia,cross1993pattern,akhmediev2008dissipative}. Such slow decay suggests that these waves can be considered analogs to weakly dissipative solitary waves \cite{chong2018coherent, akhmediev2008dissipative}. 

Augmenting the complexity of the two solitary waves generated from a single point perturbation in the examples of Fig.~\ref{a2_soliton_wave}, we simulate the response of the TMM to two point perturbations of differing signs. As can be expected from the prior results, four solitary waves are generated, however, here we see the unexpected phenomenon that the two long-lived solitary waves appear to repel one another and propagate with similar speeds thereafter. This change of behavior can be seen by comparing to their predicted intersection is denoted by the white solid and dashed lines, which use the wave-speeds identified from the single perturbation case in the SI. Such repulsion has been seen for solitary waves in other nonlinear systems, for instance that of two kinks or two anti-kinks (topological solitons) in Sine-Gordon systems \cite{dauxois2006physics,gao2020underwater} or optical spatial solitons \cite{stegeman1999optical}.

%%%%%%%%%%%%%%%%%%%%%%%%%%%%%%%%%%%%%%%%%%%%%
\subsection*{Experimental validation}
\begin{figure}[h!]
\centering
  \includegraphics[width=6.8in]{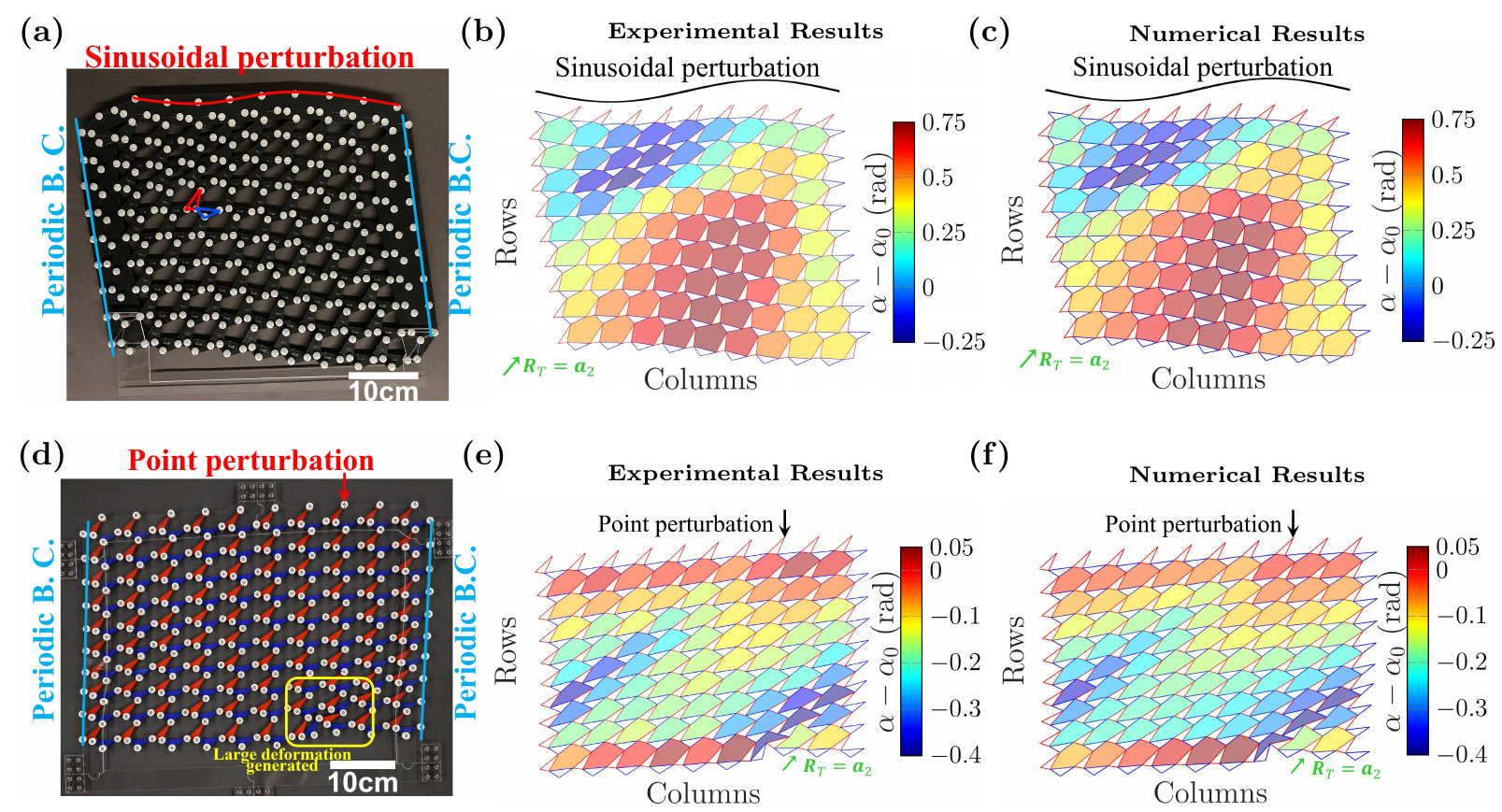}%
\caption{\textbf{Physical realization of $\mathbf{a_2}$ polarized Maxwell lattices with laser cut triangles, pinned hinges, and three prescribed boundaries, along with comparison to numerical predictions.} (a-c) Sinusoidal perturbation for $\alpha_0=1.3144$, $\varepsilon=0.1$ rad, $k_x=0.6283$ rad/unit cell. (d-f) Point perturbation for $\alpha_0=1.3144$ rad, $\varepsilon=45$ mrad applied at column 3. (a,d) Photographs of the deformed lattice, where the left and right boundaries are prescribed to follow the computed periodic boundary configuration, (b,e) measured angles, and (c,f) simulated angles.}\label{experiments1}
\end{figure}

\noindent To validate the numerical simulation results, we built physical models composed of laser cut acrylic triangles pinned together such that they are free to rotate. Further details of the experimental setup are included in the SI. In summary, boundary conditions are set by pinning the edges to the prescribed periodic angles found from the simulation on the left and right, and to the top edge where the perturbation was initiated. The angles of each triangle in the resulting deformed configuration are then measured by using image processing to identify the position of each pinned hinge. In Fig.~\ref{experiments1}, we show two lattice configurations with different sized unit cells under two different perturbations. The experimental lattices closely match the numerical predictions (exact error calculations given in the SI). We note that in our experimental realization for the configuration showing the solitary wave propagation Fig.~\ref{experiments1}(d-f), multiple unit cells with $\theta$ angles close to $\pi$ with sufficient pressure can be forced to snap from concave to convex configurations with the given prescribed boundaries, as well as slight variability in the experimental configuration due to manufacturing tolerances of the hinges.

%%%%%%%%%%%%%%%%%%%%%%%%%%%%%%%
\section*{Discussion}

\noindent The two central contributions of this work are: i) demonstrating, to the linear order, that the ZMs of 2D TMMs (Maxwell lattices) can be mapped to waves in 1D non-Hermitian, non-reciprocal, dynamical systems; and ii) extending the study of 2D TMMs to the nonlinear regime. Within that scope, we showed that an array of nonlinear wave-like phenomena exists, including harmonic generation, localized topological domain switching, amplification enhanced frequency conversion, and solitary wave generation. Each of these phenomena has its own unique implications for designing stress and deformation fields in materials that extends significantly past what has hitherto been achievable in linear regimes of TMMs, and, more broadly, via elasticity. Amongst these, localized intrinsic domain formation has the potential to tailor regions of SSSs with implications to fracture mechanics \cite{zhang2018fracturing}, and solitary wave generation has implications for compact, amplitude-dependent spatially-addressable signal transmission \cite{garbin2015topological} and remote-controlled localization of stress and deformation, both of which may find interesting use in the context of mechano-responsive metamaterials \cite{ghanem2021role}. In the context of the analogy to the 1D spatiotemporal system, these 2D lattices offer a convenient emulator for nonlinear waves and shares similarities with the mapping between $d$-dimensional quantum systems with $(d+1)$-dimensional classical systems, which led to important advances in understanding quantum phase transitions~\cite{Sondhi1997}. Finally, we envision a potentially intriguing scenario stemming from this work, wherein elasticity and inertial effects are incorporated into the lattice, such that perturbations are restricted by the underlying topology of the lattice as shown herein, but evolve in time.

%%%%%%%%%%%%%%%%%%%%%%%%%%%%%%%
\section*{Materials and Methods}
\noindent %The experimental lattices were fabricated using a Glowforge Basic laser cutter and 1/8" thick acrylic, and then pinned together. Each boundary condition was created by taking the positions at the edge with the applied perturbation and the two periodic boundaries for the simulated results and creating a DXF image of the hinge points. Photographs were taken and MATLAB R2021a was used to post process images. The hinge points were found using the imfindcircles function, which allows the lattice to be reconstructed and analyzed. The lattices with triangles composed of black (red and blue) acrylic

%The unit cell coordinates are numerically solved and post-processed in MATLAB R2021a using a the strain free geometric relationship of the triangles (details of which are in the SI).  

\noindent The fabricated lattice structures are created by laser cutting (using a Glowforge Basic 3D laser cutter) 1/8-inch-thick acrylic layers assembled with barrels (Fig.7(a)) or dowel pins (Fig.7(d)) and screws. For the experiment in Fig.7(a), each of the barrels (6063 Aluminum Low-Profile Binding Barrels from McMaster-Carr) has a diameter of 13/64 inch and length of 1/4 inch (3/8 inch at the boundaries due to an extra layer of acrylic for prescribed periodic left-right boundary condition). For the experiment in Fig.7(d), each of the dowel pins (Alloy Steel Pull-Out Dowel Pin from McMaster-Carr) has a diameter of 1/4 inch and a length of 1/2 inch. Geometries of triangle unit cells are shown in the SI. The lattices are assembled by pinning down the triangles in two layers and to the laser cut acrylic boundaries. Left and right boundaries are connected at the bottom without interfering with the bottom row of triangles mitigate bending of the boundary. Boundary pieces are connected using M3 screws and nuts.

%%%%%%%%%%%%%%%%%%%%%%%%%%%%%%%%%%%%%%%%%%%%%%%

\bibliography{scibib}

\begin{thebibliography}{10}

\bibitem{hasan2010colloquium}
M.~Z. Hasan, C.~L. Kane, Colloquium: topological insulators.
\newblock {\it Reviews of modern physics\/} {\bf 82}, 3045 (2010).

\bibitem{zangeneh2020topological}
F.~Zangeneh-Nejad, A.~Al{\`u}, R.~Fleury, Topological wave insulators: a
  review.
\newblock {\it Comptes Rendus. Physique\/} {\bf 21}, 467--499 (2020).

\bibitem{qi2011topological}
X.-L. Qi, S.-C. Zhang, Topological insulators and superconductors.
\newblock {\it Rev. Mod. Phys.\/} {\bf 83}, 1057--1110 (2011).

\bibitem{asboth2016short}
J.~K. Asb{\'o}th, L.~Oroszl{\'a}ny, A.~P{\'a}lyi, A short course on topological
  insulators: band structure and edge states in one and two dimensions.
\newblock {\it Lecture notes in physics\/} {\bf 919}, 166 (2016).

\bibitem{beenakker2016road}
C.~Beenakker, L.~Kouwenhoven, A road to reality with topological
  superconductors.
\newblock {\it Nature Physics\/} {\bf 12}, 618--621 (2016).

\bibitem{aasen2016milestones}
D.~Aasen, M.~Hell, R.~V. Mishmash, A.~Higginbotham, J.~Danon, M.~Leijnse, T.~S.
  Jespersen, J.~A. Folk, C.~M. Marcus, K.~Flensberg, {\it et~al.\/}, Milestones
  toward majorana-based quantum computing.
\newblock {\it Physical Review X\/} {\bf 6}, 031016 (2016).

\bibitem{PhysRevB.102.125407}
C.~Tutschku, R.~W. Reinthaler, C.~Lei, A.~H. MacDonald, E.~M. Hankiewicz,
  Majorana-based quantum computing in nanowire devices.
\newblock {\it Phys. Rev. B\/} {\bf 102}, 125407 (2020).

\bibitem{gilbert2021topological}
M.~J. Gilbert, Topological electronics.
\newblock {\it Communications Physics\/} {\bf 4}, 1--12 (2021).

\bibitem{chorsi2022topological}
H.~Chorsi, B.~Cheng, B.~Zhao, J.~Toudert, V.~Asadchy, O.~F. Shoron, S.~Fan,
  R.~Matsunaga, Topological materials for functional optoelectronic devices.
\newblock {\it Advanced Functional Materials\/} p. 2110655 (2022).

\bibitem{yang2015topological}
Z.~Yang, F.~Gao, X.~Shi, X.~Lin, Z.~Gao, Y.~Chong, B.~Zhang, Topological
  acoustics.
\newblock {\it Physical review letters\/} {\bf 114}, 114301 (2015).

\bibitem{xin2020topological}
X.~Li, S.~Yu, H.~Liu, M.~Lu, Y.~Chen, Topological mechanical metamaterials: A
  brief review.
\newblock {\it Current Opinion in Solid State and Materials Science\/} {\bf
  24}, 100853 (2020).

\bibitem{huber2016topological}
S.~D. Huber, Topological mechanics.
\newblock {\it Nature Physics\/} {\bf 12}, 621--623 (2016).

\bibitem{bertoldi2017flexible}
K.~Bertoldi, V.~Vitelli, J.~Christensen, M.~Van~Hecke, Flexible mechanical
  metamaterials.
\newblock {\it Nature Reviews Materials\/} {\bf 2}, 1--11 (2017).

\bibitem{singh2021design}
N.~Singh, M.~van Hecke, Design of pseudo-mechanisms and multistable units for
  mechanical metamaterials.
\newblock {\it Physical Review Letters\/} {\bf 126}, 248002 (2021).

\bibitem{kane2014topological}
C.~Kane, T.~Lubensky, Topological boundary modes in isostatic lattices.
\newblock {\it Nature Physics\/} {\bf 10}, 39--45 (2014).

\bibitem{mao2018maxwell}
X.~Mao, T.~C. Lubensky, Maxwell lattices and topological mechanics.
\newblock {\it Annual Review of Condensed Matter Physics\/} {\bf 9}, 413--433
  (2018).

\bibitem{Su1979}
W.~P. Su, J.~R. Schrieffer, A.~J. Heeger, Solitons in polyacetylene.
\newblock {\it Phys. Rev. Lett.\/} {\bf 42}, 1698--1701 (1979).

\bibitem{rocklin2017transformable}
D.~Rocklin, S.~Zhou, K.~Sun, X.~Mao, Transformable topological mechanical
  metamaterials.
\newblock {\it Nature communications\/} {\bf 8}, 1--9 (2017).

\bibitem{paulose2015topological}
J.~Paulose, B.~G.-g. Chen, V.~Vitelli, Topological modes bound to dislocations
  in mechanical metamaterials.
\newblock {\it Nature Physics\/} {\bf 11}, 153--156 (2015).

\bibitem{sun2012surface}
K.~Sun, A.~Souslov, X.~Mao, T.~Lubensky, Surface phonons, elastic response, and
  conformal invariance in twisted kagome lattices.
\newblock {\it Proceedings of the National Academy of Sciences\/} {\bf 109},
  12369--12374 (2012).

\bibitem{lubensky2015phonons}
T.~Lubensky, C.~Kane, X.~Mao, A.~Souslov, K.~Sun, Phonons and elasticity in
  critically coordinated lattices.
\newblock {\it Reports on Progress in Physics\/} {\bf 78}, 073901 (2015).

\bibitem{zhang2021hierarchical}
H.~Zhang, J.~Wu, D.~Fang, Y.~Zhang, Hierarchical mechanical metamaterials built
  with scalable tristable elements for ternary logic operation and amplitude
  modulation.
\newblock {\it Science Advances\/} {\bf 7}, eabf1966 (2021).

\bibitem{zhou2020switchable}
D.~Zhou, J.~Ma, K.~Sun, S.~Gonella, X.~Mao, Switchable phonon diodes using
  nonlinear topological maxwell lattices.
\newblock {\it Physical Review B\/} {\bf 101}, 104106 (2020).

\bibitem{silverberg2015origami}
J.~L. Silverberg, J.-H. Na, A.~A. Evans, B.~Liu, T.~C. Hull, C.~D. Santangelo,
  R.~J. Lang, R.~C. Hayward, I.~Cohen, Origami structures with a critical
  transition to bistability arising from hidden degrees of freedom.
\newblock {\it Nature materials\/} {\bf 14}, 389--393 (2015).

\bibitem{zunker2021soft}
W.~Zunker, S.~Gonella, Soft topological lattice wheels.
\newblock {\it Extreme Mechanics Letters\/} {\bf 46}, 101344 (2021).

\bibitem{mcinerney2020origami}
J.~McInerney, B.~G. ge~Chen, L.~Theran, C.~D. Santangelo, D.~Z. Rocklin, Hidden
  symmetries generate rigid folding mechanisms in periodic origami.
\newblock {\it Proceedings of the National Academy of Sciences\/} {\bf 117},
  30252-30259 (2020).

\bibitem{bossart2021oligomodal}
A.~Bossart, D.~M. Dykstra, J.~Van~der Laan, C.~Coulais, Oligomodal
  metamaterials with multifunctional mechanics.
\newblock {\it Proceedings of the National Academy of Sciences\/} {\bf 118},
  e2018610118 (2021).

\bibitem{bilal2017intrinsically}
O.~R. Bilal, R.~S{\"u}sstrunk, C.~Daraio, S.~D. Huber, Intrinsically polar
  elastic metamaterials.
\newblock {\it Advanced Materials\/} {\bf 29}, 1700540 (2017).

\bibitem{rocklin2017directional}
D.~Z. Rocklin, Directional mechanical response in the bulk of topological
  metamaterials.
\newblock {\it New Journal of Physics\/} {\bf 19}, 065004 (2017).

\bibitem{chen2014nonlinear}
B.~G.-g. Chen, N.~Upadhyaya, V.~Vitelli, Nonlinear conduction via solitons in a
  topological mechanical insulator.
\newblock {\it Proceedings of the National Academy of Sciences\/} {\bf 111},
  13004--13009 (2014).

\bibitem{lo2021topology}
P.~W. Lo, C.~D. Santangelo, B.~G.-g. Chen, C.-M. Jian, K.~Roychowdhury, M.~J.
  Lawler, Topology in nonlinear mechanical systems.
\newblock {\it Physical Review Letters\/} {\bf 127}, 076802 (2021).

\bibitem{patil2021review}
G.~U. Patil, K.~H. Matlack, Review of exploiting nonlinearity in phononic
  materials to enable nonlinear wave responses.
\newblock {\it Acta Mechanica\/} {\bf 233}, 1--46 (2021).

\bibitem{lapine2014colloquium}
M.~Lapine, I.~V. Shadrivov, Y.~S. Kivshar, Colloquium: nonlinear metamaterials.
\newblock {\it Reviews of Modern Physics\/} {\bf 86}, 1093 (2014).

\bibitem{scott2006encyclopedia}
A.~Scott, {\it Encyclopedia of nonlinear science\/} (Routledge, 2006).

\bibitem{chen2012optical}
Z.~Chen, M.~Segev, D.~N. Christodoulides, Optical spatial solitons: historical
  overview and recent advances.
\newblock {\it Reports on Progress in Physics\/} {\bf 75}, 086401 (2012).

\bibitem{dauxois2006physics}
T.~Dauxois, M.~Peyrard, {\it Physics of solitons\/} (Cambridge University
  Press, 2006).

\bibitem{boyd2020nonlinear}
R.~W. Boyd, {\it Nonlinear optics\/} (Academic press, 2020).

\bibitem{nayfeh2008nonlinear}
A.~H. Nayfeh, D.~T. Mook, {\it Nonlinear oscillations\/} (John Wiley \& Sons,
  2008).

\bibitem{strogatz2018nonlinear}
S.~H. Strogatz, {\it Nonlinear dynamics and chaos: with applications to
  physics, biology, chemistry, and engineering\/} (CRC press, 2018).

\bibitem{manda2021nonlinear}
B.~M. Manda, R.~Chaunsali, G.~Theocharis, C.~Skokos, Nonlinear topological edge
  states: from dynamic delocalization to thermalization.
\newblock {\it arXiv preprint arXiv:2112.12997\/}  (2021).

\bibitem{smirnova2020nonlinear}
D.~Smirnova, D.~Leykam, Y.~Chong, Y.~Kivshar, Nonlinear topological photonics.
\newblock {\it Applied Physics Reviews\/} {\bf 7}, 021306 (2020).

\bibitem{chaunsali2021stability}
R.~Chaunsali, H.~Xu, J.~Yang, P.~G. Kevrekidis, G.~Theocharis, Stability of
  topological edge states under strong nonlinear effects.
\newblock {\it Physical Review B\/} {\bf 103}, 024106 (2021).

\bibitem{tempelman2021topological}
J.~R. Tempelman, K.~H. Matlack, A.~F. Vakakis, Topological protection in a
  strongly nonlinear interface lattice.
\newblock {\it Physical Review B\/} {\bf 104}, 174306 (2021).

\bibitem{pernet2022gap}
N.~Pernet, P.~St-Jean, D.~D. Solnyshkov, G.~Malpuech, N.~Carlon~Zambon,
  Q.~Fontaine, B.~Real, O.~Jamadi, A.~Lema{\^\i}tre, M.~Morassi, {\it
  et~al.\/}, Gap solitons in a one-dimensional driven-dissipative topological
  lattice.
\newblock {\it Nature Physics\/} pp. 1--7 (2022).

\bibitem{zhou2022topological}
D.~Zhou, D.~Rocklin, M.~Leamy, Y.~Yao, Topological invariant and anomalous edge
  modes of strongly nonlinear systems.
\newblock {\it Nature Communications\/} {\bf 13}, 1--9 (2022).

\bibitem{nesterenko2013dynamics}
V.~Nesterenko, {\it Dynamics of heterogeneous materials\/} (Springer Science \&
  Business Media, 2013).

\bibitem{markovic2020physics}
D.~Markovi{\'c}, A.~Mizrahi, D.~Querlioz, J.~Grollier, Physics for neuromorphic
  computing.
\newblock {\it Nature Reviews Physics\/} {\bf 2}, 499--510 (2020).

\bibitem{raja2021ultrafast}
A.~S. Raja, S.~Lange, M.~Karpov, K.~Shi, X.~Fu, R.~Behrendt, D.~Cletheroe,
  A.~Lukashchuk, I.~Haller, F.~Karinou, {\it et~al.\/}, Ultrafast optical
  circuit switching for data centers using integrated soliton microcombs.
\newblock {\it Nature communications\/} {\bf 12}, 1--7 (2021).

\bibitem{kippenberg2018dissipative}
T.~J. Kippenberg, A.~L. Gaeta, M.~Lipson, M.~L. Gorodetsky, Dissipative kerr
  solitons in optical microresonators.
\newblock {\it Science\/} {\bf 361}, eaan8083 (2018).

\bibitem{el2018non}
R.~El-Ganainy, K.~G. Makris, M.~Khajavikhan, Z.~H. Musslimani, S.~Rotter, D.~N.
  Christodoulides, Non-hermitian physics and pt symmetry.
\newblock {\it Nature Physics\/} {\bf 14}, 11--19 (2018).

\bibitem{ashida2020non}
Y.~Ashida, Z.~Gong, M.~Ueda, Non-hermitian physics.
\newblock {\it Advances in Physics\/} {\bf 69}, 249--435 (2020).

\bibitem{gennes1968soluble}
P.-G.~d. Gennes, Soluble model for fibrous structures with steric constraints.
\newblock {\it The Journal of Chemical Physics\/} {\bf 48}, 2257--2259 (1968).

\bibitem{Sondhi1997}
S.~L. Sondhi, S.~M. Girvin, J.~P. Carini, D.~Shahar, Continuous quantum phase
  transitions.
\newblock {\it Rev. Mod. Phys.\/} {\bf 69}, 315--333 (1997).

\bibitem{kolner1994space}
B.~H. Kolner, Space-time duality and the theory of temporal imaging.
\newblock {\it IEEE Journal of Quantum Electronics\/} {\bf 30}, 1951--1963
  (1994).

\bibitem{wilczek2012quantum}
F.~Wilczek, Quantum time crystals.
\newblock {\it Physical review letters\/} {\bf 109}, 160401 (2012).

\bibitem{chong2018coherent}
C.~Chong, P.~G. Kevrekidis, {\it Coherent structures in granular crystals: From
  experiment and modelling to computation and mathematical analysis\/}
  (Springer, 2018).

\bibitem{xia2021nonlinear}
S.~Xia, D.~Kaltsas, D.~Song, I.~Komis, J.~Xu, A.~Szameit, H.~Buljan, K.~G.
  Makris, Z.~Chen, Nonlinear tuning of pt symmetry and non-hermitian
  topological states.
\newblock {\it Science\/} {\bf 372}, 72--76 (2021).

\bibitem{cross1993pattern}
M.~C. Cross, P.~C. Hohenberg, Pattern formation outside of equilibrium.
\newblock {\it Reviews of modern physics\/} {\bf 65}, 851 (1993).

\bibitem{akhmediev2008dissipative}
N.~Akhmediev, A.~Ankiewicz, {\it Dissipative solitons: from optics to biology
  and medicine\/}, vol. 751 (Springer Science \& Business Media, 2008).

\bibitem{calladine1978buckminster}
C.~R. Calladine, Buckminster fuller's “tensegrity” structures and clerk
  maxwell's rules for the construction of stiff frames.
\newblock {\it International journal of solids and structures\/} {\bf 14},
  161--172 (1978).

\bibitem{guest2003determinacy}
S.~Guest, J.~Hutchinson, On the determinacy of repetitive structures.
\newblock {\it Journal of the Mechanics and Physics of Solids\/} {\bf 51},
  383--391 (2003).

\bibitem{zheng2022continuum}
Y.~Zheng, I.~Niloy, P.~Celli, I.~Tobasco, P.~Plucinsky, Continuum field theory
  for the deformations of planar kirigami.
\newblock {\it Physical Review Letters\/} {\bf 128}, 208003 (2022).

\bibitem{PhysRevE.99.013002}
M.~Moshe, E.~Esposito, S.~Shankar, B.~Bircan, I.~Cohen, D.~R. Nelson, M.~J.
  Bowick, Nonlinear mechanics of thin frames.
\newblock {\it Phys. Rev. E\/} {\bf 99}, 013002 (2019).

\bibitem{czajkowski2022conformal}
M.~Czajkowski, C.~Coulais, M.~van Hecke, D.~Rocklin, Conformal elasticity of
  mechanism-based metamaterials.
\newblock {\it Nature Communications\/} {\bf 13}, 1--9 (2022).

\bibitem{zheng2022modeling}
Y.~Zheng, I.~Tobasco, P.~Celli, P.~Plucinsky, Modeling planar kirigami
  metamaterials as generalized elastic continua.
\newblock {\it Soft Condensed Matter, arXiv preprint arXiv:2206.00153\/}
  (2022).

\bibitem{czajkowski2022duality}
M.~Czajkowski, D.~Rocklin, Duality and sheared analytic response in
  mechanism-based metamaterials.
\newblock {\it Soft Condensed Matter, arXiv preprint arXiv:2205.10751\/}
  (2022).

\bibitem{zhang2018fracturing}
L.~Zhang, X.~Mao, Fracturing of topological maxwell lattices.
\newblock {\it New Journal of Physics\/} {\bf 20}, 063034 (2018).

\bibitem{zabusky1965interaction}
N.~J. Zabusky, M.~D. Kruskal, Interaction of" solitons" in a collisionless
  plasma and the recurrence of initial states.
\newblock {\it Physical review letters\/} {\bf 15}, 240 (1965).

\bibitem{gao2020underwater}
N.~Gao, K.~Lu, An underwater metamaterial for broadband acoustic absorption at
  low frequency.
\newblock {\it Applied Acoustics\/} {\bf 169}, 107500 (2020).

\bibitem{stegeman1999optical}
G.~I. Stegeman, M.~Segev, Optical spatial solitons and their interactions:
  universality and diversity.
\newblock {\it Science\/} {\bf 286}, 1518--1523 (1999).

\bibitem{garbin2015topological}
B.~Garbin, J.~Javaloyes, G.~Tissoni, S.~Barland, Topological solitons as
  addressable phase bits in a driven laser.
\newblock {\it Nature communications\/} {\bf 6}, 1--7 (2015).

\bibitem{ghanem2021role}
M.~A. Ghanem, A.~Basu, R.~Behrou, N.~Boechler, A.~J. Boydston, S.~L. Craig,
  Y.~Lin, B.~E. Lynde, A.~Nelson, H.~Shen, {\it et~al.\/}, The role of polymer
  mechanochemistry in responsive materials and additive manufacturing.
\newblock {\it Nature Reviews Materials\/} {\bf 6}, 84--98 (2021).

\end{thebibliography}
\bibliographystyle{ScienceAdvances}

\noindent \textbf{Acknowledgements:} 
The authors thank N. Gravish for useful discussions regarding active systems. \\

\noindent \textbf{Funding:} H.L., K.Q., S.S., N.B., and X.M. acknowledge support from the US Army Research Office (Grant No. W911NF-20-2-0182). I.F. acknowledges support from the Department of Defense (DoD) through the National Defense Science \& Engineering Graduate (NDSEG) Fellowship Program. B.M. acknowledges support from the U.S. Department of Energy (DOE) National Nuclear Security Administration (NNSA) Laboratory Graduate Residency Fellowship (LRGF) under Cooperative Agreement DE-NA0003960. H.X., Z.C., and X.M. acknowledge support from the Office of Naval Research (Grant No. ONR MURI N00014-20-1-2479). \\

\noindent \textbf{Author Contributions:} Each author’s contribution(s) to the paper are listed below:

Conceptualization: ZC, NB, XM

Methodology: HX, IF, HL, SS, BCM
	
Investigation: HX, IF, HL, KQ, SS, BCM, ZC, NB, XM

Experiment realization: IF, KQ
	
%Visualization: HX, IF, HL
	
%Supervision: ZC, NB, XM
	
%Writing—original draft: HX, IF, HL
	
Writing—review \& editing: HX, IF, HL, KQ, SS, BCM, ZC, NB, XM\\

\noindent \textbf{Competing Interests:} The authors declare that they have no competing interests.\\

\noindent \textbf{Data and materials availability:} All data are available on the request to the authors.
%in the main text or the supplementary materials.

%The data presented herein is stored on \nb{TBD online repository}.\\

%Here you should list the contents of your Supplementary Materials -- below is an example. 
%You should include a list of Supplementary figures, Tables, and any references that appear only in the SM. 
%Note that the reference numbering continues from the main text to the SM.
% In the example below, Refs. 4-10 were cited only in the SM.     
\section*{Supplementary materials}
%   Materials and Methods\\
Supplementary Text\\
Figs. S1 to S14\\
% Tables S1 to S4\\
% References \textit{(4-10)}

\end{document}

%\begin{frontmatter}
% \setcounter{page}{19}
\baselineskip16pt

\begin{center} 
{\LARGE{\textbf{Supplementary Information: Synthetically Non-Hermitian Nonlinear Wave-like Behavior in a Topological Mechanical Metamaterial}}}

\vspace{0.1in} 

\end{center}

%%%%%%%%%%%%%%%%%%%%%%%%%%%%%
%%%%%%%%%%%%%%%%%%%%%%%%%%%%%

\section{Floppy Modes of Maxwell Lattice and Control Variables}

\begin{figure}[h!]
\centering
  \includegraphics[height=2.5in]{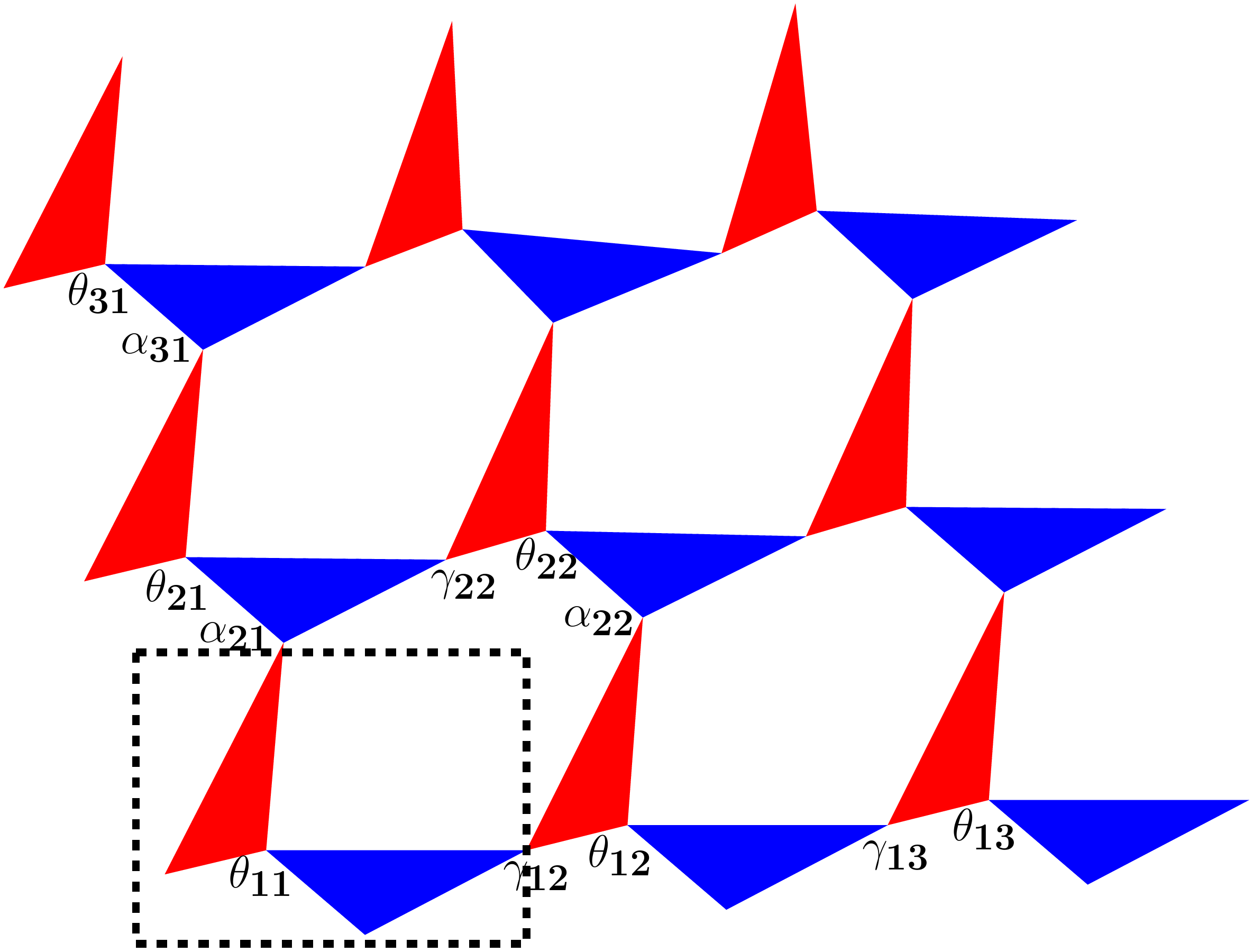}%
\caption{Schematic of a deformed kagome lattice with $3\times 3$ unit cells. Each unit cell consists of a rigid red and blue triangle connected by hinges. The configuration of the lattice is represented by the angles between the triangles, $\theta$, $\alpha$, and $\gamma$.}\label{schematic_deformed_kagome_lattice}
\end{figure}

\noindent For a general mechanical structural frame (a Maxwell lattice herein) in a $d$-dimensional domain that has point masses hinge-connected by central-force bonds, one can apply the Calladine index theorem %(\textit{15, 16, 20, 60, 67}) 
\cite{calladine1978buckminster,kane2014topological,sun2012surface,mao2018maxwell,zhang2018fracturing}\footnote{The numbers of References in the SI are independent of those the main text.}
to count the number of zero energy lattice modes (ZMs). The ZMs are given by
\begin{equation}\label{maxwell_calladine}
  d N-N_b = N_0-N_s,
\end{equation}
where $N_0$ is the number of ZMs, $N_s$ is  the number of states of self-stress (SSSs), $N$ is the number of nodes, and $N_b$ is the number of bonds in the lattice. A periodic Maxwell lattice with no boundaries always yields $dN-N_b=0$, resulting in no ZMs in the lattice unless pairs of SSS and ZM arise due to the geometric singularity, which is the case at topological transitions. However, the generation of ZMs can be achieved by selecting a finite piece from an infinite lattice. In that case, $dN-N_b>0$, and the number of floppy modes are related to the length of the boundary. Considering a two-dimensional (2D) finite Maxwell lattice consisting of $N_x$ columns by $N_y$ rows of unit cells (the dashed box area in Fig.~\ref{schematic_deformed_kagome_lattice} represents a unit cell), the total number of nodes and bonds are $N=3N_xN_y+N_x+N_y$ and $N_b=6N_xN_y$, respectively, and consequently, the number of ZMs is $N_0=2N_x+2N_y+N_s$. Removing the number of rigid body planar degrees of freedom (DOFs), the remaining floppy modes (FMs) of the finite lattice is $2N_x+2N_y+N_s-3$ (note in the main text FMs and ZMs are used almost interchangeable since for a large lattice they are approximately equal to each other and only differ by three rigid body rotations). In other words, by controlling these $2N_x+2N_y-3$ floppy modes (FMs) of a Maxwell lattice with $N_x \times N_y$ unit cells (here $N_s=0$), the configuration of the lattice is fully determined.

A deformed kagome lattice (the type of Maxwell lattice considered herein) is configured by the internal angles of $\theta$, $\alpha$, and $\gamma$ between triangles, each one of which is formed by three bonds. The angle, $\theta$, is the one between the red and blue triangles in the unit cell. The notation of $\alpha$ represents the angle between triangles from different unit cells connected in adjacent rows, while $\gamma$ shows the angle between triangles from different unit cells connected in adjacent columns. The Maxwell-Calladine counting rule, for a deformed kagome lattice with periodic left-right boundaries has $2(N_x+N_y)-3$ floppy modes originally, then periodic boundary conditions give additional constraints on each row (except the bottom row) removing the $2N_y-2$ modes. Additionally connecting the right-most node to the left-most node in the bottom row removes $\theta_{1,N_x}$ as a free variable leaving $2N_x-2$ floppy modes which can be fully defined by choosing $\theta$, and $\gamma$ angles along the bottom. Each $\alpha_{i,1}=\alpha_{i,N_x}$ angle is numerically determined such that periodicity is met (see Fig.~\ref{schematic_deformed_kagome_lattice}). For example, a lattice consisting of $3\times3$ unit cells can be fully determined by 4 variables including $\theta_{11}$, $\theta_{12}$, $\gamma_{12}$, and $\gamma_{13}$. Here it is assumed $\theta_{13}=\theta_{11}$ from periodic boundary conditions. All other angles of $\theta_{i,j}$, $\alpha_{i,j}$, and $\gamma_{i,j}$ inside the lattice can be solved by these variables, and the remaining boundary variables $\theta_{21},\theta_{31}$, $\alpha_{21}$ and $\alpha_{31}$ can be numerically solved using the periodic boundary conditions as constraints. The non-linear configuration of the Maxwell lattice in the following sections will be solved using these defined independent variables.

%%%%%%%%%%%%%%%%%%%%%%%%%%%%%
%%%%%%%%%%%%%%%%%%%%%%%%%%%%%
\pagebreak

\section{Nonlinear Exact Solution of Zero Energy Configurations}\label{theory}

\noindent The internal geometry formed by unit cells in the lattice is investigated and the corresponding structure of the lattice is then studied and presented. As can be seen in Fig.~\ref{schematic_deformed_kagome_lattice}, a hexagon is generated by the adjacent four unit cells, including two entire unit cells of $\theta_{i,j}$ and $\theta_{i+1,j+1}$, a blue triangle from the unit cell of $\theta_{i+1,j}$, and a red triangle from the one of $\theta_{i,j+1}$, where the first subscript represents the row of the lattice while the second subscript denotes the column. The nonlinear configuration of the Maxwell lattice can be determined based on all angles in the lattice.

%%%%%%%%%%%%%%%%%%%%%%%%%%%%%
\subsection{Geometry of a General Hexagon}

\begin{figure}[h!]
\centering
\subfloat[]{%
  \includegraphics[height=2.8in]{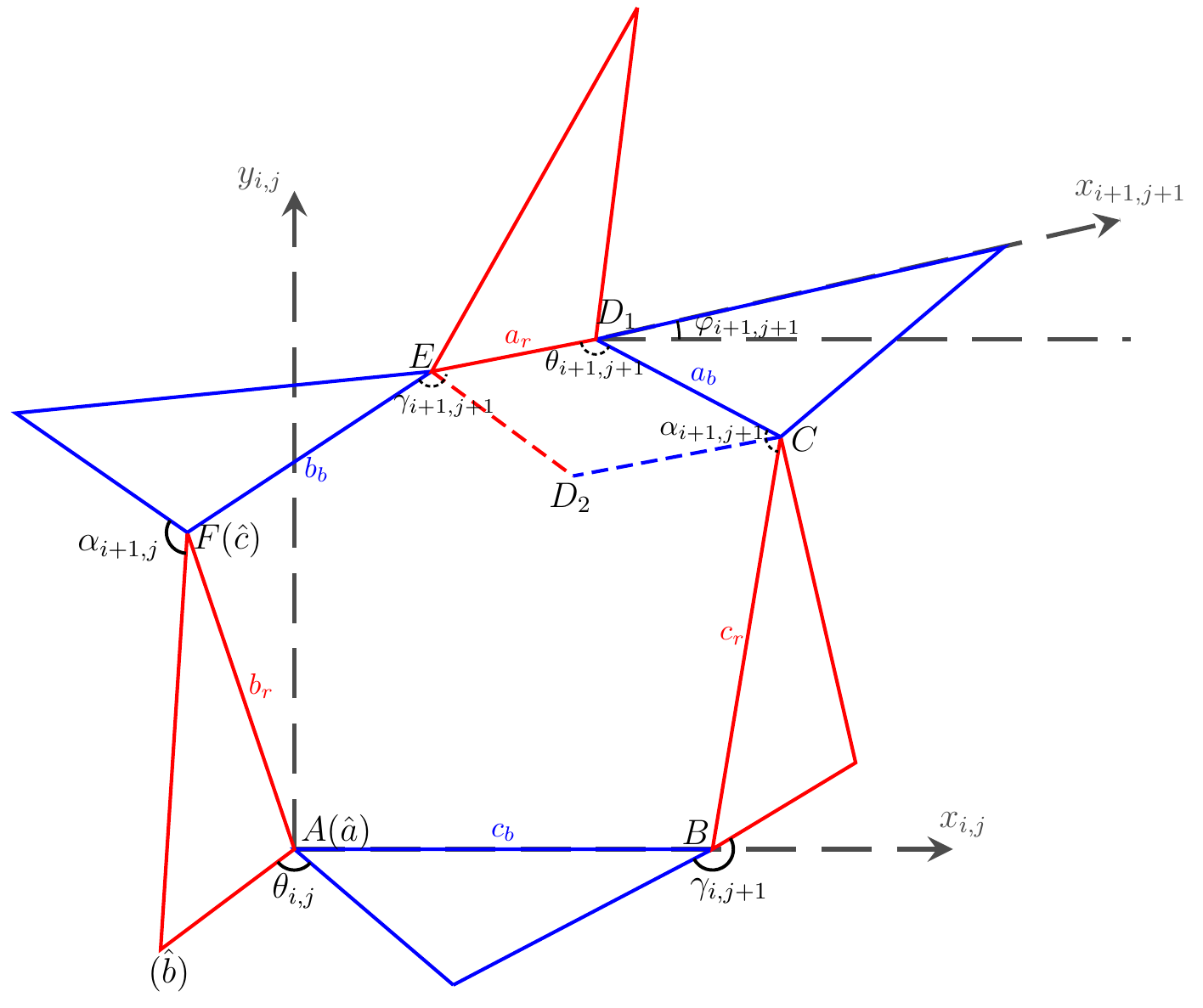}%
}
\subfloat[]{%
  \includegraphics[height=2.8in]{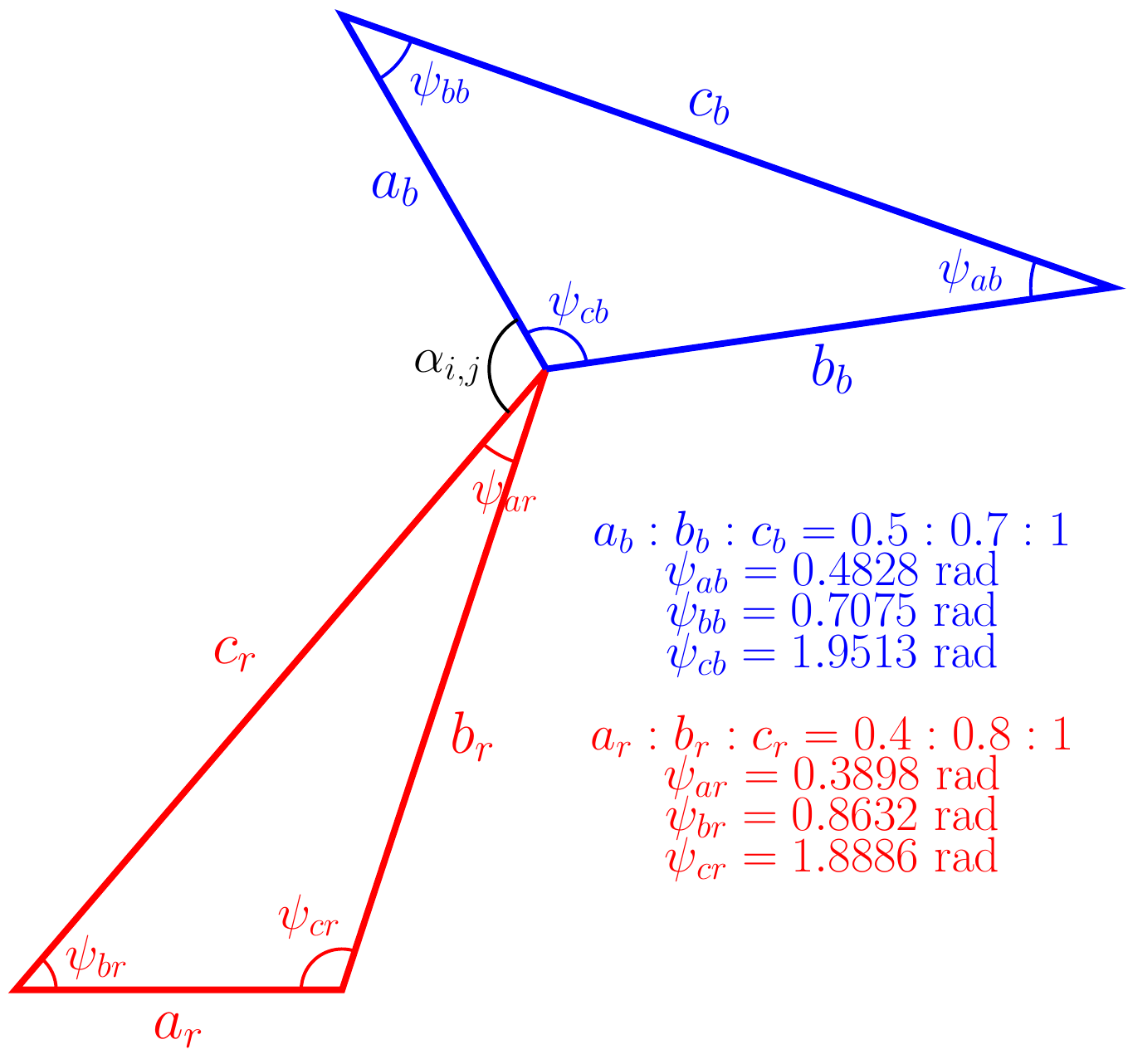}%
}

\subfloat[]{%
  \includegraphics[height=4in]{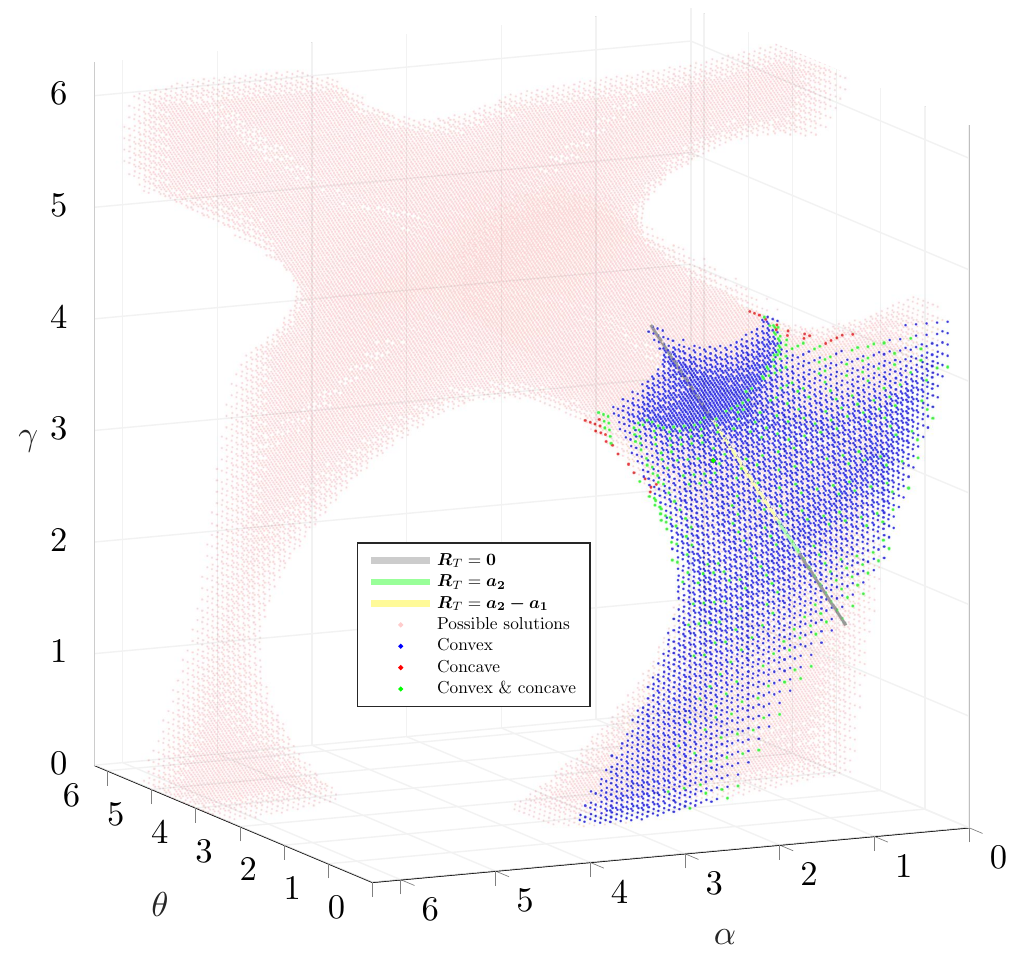}%
}%\hspace*{3.5em}
%\subfloat[]{%
%  \includegraphics[height=2.8in]{angle_surf_alpha_gamma.pdf}%
%}
\caption{(a) Schematic of a general hexagon in the lattice. The hexagon is generated by sides from three red triangles and three blue triangles. Site $A$, $B$, and $F$ correspond to three known angles, $\theta_{i,j}$, $\gamma_{i,j+1}$, and $\alpha_{i+1,j}$, respectively, while site $C$, $D$, and $E$ are related to the other three angles ($\alpha_{i+1,j+1}$, $\theta_{i+1,j+1}$, and $\gamma_{i+1,j+1}$) to be solved. (b) Dimensions of a unit cell from (a). (c) Possible given angles that can generate hexagons shown in (a). Pink dots represent all possible given angles, which can form a hexagon without constraints (i.e. triangles can overlap indicating not a strain free condition or ZM), while blue, red, and green dots indicate all feasible angles under a strain free constraint. Blue and red dots show the convex ($D_1$) and concave ($D_2$) cases, respectively, and green dots imply that the given angles can create both convex and concave solutions.} \label{schematic_hexagon}
\end{figure}

\noindent Consider a hexagon defined in a Cartesian coordinate system (Fig. \ref{schematic_hexagon}(a)) with the origin at site $A$, which is located at the angle of $\theta_{i,j}$. The $x_{ij}$-axis is aligned with side $c_b$ of the blue triangle of the bottom row. Sites $B$ to $F$ correspond to the angles of $\gamma_{i,j+1}$, $\alpha_{i+1,j+1}$, $\theta_{i+1,j+1}$, $\gamma_{i+1,j+1}$ , and $\alpha_{i+1,j}$, respectively. If initial conditions are given for the left and bottom edges of the lattice, all other angles can be obtained by solving hexagons from the bottom left corner to the top right corner in the lattice. Starting from a single hexagon, using known angles, $\theta_{i,j}$, $\gamma_{i,j+1}$ and $\alpha_{i+1,j}$, and related sides of triangles, the other unknown angles ($\theta_{i+1,j+1}$, $\gamma_{i+1,j+1}$ and $\alpha_{i+1,j+1}$) can be determined by calculating the intersection coordinate of site $D$. The coordinates of the remaining vertexes are given:
\begin{equation}\label{site_coordinates}
\begin{split}
  &x_A=0, \quad y_A=0,\quad x_B=c_b,\quad y_B=0,\\
  &x_C=c_b-c_r\cos(\gamma_{i,j+1}+\psi_{ab}+\psi_{br}),\quad y_C=-c_r\sin(\gamma_{i,j+1}+\psi_{ab}+\psi_{br}),\\
  &x_F=b_r\cos(\theta_{ij}+\psi_{bb}+\psi_{cr}),\quad y_F=-b_r\sin(\theta_{ij}+\psi_{bb}+\psi_{cr}),\\
  &x_E=b_r\cos(\theta_{ij}+\psi_{bb}+\psi_{cr})-b_b\cos(\theta_{ij}+\alpha_{i+1,j}+\psi_{bb}+\psi_{cr}+\psi_{cb}+\psi_{ar}),\\
  &y_E= -b_r\sin(\theta_{ij}+\psi_{bb}+\psi_{cr})+b_b\sin(\theta_{ij}+\alpha_{i+1,j}+\psi_{bb}+\psi_{cr}+\psi_{cb}+\psi_{ar}),
\end{split}
\end{equation}
where both the sides of the triangles $c_b$, $b_b$, $c_r$, and $b_r$, as well as the internal angles of the triangles  $\psi_{ab}$, $\psi_{bb}$, $\psi_{cb}$, $\psi_{ar}$, $\psi_{br}$ and $\psi_{cr}$ are given in Fig.~ \ref{schematic_hexagon}. Based on the geometry of the hexagon, the last site $D$ should satisfy the following constraints:
\begin{equation}\label{site_D}
\begin{cases}
  (x_D-x_E)^2+(y_D-y_E)^2=a_r^2,\\
  (x_D-x_C)^2+(y_D-y_C)^2=a_b^2.
 \end{cases}
\end{equation}
Equations~(\ref{site_D}) result in two solutions, one of which ($x_{D_2}$, $y_{D_2}$) forms $\theta_{i+1,j+1}>\pi$ (site $D_2$ in Fig. \ref{schematic_hexagon}(a), which we call the concave solution), and the other of which ($x_{D_1}$, $y_{D_1}$) corresponds to $\theta_{i+1,j+1}\leqslant \pi$ (site $D_1$ in Fig.~\ref{schematic_hexagon}(a), which we call the convex solution). A hexagon is considered concave or convex based on the $\theta_{i+1,j+1}$ angle, or in coordinate basis the angle $\Bar{EDC}$. There exists a particular situation where two solutions overlap and hence $\theta_{i+1,j+1}= \pi$. The concave solution $D_1$ is chosen if the previous $\theta_{i,j}<\pi$, and the convex solution is chosen if $\theta_{i,j}>\pi$.  

%Projections of feasible angle solutions that can generate hexagons shown in (a) onto (b) $\theta$-$\alpha$, (c) $\theta$-$\gamma$, and (d) $\alpha$-$\gamma$ planes, respectively. The red line indicates the family of angles forming homogeneous lattices.

The coordinate of site $D$ can be presented in two ways using known angles defined above and written as 
\begin{equation}\label{gamma_i1j1}
\begin{split}
  & x_{D}=b_r\cos(\theta_{i,j}+\psi_{bb}+\psi_{cr})-b_b\cos(\theta_{i,j}+\alpha_{i+1,j}+\psi_{bb}+\psi_{cr}+\psi_{cb}+\psi_{ar})\\
  &\qquad+a_r\cos(\gamma_{i+1,j+1}-\theta_{i,j}-\alpha_{i+1,j}-\psi_{bb}-\psi_{cr}-\psi_{cb}-\psi_{ar}),\\
  &y_{D}=-b_r\sin(\theta_{i,j}+\psi_{bb}+\psi_{cr})+b_b\sin(\theta_{i,j}+\alpha_{i+1,j}+\psi_{bb}+\psi_{cr}+\psi_{cb}+\psi_{ar})\\
  &\qquad+a_r\sin(\gamma_{i+1,j+1}-\theta_{i,j}-\alpha_{i+1,j}-\psi_{bb}-\psi_{cr}-\psi_{cb}-\psi_{ar}).
\end{split}
\end{equation}
\begin{equation}\label{alpha_i1j1}
\begin{split}
  & x_{D}=c_b-c_r\cos(\gamma_{i,j+1}+\psi_{ab}+\psi_{br})+a_b\cos(\gamma_{i,j+1}-\alpha_{i+1,j+1}+\psi_{ab}+\psi_{br}),\\
  &y_{D}=-c_r\sin(\gamma_{i,j+1}+\psi_{ab}+\psi_{br})+a_b\sin(\gamma_{i,j+1}-\alpha_{i+1,j+1}+\psi_{ab}+\psi_{br}).
\end{split}
\end{equation}
By solving Eqs.~(\ref{gamma_i1j1}) and (\ref{alpha_i1j1}), angles of $\gamma_{i+1,j+1}$ and $\alpha_{i+1,j+1}$ can be achieved. The last angle $\theta_{i+1,j+1}$ is then obtained by the sum of the inner angles of the hexagon subtracting the other five angles. Figure~\ref{schematic_hexagon}(c) illustrates possible three known angles, which can generate hexagons using the configuration and parameters given from Fig.~\ref{schematic_hexagon}(b). Constraints, such as the strain free condition, are shown by blue, red, and green dots. The multicolored line (grey for $\mathbf{R}_T=0$, light green for $\mathbf{R}_T=\mathbf{a_2}$, and light yellow for $\mathbf{R}_T=\mathbf{a_2-a_1}$) represents the family of angles generating homogeneous Maxwell lattices. Processing along this line represents a uniform twisting synonymous with the Guest-Hutchinson modes in the system. The choice of solutions $D_1$ and $D_2$ to calculate the value of $\theta_{i+1,j+1}$ for the current hexagon is determined by the value of $\theta_{i,j}$. For the topological Maxwell lattice, we assume the more favorable solution for a lattice configuration is one with less discontinuous jumps in perturbation, thus we choose either the concave or convex $\theta_{i+1,j+1}$ based on the previous result $\theta_{i,j}$. If $\theta_{i,j}>\pi$, $\theta_{i+1,j+1}$ will still be greater than $\pi$, and vice versa. However, we note that for all cases shown herein, the hexagons do not transition between different convexities. 

Note that if a lattice is perturbed at the hard edge, we solve all unknown angles of hexagons ($\theta_{i+1,j+1}$, $\gamma_{i+1,j+1}$ and $\alpha_{i+1,j+1}$) using known angles of $\theta_{i,j}$, $\gamma_{i,j+1}$ and $\alpha_{i+1,j}$, shown in Fig.~\ref{schematic_hexagon}(a). If a lattice is perturbed at the soft edge, we start solving angles of $\bar{\theta}_{i,j}$, $\bar{\gamma}_{i,j+1}$, and $\bar{\alpha}_{i+1,j}$, which are complementary angles of $\theta_{i,j}$, $\gamma_{i,j+1}$, and $\alpha_{i+1,j}$ subtracting two neighboring inner angles of triangles, using $\bar{\theta}_{i+1,j+1}$, $\bar{\gamma}_{i+1,j+1}$, and $\bar{\alpha}_{i+1,j+1}$.

%%%%%%%%%%%%%%%%%%%%%%%%%%%%%
\subsection{Transformation of Coordinates for Unit Cells}

\noindent The positions of each unit cell can be solved once all necessary angles in the lattice are obtained. Here, local coordinate systems from each unit cell associated with $\theta_{i,j}$ must be rotated to align within a global coordinate system, set by the original unit cell and the value of $\theta_{11}$. 

To simplify the notations of the lattice, where all nodes except those on boundaries are shared with two surrounding unit cells, only three nodes are counted for one unit cell, noted by $\hat{a}$, $\hat{b}$, and $\hat{c}$ (symbols in the parentheses in Fig.~\ref{schematic_hexagon}(a) from red triangles). The coordinates of nodes $\hat{a}_{n,m}$, $\hat{b}_{n,m}$ and $\hat{c}_{n,m}$ in the unit cell at $n$th row and $m$th column (the lattice consists of $N_x\times N_y$ unit cells) are shown below:
\begin{equation}\label{global_coor_n_m}
\begin{split}
  &\text{Unit cell of }\theta_{n,m} \text{ } (n\leqslant m) \\
  &\begin{pmatrix}
	x \\
	y	
	\end{pmatrix}_{\hat{a}_{n,m}}=\sum_{p=1}^{m-n}\left(\prod_{i=1}^{p}\bm{R}(\varphi_{1i})
	\begin{pmatrix}
	x_{\theta_{1,p}\rightarrow\theta_{1,p+1}}\\	
	y_{\theta_{1,p}\rightarrow\theta_{1,p+1}}
	\end{pmatrix}\right)+\sum_{p=1}^{n-1}\left(\prod_{i=1}^{m-n}\bm{R}(\varphi_{1i})\prod_{i=1}^{p}\bm{R}(\varphi_{i,i+m-n})
	\begin{pmatrix}
	x_{\theta_{p,p+m-n}\rightarrow\theta_{p+1,p+m-n+1}}\\	
	y_{\theta_{p,p+m-n}\rightarrow\theta_{p+1,p+m-n+1}}
	\end{pmatrix}\right),\\
	&\begin{pmatrix}
	x \\
	y	
	\end{pmatrix}_{\hat{b}_{n,m}}=\begin{pmatrix}
	x \\
	y	
	\end{pmatrix}_{\hat{a}_{n,m}}+\prod_{i=1}^{m-n}\bm{R}(\varphi_{1i})\prod_{i=1}^{n}\bm{R}(\varphi_{i,i+m-n})
	\begin{pmatrix}
	a_r\cos(\theta_{n,m}+\psi_{bb})\\	
	-a_r\sin(\theta_{n,m}+\psi_{bb})
	\end{pmatrix},\\
	&\begin{pmatrix}
	x \\
	y	
	\end{pmatrix}_{\hat{c}_{n,m}}=\begin{pmatrix}
	x \\
	y	
	\end{pmatrix}_{\hat{a}_{n,m}}+\prod_{i=1}^{m-n}\bm{R}(\varphi_{1i})\prod_{i=1}^{n}\bm{R}(\varphi_{i,i+m-n})
	\begin{pmatrix}
	b_r\cos(\theta_{n,m}+\psi_{bb}+\psi_{cr})\\	
	-b_r\sin(\theta_{n,m}+\psi_{bb}+\psi_{cr})
	\end{pmatrix},\\
\end{split}
\end{equation}
\begin{equation}\label{global_coor_m_n}
\begin{split}
  &\text{Unit cell of }\theta_{n,m} \text{ } (n>m) \\
  &\begin{pmatrix}
	x \\
	y	
	\end{pmatrix}_{\hat{a}_{n,m}}=\sum_{p=1}^{n-m}\left(\prod_{i=1}^{p}\bm{R}(\varphi_{i1})
	\begin{pmatrix}
	x_{\theta_{p,1}\rightarrow\theta_{p+1,1}}\\	
	y_{\theta_{p,1}\rightarrow\theta_{p+1,1}}
	\end{pmatrix}\right)+\sum_{p=1}^{m-1}\left(\prod_{i=1}^{n-m}\bm{R}(\varphi_{i1})\prod_{i=1}^{p}\bm{R}(\varphi_{i,i+n-m})
	\begin{pmatrix}
	x_{\theta_{p,p+n-m}\rightarrow\theta_{p+1,p+n-m+1}}\\	
	y_{\theta_{p,p+n-m}\rightarrow\theta_{p+1,p+n-m+1}}
	\end{pmatrix}\right),\\
	&\begin{pmatrix}
	x \\
	y	
	\end{pmatrix}_{\hat{b}_{n,m}}=\begin{pmatrix}
	x \\
	y	
	\end{pmatrix}_{\hat{a}_{n,m}}+\prod_{i=1}^{n-m}\bm{R}(\varphi_{i1})\prod_{i=1}^{m}\bm{R}(\varphi_{i,i+n-m})
	\begin{pmatrix}
	a_r\cos(\theta_{n,m}+\psi_{bb})\\	
	-a_r\sin(\theta_{n,m}+\psi_{bb})
	\end{pmatrix},\\
	&\begin{pmatrix}
	x \\
	y	
	\end{pmatrix}_{\hat{c}_{n,m}}=\begin{pmatrix}
	x \\
	y	
	\end{pmatrix}_{\hat{a}_{n,m}}+\prod_{i=1}^{n-m}\bm{R}(\varphi_{i1})\prod_{i=1}^{m}\bm{R}(\varphi_{i,i+n-m})
	\begin{pmatrix}
	b_r\cos(\theta_{n,m}+\psi_{bb}+\psi_{cr})\\	
	-b_r\sin(\theta_{n,m}+\psi_{bb}+\psi_{cr})
	\end{pmatrix},\\
\end{split}
\end{equation}
where $\sum$ and $\prod$ are cumulative sum and product operators. The coordinate transformation matrix, $\bm{R}(\varphi_{i,j})$, is given by 
\begin{equation}\label{transformation_matrix}
\bm{R}(\varphi_{i,j})=
\begin{bmatrix}
\cos{\varphi_{i,j}} & -\sin{\varphi_{i,j}} \\
\sin{\varphi_{i,j}} & \cos{\varphi_{i,j}} 
\end{bmatrix},
\end{equation}
where $\varphi_{i,j}$ is the deflection angle between $x_{i,j}$-axis and the previous $x$-axis ($\varphi_{11}=0$), and: 
\begin{equation}\label{deflection_angle}
\begin{split}
  \varphi_{1,j}&=\gamma_{1,j}+\theta_{1,j-1}+\psi_{ab}+\psi_{bb}-2\pi,\\
  \varphi_{i,1}&=2\pi-\alpha_{i,1}-\theta_{i-1,1}-\psi_{ar}-\psi_{cr},\\
   \varphi_{i,j}&=\gamma_{i-1,j}-\alpha_{i,j}+\psi_{bb}+\psi_{ab}+\psi_{br}-\pi \quad (i, j>1).
\end{split}
\end{equation}

\begin{figure}[h!]
\centering
\subfloat[]{%
  \includegraphics[height=1.5in]{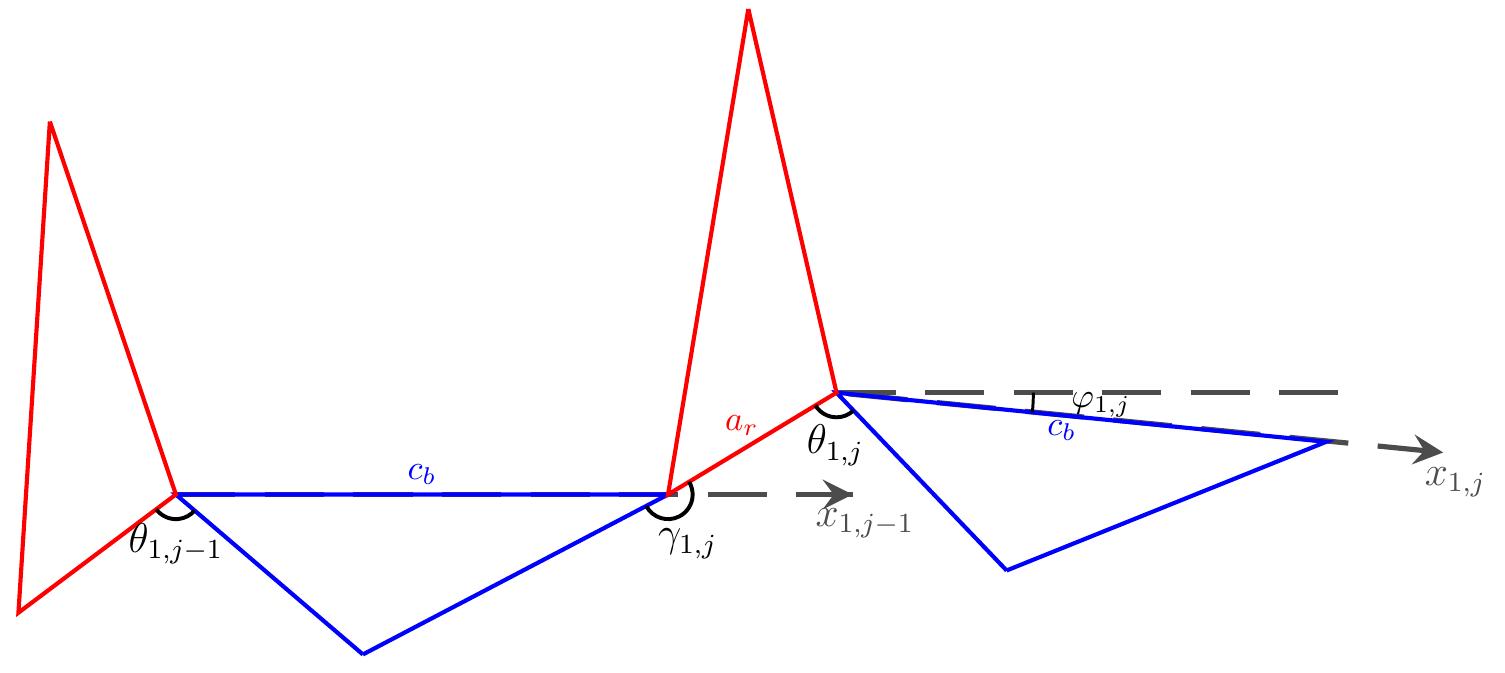}%
}
\subfloat[]{%
  \includegraphics[height=2.5in]{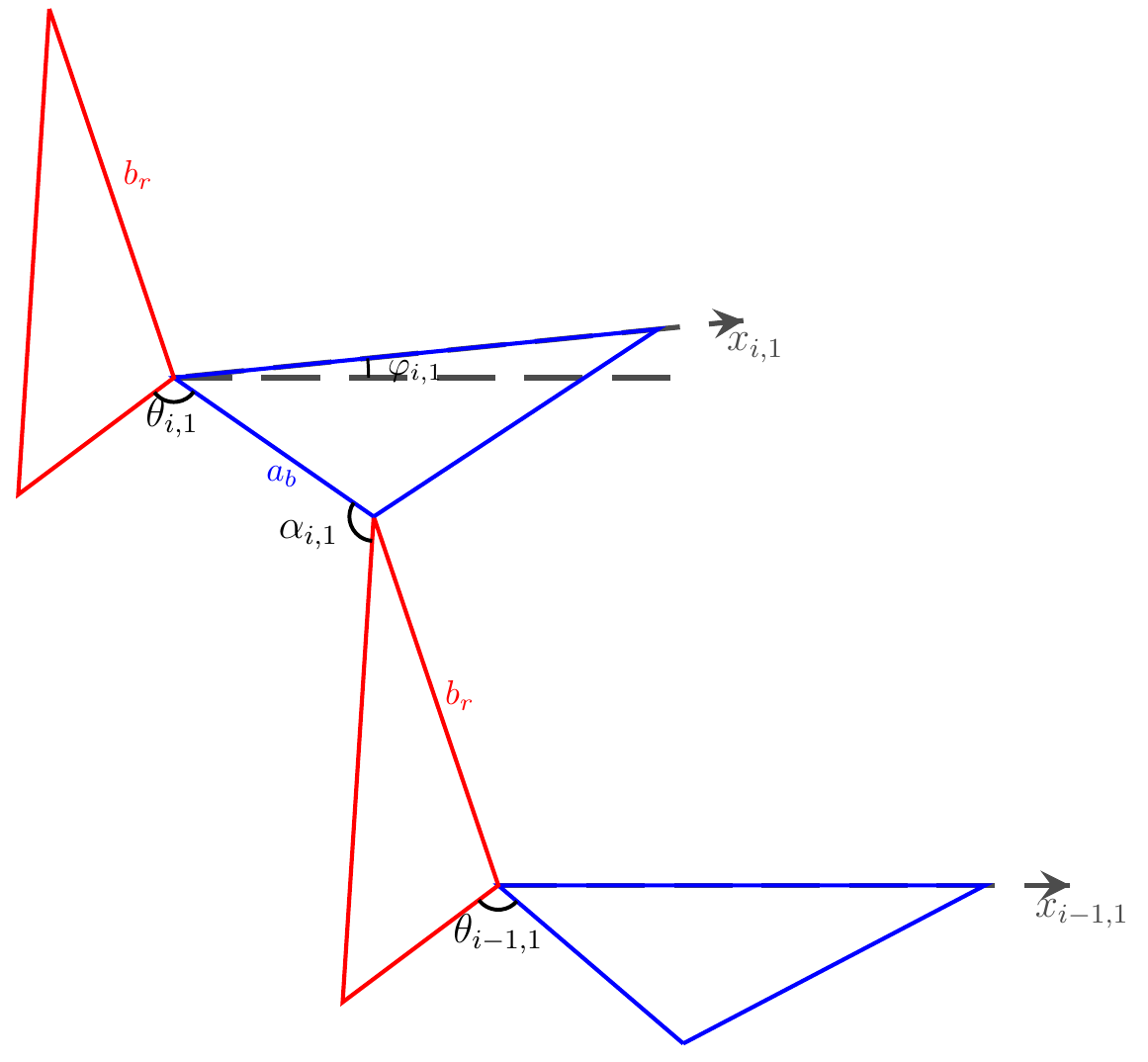}%
}
\caption{Schematic of two unit cells from (a) the bottom edge and (b) the left edge of the lattice.}\label{schematic_hexagon_edge}
\end{figure}

The cumulative coordinates with subscripts $\theta_{1,j}\rightarrow\theta_{1,j+1}$ (in Fig.~\ref{schematic_hexagon_edge}(a)), $\theta_{i,1}\rightarrow\theta_{i+1,1}$ (in Fig.~\ref{schematic_hexagon_edge}(b)), and $\theta_{i,j}\rightarrow\theta_{i+1,j+1}$ (in Fig.~\ref{schematic_hexagon}(a)) are those converted from $(1,j+1)^{\mathrm{th}}$ to $(1,j)^{\mathrm{th}}$ unit cell, from $(i+1,1)^{\mathrm{th}}$ to $(i,1)^{\mathrm{th}}$ unit cell and from $(i+1,j+1)^{\mathrm{th}}$ to $(i,j)^{\mathrm{th}}$ unit cell, respectively:
\begin{equation}
\begin{split}
  & x_{\theta_{1,j}\rightarrow\theta_{1,j+1}}=c_b-a_r\cos(\psi_{ab}+\gamma_{1,j+1}), \\
  &y_{\theta_{1,j}\rightarrow\theta_{1,j+1}}=-a_r\sin(\psi_{ab}+\gamma_{1,j+1}),\\
  & x_{\theta_{i,1}\rightarrow\theta_{i+1,1}}=b_r\cos(\theta_{i,1}+\psi_{bb}+\psi_{cr})-a_b\cos(\alpha_{i+1,1}+\theta_{i,1}+\psi_{ar}+\psi_{cr}+\psi_{bb}), \\
  &y_{\theta_{i,1}\rightarrow\theta_{i+1,1}}=-b_r\sin(\theta_{i,1}+\psi_{bb}+\psi_{cr})+a_b\sin(\alpha_{i+1,1}+\theta_{i,1}+\psi_{ar}+\psi_{cr}+\psi_{bb})),\\
  & x_{\theta_{i,j}\rightarrow\theta_{i+1,j+1}}=c_b-c_r\cos(\gamma_{i,j+1}+\psi_{ab}+\psi_{br})+a_b\cos(\gamma_{i,j+1}-\alpha_{i+1,j+1}+\psi_{ab}+\psi_{br}),\\
  &y_{\theta_{i,j}\rightarrow\theta_{i+1,j+1}}=-c_r\sin(\gamma_{i,j+1}+\psi_{ab}+\psi_{br})+a_b\sin(\gamma_{i,j+1}-\alpha_{i+1,j+1}+\psi_{ab}+\psi_{br}).
\end{split}
\end{equation}

Equations (\ref{global_coor_n_m}) and (\ref{global_coor_m_n}) solve for the number of nodes of $3N_xN_y$ related to all red triangles, while the remaining ($N_x+N_y$) nodes come from the first row $\theta_{1,m}$ and last column $\theta_{n,N_x}$, which are respectively given by:
\begin{equation}\label{global_coor_1_m}
\begin{split}
  %&\text{Node of the blue triangle from the unit cell of }\theta_{1,m} \\
  &	\begin{pmatrix}
	x \\
	y	
	\end{pmatrix}_{\hat{c}_{0,m}}=\begin{pmatrix}
	x \\
	y	
	\end{pmatrix}_{\hat{a}_{1,m}}+\prod_{i=1}^{m}\bm{R}(\varphi_{1i})
	\begin{pmatrix}
	a_b\cos(\psi_{bb})\\	
	-a_b\sin(\psi_{bb})
	\end{pmatrix}.\\
\end{split}
\end{equation}
\begin{equation}\label{global_coor_n_Nx}
\begin{split}
 % &\text{Node of the blue triangle from the unit cell of }\theta_{n,N_x} \\
  &	\begin{pmatrix}
	x \\
	y	
	\end{pmatrix}_{\hat{b}_{n,N_x+1}}=\begin{pmatrix}
	x \\
	y	
	\end{pmatrix}_{\hat{a}_{n,N_x}}+\prod_{i=1}^{N_x-n}\bm{R}(\varphi_{1i})\prod_{i=1}^{n}\bm{R}(\varphi_{i,i+N_x-n})
	\begin{pmatrix}
	c_b\\	
	0
	\end{pmatrix},\text{ } (n\leqslant N_x)\\
	&	\begin{pmatrix}
	x \\
	y	
	\end{pmatrix}_{\hat{b}_{n,N_x+1}}=\begin{pmatrix}
	x \\
	y	
	\end{pmatrix}_{\hat{a}_{n,N_x}}+\prod_{i=1}^{n-N_x}\bm{R}(\varphi_{i1})\prod_{i=1}^{N_x}\bm{R}(\varphi_{i,i+n-N_x})
	\begin{pmatrix}
	c_b\\	
	0
	\end{pmatrix},\text{ } (n>N_x).\\
\end{split}
\end{equation}

\subsection{Periodic Boundary Conditions}

Newton's method, an algorithm for finding roots of a function $f(x)$ such that $f(x)=0$, is used in order to achieve periodic boundary conditions. Herein, the goal is to find periodic boundary conditions such that $f(x)$ is a function of $\alpha$, and periodic boundary conditions are met when $f(\alpha_{i,1})=\alpha_{i,n}-\alpha_{i,1}=0$, for the $i$-th row of an $M\times N$ lattice. The general Newton's method algorithm is given below, where: 

\begin{equation}
    \alpha_{n+1}=\alpha_n-\frac{f(\alpha_n)}{f'(\alpha_n)}.
\end{equation}

\noindent The input variable $\alpha_n$ is the initial guess for $\alpha_{i,1}$. The derivative $f'(\alpha_n)$ is calculated numerically using a central difference method, such that: 

\begin{equation}
    f'(\alpha_n) = \frac{f(\alpha+h/2)-f(\alpha-h/2)}{h},
\end{equation}
where $h/2$ is an angular increment $\delta \alpha$ that is small in comparison to $\alpha$. In this paper, a value of $\delta \alpha=0.1$ $\mu$rad was used. For each value of $f(\alpha)$, $f(\alpha+\delta \alpha)$, and $f(\alpha-\delta \alpha)$, the corresponding configuration of the row must be calculated. The algorithm continues to update the initial guess of $\alpha$, until the residual is less than a specified tolerance value. In this paper, we set the tolerance value as 0.01 $\mu$rad.

\subsection{Linear Mode Analysis}
\begin{figure}[h!]
\centering
\subfloat[]{%
  \includegraphics[height=2in]{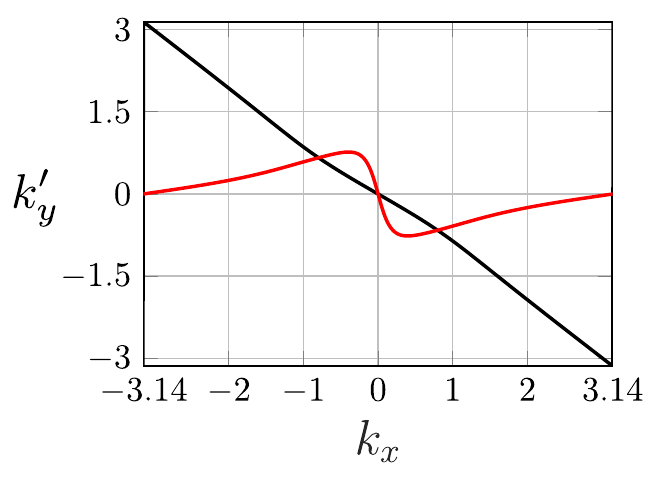}%
}
\subfloat[]{%
  \includegraphics[height=2in]{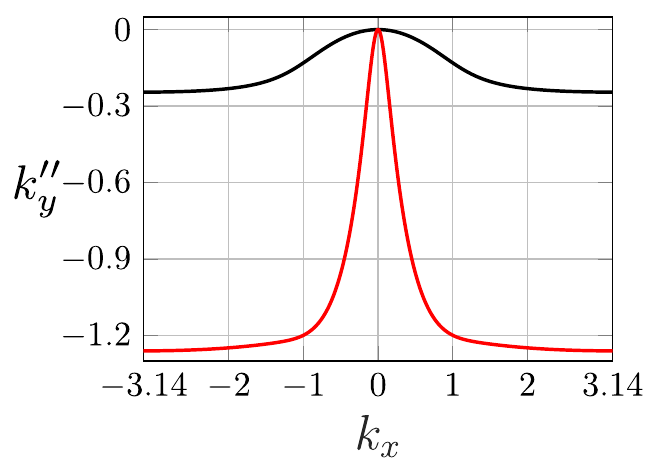}%
}
\caption{``Dispersion relation.'' (a) Real component $k'_y = \Re(k_y)$ vs $k_x$ in the first Brillouin Zone for an $\mathbf{a_2}$ polarized lattice. (b) Imaginary component $k''_y = \Im(k_y)$ vs $k_x$  for an $\mathbf{a_2}$ polarized lattice. The red branch has a much higher decay rate, and ``velocity'' at long wavelength (low $k_x$ values).}\label{linear_kxky}
\end{figure}

\noindent  In the small amplitude perturbation limit, where $\varepsilon$ is small (on the order of $10^{-6}$ rad, for our simulations $10^{-3}$ rad was considered "large perturbations"), the numerically solved wave profile in the static limit can be approximated by a linear mode analysis. In our case, of a finite boundary in $y$ and a periodic boundary in $x$, one can prescribe a real $k_x$ and solve for a complex $k_y$ that satisfies the condition $\det C (\vec{k}) = 0$ to find the ZMs %(\textit{15}).
\cite{kane2014topological}. 
The resulting complex $k_y = k'_y(k_x) + i k''_y (k_x)$ has $k'_y = \Re (k_y)$ being the real spatial wave profile in the lattice and $k''_y = \Im (k_y)$ being the decay rate of the ZMs. The relation of $k_x$ and $k_y$ for the given homogeneous lattice ($\alpha_0 = 1.3144$ rad) is shown in Fig.~\ref{linear_kxky} and Fig.~2 of the main text. The numerical results shown in Fig.~2(a) are in close agreement with the linear ZM analysis.

%%%%%%%%%%%%%%%%%%%%%%%%%%%%%
%%%%%%%%%%%%%%%%%%%%%%%%%%%%%

\pagebreak

\section{Space-Time Mapping in Maxwell Lattices}
\noindent In this section, we discuss the general elasticity theory of 2D materials with a soft strain (or ``mechanism''), and how their static zero energy configurations map to 1D dynamical systems in linear and weakly nonlinear cases.

\subsection{From soft strain to partial differential equations of ZMs in linear theory}
\noindent We start by considering the general case of elasticity in 2D, where, to leading order in strain and spatial gradient of deformation, the elastic energy can be written as: 
%The continuum limit of the elastic energy of a general 2D media can be written as
\begin{equation}\label{EQ:EL}
    f_{el}=\frac{1}{2}\epsilon_a \mathbb{K}_{ab} \epsilon_b,
\end{equation}
where $\epsilon=\{\epsilon_{xx},\epsilon_{yy},\epsilon_{xy}\}$ are the strains in Voigt notation, forming a 3-dimensional vector, and $\mathbb{K}$ is the (real symmetric) elastic constant matrix. This elastic energy can be generally applied to any 2D elastic structure.    
For lattices considered in this paper, this continuum elastic energy can be obtained by properly taking the long-wavelength limit from the lattice elastic energy %(\textit{21, 73}).
~\cite{lubensky2015phonons,sun2020comtinuum}.

Now, let's consider the case when this material has a soft strain, which means the $\mathbb{K}$ matrix has an eigenvalue that is zero or much smaller than the other two eigenvalues.  This can arise from microscopic mechanisms of the material, e.g. auxetic foams or pentamode metamaterials %(\textit{74, 75}).
\cite{milton1995elasticity,lakes2017negative}. 
In this paper, we are particularly interested in the case of Maxwell lattices, where this soft mode is \emph{guaranteed to arise}, taking the form of Guest-Hutchinson modes %(\textit{21, 61}). 
~\cite{guest2003determinacy,lubensky2015phonons}.

%It is known that all Maxwell lattices exhibit soft homogeneous deformations (Guest-Hutchinson modes).  
This soft strain has far-reaching effects on the elasticity of these lattices. To see this, we diagonalize the $\mathbb{K}$ matrix, $\mathbb{K}=V^T \Lambda V$, where $V$ is a $3\times 3$ orthogonal matrix, the row vectors of which formed by eigenvectors of $\mathbb{K}$, and $\Lambda$ is a diagonal matrix with two positive eigenvalues:
\begin{equation}\label{EQ:VL}
    V =
    \begin{pmatrix}
\epsilon_1\\
\epsilon_2\\
\epsilon_0
\end{pmatrix}, \quad
\Lambda =
    \begin{pmatrix}
\lambda_1 & 0 & 0 \\
0 & \lambda_2 & 0\\
0 & 0& 0
\end{pmatrix},
\end{equation}
where the last strain vector $\epsilon_0$ is the soft deformation, which comes with the zero eigenvalue in $\Lambda$. 

To see the consequence of this elastic energy on spatially varying ZMs in the material, we rewrite Eq.~\eqref{EQ:EL} in terms of displacement field $u=\{u_x,u_y\}$:
\begin{equation}\label{EQ:Eu}
    f_{el}=\frac{1}{2}\epsilon_a \mathbb{K}_{ab} \epsilon_b
    = \frac{1}{2} u_i \overleftarrow{S^{T}_{ia}} \mathbb{K}_{ab}
    \overrightarrow{S_{bj}} u_j \equiv \frac{1}{2} u_i K_{ij} u_j,
\end{equation}
where
\begin{equation}
    \overrightarrow{S} \equiv 
  i  \begin{pmatrix}
k_x & 0\\
0 & k_y \\
\frac{k_y}{2} & \frac{k_x}{2}
\end{pmatrix},
\end{equation}
with $k_x$, $k_y$ representing x and y directional wave vectors and the overhead arrow denoting that these wave vectors are differential operators (in real space) acting on the wave to the right. Similarly, the left overhead arrow on $\overleftarrow{S^{T}_{ia}}$ acts on $u_i$ to its left.

With the diagonalization in Eq.~\eqref{EQ:VL}, the stiffness matrix $K$ is given by:
\begin{equation}
    K = \overleftarrow{S^T}\mathbb{K}\overrightarrow{S}
    = \overleftarrow{S^T}V^T \Lambda V \overrightarrow{S}.
\end{equation}
Because the last eigenvalue of $\Lambda$ vanishes, we can define $\tilde{V}$ to be the $2\times 3$ matrix with the last row removed from $V$, and $\tilde{\Lambda}$ to the $2\times 2$ matrix with the last row removed in $\Lambda$.  Thus, we have a $2\times2$ matrix $C$ that maps displacement fields to the ``nonsoft'' strains:
\begin{equation}
    \overrightarrow{C} = \tilde{V} \overrightarrow{S},
\end{equation}
and correspondingly
\begin{equation}
    K = \overleftarrow{C^T}\tilde{\Lambda}\overrightarrow{C}. 
\end{equation}

Therefore, zero modes in this media are waves that satisfy:
\begin{equation}\label{EQ:Cu}
    \overrightarrow{C} u =0. 
\end{equation}
We can eliminate $u_y$ and write this as a differential equation of $u_x$ only, which becomes:
\begin{equation}
    \det \overrightarrow{C} u_x =0 .
\end{equation}
This $C$ matrix is the same as the final $2\times 2$ matrix in the continuum theory (after integrating out high frequency modes and expanding to the linear order in wavenumbers $k_x$ and $k_y$) for topological ZMs in Maxwell lattices %(\textit{73}).
~\cite{sun2020comtinuum}.

Using the fact that $\{ \epsilon_1,\epsilon_2,\epsilon_0 \}$ form an orthogonal triplet, we have:
\begin{equation}
\label{EQ:detCquad}
    \det \overrightarrow{C} = -\frac{1}{2}
    \left(
     \epsilon_{02} k_x^2 
     - 2 \epsilon_{03} k_x k_y  
     +\epsilon_{01}k_y^2 \right) .
\end{equation}

\noindent It is perhaps more transparent if we turn $q$ into differential operators, where Eq.~\eqref{EQ:Cu} becomes:
\begin{equation}
\label{Eq:EoMQuadratic}
    \left(
     \epsilon_{02} \partial_x^2 
     - 2 \epsilon_{03} \partial_x \partial_y  
     +\epsilon_{01}\partial_y^2 \right) u_x =0 , 
\end{equation}
where $\{\epsilon_{01},\epsilon_{02},\epsilon_{03}\}$ are the 3 components of the soft strain $\epsilon_{0}$ (using the Voigt notation).  
When $\det \epsilon_0 >0$, the material is dilation dominated (auxetic), and this equation is elliptic; when $\det \epsilon_0 <0$, the material is shear dominated (non-auxetic), and this equation is hyperbolic. The relationship between the real and imaginary components of the x and y component wavenumbers for the first Brillouin zone of an $a_2$ polarized lattice are shown in Fig.~\ref{linear_kxky}.

\subsection{Interpreting Eq. \eqref{Eq:EoMQuadratic} as a wave equation when $\det \epsilon_0 < 0$}

\noindent Since for  $\det \epsilon_0 < 0$, Eq.~\eqref{Eq:EoMQuadratic} is hyperbolic, it can be thought of as a wave equation in a 1D medium if we interpret spatial direction $y$ as time $t$. With this interpretation, Eq.~\eqref{Eq:EoMQuadratic} can be rewritten as:
\begin{equation}
\label{Eq:WaveEquation}
\left(\epsilon_{01}\partial_t^2
     - 2 \epsilon_{03} \partial_x \partial_t  
     +\epsilon_{02} \partial_x^2  \right) u_x =0. 
\end{equation}
The presence of the second term means that the equation is not time reversal invariant since under the transformation $t\rightarrow -t$, $\partial_x \partial_t \rightarrow -\partial_x \partial_t$. This manifests itself in the nonsymmetric dispersion relation once we plug in the plane wave ansatz $u_x \sim e^{i(kx-\omega t)}$:
\begin{equation}
\omega = -\frac{\epsilon_{03}}{\epsilon_{01}}k\pm k \sqrt{\frac{\epsilon_{03}^2}{\epsilon_{01}^2}-\frac{\epsilon_{02}}{\epsilon_{01}}} = -\frac{\epsilon_{03}}{\epsilon_{01}}k \pm k\frac{\sqrt{-\det \epsilon_0}}{|\epsilon_{01}|};
\end{equation}
i.e., $\omega(k) \neq \omega(-k)$. Note that $\omega$ is real since $\det \epsilon_0 < 0$. And since the equations of motion is not time reversal symmetric, it is non-reciprocal.

Even though Eq.~\eqref{Eq:WaveEquation} is not time reversal symmetric, it is energy conserving. This can be seen from the following argument. This wave equation can be written as the Euler-Lagrange equation of the action:
\begin{equation}
S = \int dt dx\, \mathcal{L}= \int dt dx \left[ \frac{1}{2}\epsilon_{01} (\partial_t u_x)^2 + \frac{1}{2}\epsilon_{02} (\partial_x u_x)^2 - \epsilon_{03} (\partial_t u_x)(\partial_x u_x)\right],
\end{equation}
where $\mathcal{L}$ is the Lagrangian density. The Euler-Lagrange equation of this action is:
\begin{equation}
\partial_t \frac{\partial \mathcal{L}}{\partial (\partial_t u_x)} + \partial_x \frac{\partial \mathcal{L}}{\partial (\partial_x u_x)}  = \frac{\partial \mathcal{L}}{\partial u_x} \Rightarrow \left(\epsilon_{01}\partial_t^2
     - 2 \epsilon_{03} \partial_x \partial_t  
     +\epsilon_{02} \partial_x^2  \right) u_x =0,
\end{equation}
which confirms the validity of the action. Since this Lagrangian is invariant under time translation $t \rightarrow t + t_0$, according to Noether's theorem, the energy $E$, defined below, is conserved:
\begin{equation}
E = \int dx\,\left[ \frac{\partial \mathcal{L}}{\partial (\partial_t u_x)}\partial_t u_x - \mathcal{L} \right]= \int dx\,\frac{1}{2}\left[\epsilon_{01} (\partial_t u_x)^2 - \epsilon_{02} (\partial_x u_x)^2\right].
\end{equation}
The fact that energy $E$ is conserved can be seen directly in the following way:
\begin{equation}
\begin{split}
\frac{d E}{dt} &= \int dx\,\frac{1}{2}\partial_t\left[\epsilon_{01} (\partial_t u_x)^2 - \epsilon_{02} (\partial_x u_x)^2\right]\\
&= \int dx\,\left[\epsilon_{01}  \partial_t u_x\partial_t^2 u_x- \epsilon_{02} \partial_x u_x\partial_t\partial_x u_x\right]\\
&= \int dx\,\left[ \partial_t u_x(2\epsilon_{03}\partial_x\partial_t u_x -\epsilon_{02} \partial_x^2 u_x)- \epsilon_{02} \partial_x u_x\partial_t\partial_x u_x\right]\\
&= \int dx\,\partial_x\left[\epsilon_{03}(\partial_t u_x)^2 - \epsilon_{02}(\partial_x u_x)(\partial_t u_x)\right]\\
&= 0,
\end{split}
\end{equation}
where we used Eq.~\eqref{Eq:WaveEquation} from second to third equality. The last equality is due to periodic boundary conditions.

\subsection{Higher order corrections to Eq.~\eqref{EQ:Cu}, topological polarization, and interpretation of the resulting equation as time evolution of non-Hermitian 1D system}

\noindent So far we have kept to linear order terms in $k_x$ and $k_y$ in the effective $2 \times 2$ compatibility matrix $\overrightarrow C$, which corresponds to a classical elastic energy in terms of the strain tensor (without strain gradient terms).  In this case, the ZMs are delocalized bulk modes. This can be seem from the fact that 
%it cannot capture the localization of the zero modes on one edge of the system for $\det \epsilon_0 < 0$ since the solution to the equation 
$\det \overrightarrow{C} =  -\frac{1}{2}\left(\epsilon_{02} k_x^2 - 2 \epsilon_{03} k_x k_y +\epsilon_{01}k_y^2 \right) = 0$ results in $k_y = \frac{\epsilon_{03}}{\epsilon_{01}}k_x \pm k_x\frac{\sqrt{-\det \epsilon_0}}{|\epsilon_{01}|} \equiv k_{\pm}$, which are real---meaning that the ZMs are plane waves in the bulk. 

The localization of the ZMs to edges are captured when %It turns out that if we allow 
terms up to the second order in $k_x$ and $k_y$ in the effective $2 \times 2$ compatibility matrix $\overrightarrow C$
%, we can capture localization of the zero mode at an edge of the system
are kept %(\textit{21, 73}). 
\cite{lubensky2015phonons,sun2020comtinuum}.  
To see this, we follow the steps from Eq.~\eqref{EQ:Cu} to Eq.~\eqref{EQ:detCquad} keeping terms up to quadratic order in $q$ in $\overrightarrow C$ to get:
\begin{equation}
\label{Eq:detCcubic}
\det \overrightarrow{C} = -\frac{1}{2}
    \left(
     \epsilon_{02} k_x^2 
     - 2 \epsilon_{03} k_x k_y  
     +\epsilon_{01}k_y^2 \right) - \frac{i}{2}(C_1 k_x^3 +C_2 k_x^2 k_y + C_3 k_x k_y^2 + C_4 k_y^3) +\mathcal{O}(k^4),
\end{equation}
where the parameters $C_i$, $i=1,2,3,4$ are real (this is due to time reversal symmetry of the Maxwell lattice) and are evaluated numerically using the scheme described in Ref. %(\textit{73}).
~\cite{sun2020comtinuum}.  
The cubic terms in $k$ in $\det \overrightarrow C$ gives correction of order $k_x^2$ to solutions $k_y = k_{\pm}$ mentioned above. The real and imaginary components of $k_x$ and $k_y$ are shown in Fig.~\ref{linear_kxky} for an $\mathbf{a_2}$ polarized hyperbolic PDE. To get the corrections, we plug in the ansatz $k_y  = k_{\pm} + i \delta_{\pm} k_x^2$ in Eq.~\eqref{Eq:detCcubic}, and keep the terms up to the cubic order in $k_x$ to get:
\begin{equation}
\label{Eq:CorrectionToDispersion}
\begin{split}
\left(
     \epsilon_{02} k_x^2 
     - 2 \epsilon_{03} k_x (k_{\pm} + i \delta_{\pm} k_x^2)
     +\epsilon_{01}(k_{\pm} + i \delta_{\pm} k_x^2)^2 \right)+i(C_1 k_x^3 +C_2 k_x^2 (k_{\pm} + i \delta_{\pm} k_x^2) + C_3 k_x (k_{\pm} + i \delta_{\pm} k_x^2)^2 + C_4 (k_{\pm} + i \delta_{\pm} k_x^2)^3)&=0,\\
\Rightarrow i(2\epsilon_{01}k_{\pm}k_x^2 -2\epsilon_{03}k_x^3)\delta_{\pm}+i(C_1 k_x^3+C_2k_x^2 k_{\pm}+C_3k_x k_{\pm}^2 + C_4k_{\pm}^3)+ \mathcal{O}(k_x^4) = 0 \Rightarrow \delta_{\pm} = \frac{C_1 +C_2 \tilde{k}_{\pm}+C_3\tilde{k}_{\pm}^2 + C_4\tilde{k}_{\pm}^3}{2\epsilon_{03}-2\epsilon_{01}\tilde{k}_{\pm}},
\end{split}
\end{equation}
where we defined $\tilde{k}_{\pm} \equiv k_{\pm}/k_x =  \frac{\epsilon_{03}}{\epsilon_{01}}\pm \frac{\sqrt{-\det \epsilon_0}}{|\epsilon_{01}|}$. Note that the solutions $k_y = \tilde{k}_{\pm} k_x+i\delta_{\pm}k_x^2$ to the order $k_x^2$ are complex numbers, meaning these zero modes are actually localized at the edges of the system. $C_4$ is numerically calculated to equal zero when the lattice vector $\mathbf{a_1}$ aligns with the x-axis, and in the case of topologically polarized Maxwell lattices, $\delta_{+}$ and $\delta_{-}$ have the signs implying that both zero modes are at the same edge, hence the polarization. This is determined by the topological polarization.

Now, we can turn to the corresponding wave equation in a 1-dimensional dynamical system by replacing $k_i$ with $(-i \partial_i)$ and interpreting direction $y$ as time $t$:
\begin{equation}
\label{Eq:EoMcubic}
\left[\left(\epsilon_{01}\partial_t^2
     - 2 \epsilon_{03} \partial_x \partial_t  
     +\epsilon_{02} \partial_x^2  \right)+(C_1\partial_x^3+C_2\partial_x^2\partial_t+C_3\partial_x\partial_t^2+C_4\partial_t^3)\right] u_x = 0.
\end{equation}
We showed in the previous section that the terms with second order derivatives can be obtained from a time-invariant Lagrangian and hence energy conserving. However, the third order derivative terms cannot be obtained from a Lagrangian. The reason is the following. If there were a Lagrangian from which these terms with derivatives could be obtained, then that Lagragian would have to have three derivatives and have to be of the order $u_x^2$; hence the most generic form of the Lagrangian would be $u_x \partial_i \partial_j \partial_l u_x$ ($i,j,l \in \{x,t\}$). By taking variation of the action, we would obtain the following:
\begin{equation}
\begin{split}
\delta S &= S[u_x+\delta u_x] - S[u_x] = \int dt dx \left[(u_x+\delta u_x) \partial_i\partial_j\partial_l (u_x+\delta u_x) -u_x\partial_i\partial_j\partial_l u_x\right] =  \int dt dx \left[u_x  \partial_i\partial_j\partial_l\delta u_x+ \delta u_x \partial_i\partial_j\partial_l u_x\right] \\
&= \int dt dx \left[\delta u_x  \partial_i\partial_j\partial_l u_x- \delta u_x  \partial_i\partial_j\partial_l u_x\right]  = 0,
\end{split}
\end{equation}
where we used integration by parts from the second to the third equality and assumed that the variations at the boundary of the integration domain are zero. This shows that the Euler-Lagrange equation of terms  $u_x \partial_i \partial_j \partial_l u_x$ in the Lagrangian are zero; hence terms with third order derivatives in an equation of motion cannot be obtained from a Lagrangian. As a result the argument of energy conservation in the previous subsection does not hold anymore as we include the terms with third order derivatives in the equation of motion. Furthermore, plugging the plane wave ansatz $u_x \sim e^{i (kx-\omega t)}$ along with $\omega = -\frac{\epsilon_{03}}{\epsilon_{01}}k \pm k\frac{\sqrt{-\det \epsilon_0}}{|\epsilon_{01}|} + i\delta_{\pm}k^2$ in Eq.~\eqref{Eq:EoMcubic}, and carrying out a calculation similar to Eq.~\eqref{Eq:CorrectionToDispersion}, we would get complex $\omega$. Depending on the sign of the imaginary part of $\omega$, the wave would grow/decay exponentially with time. Hence, we can interpret the system as active/dissipative. Interestingly, the distinction between active and dissipative in this 1D problem is determined by the topological polarization of the 2D lattice.

\pagebreak

\section{Simulated Soft Edge Sinusoidal Perturbation of an $\mathbf{a_2}-\mathbf{a_1}$ Polarized Lattice}

%%%%%%%%%%%%%%%%%%%%%%%%%%%%%
% \subsection*{Simulation Validation for $\mathbf{a_2-a_1}$ Polarization}

\begin{figure}[htp!]
\centering
\subfloat[]{
\includegraphics[height=1.85in]{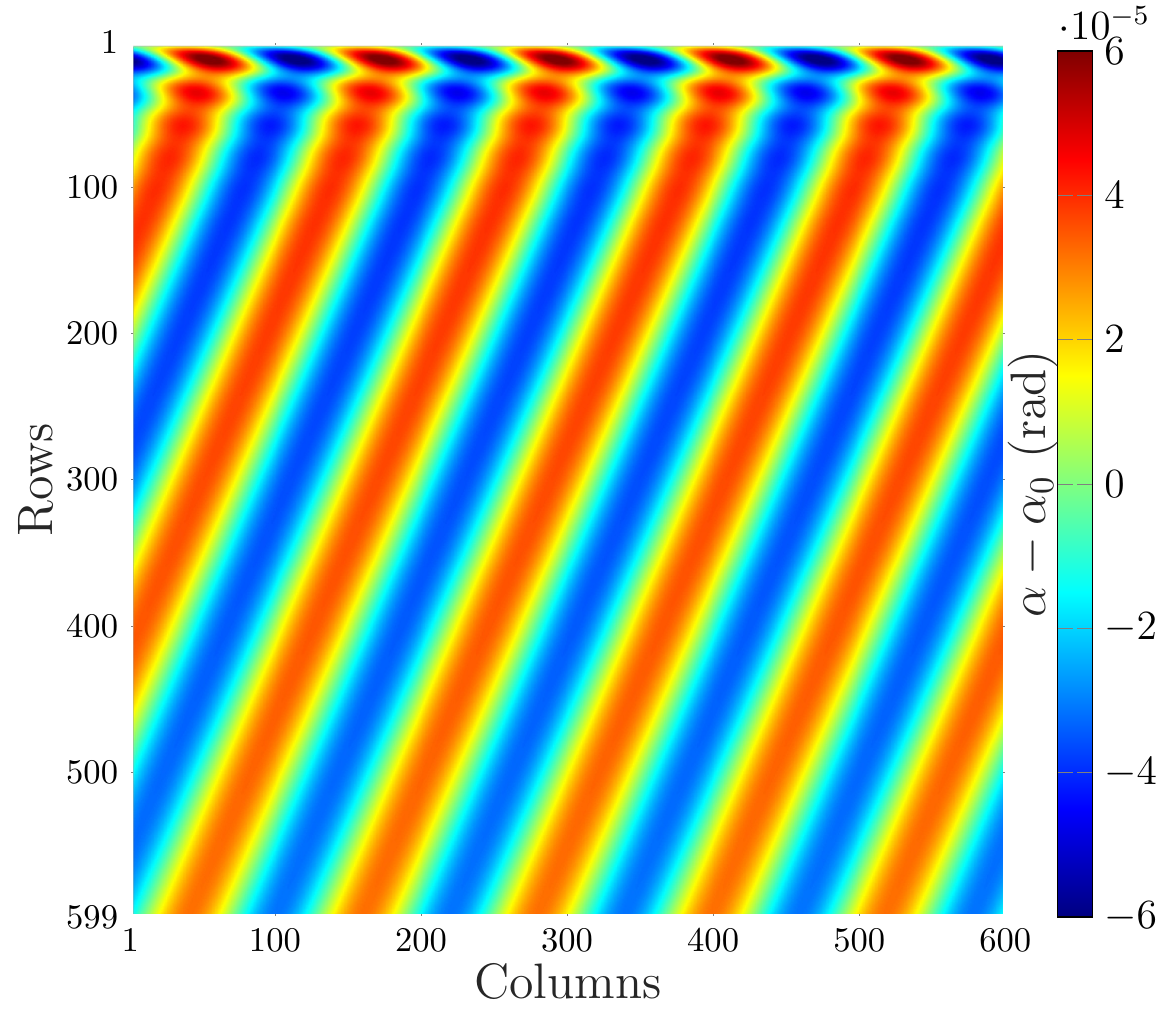}
}
\subfloat[]{
\includegraphics[height=1.45in]{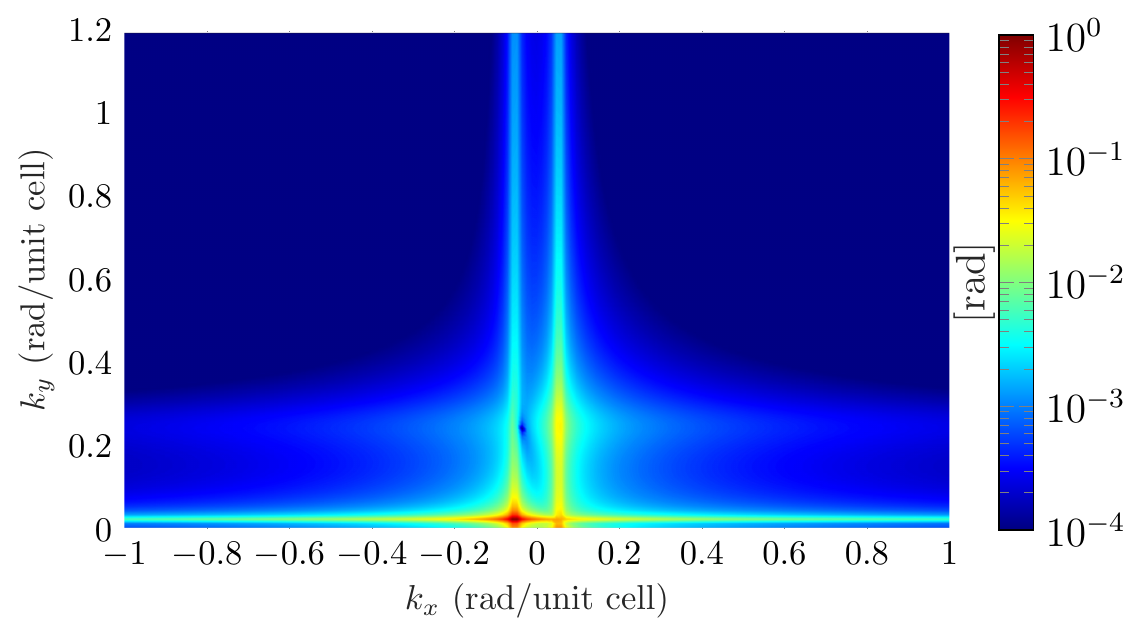}
}
\subfloat[]{
\includegraphics[height=1.8in]{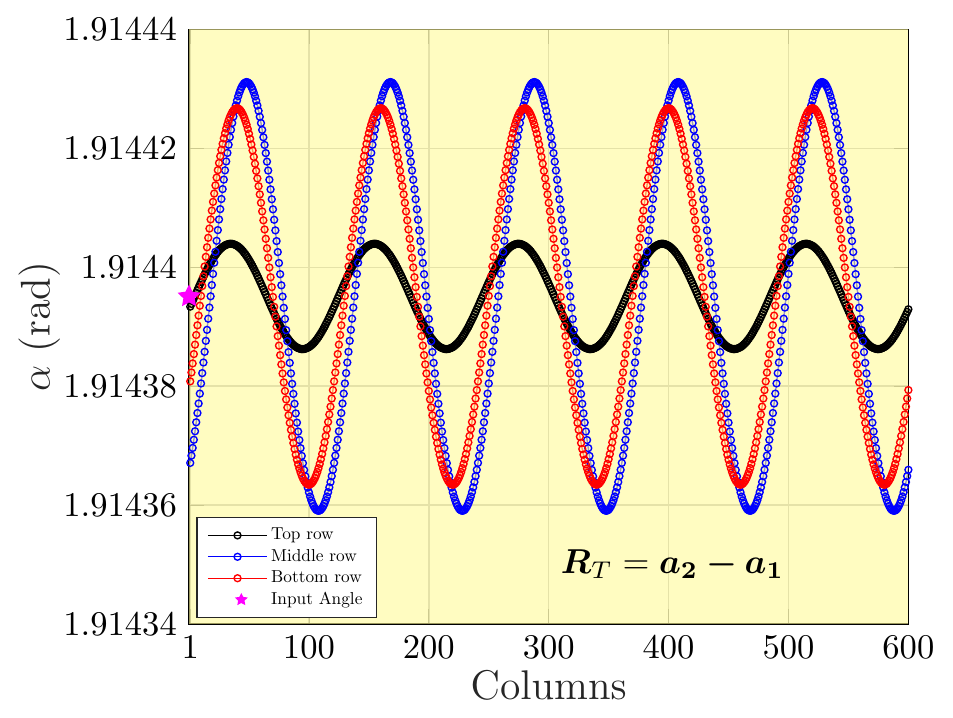}
}
\caption{(a) Perturbation from homogeneous configuration $\alpha-\alpha_0$ for a 600$\times$600 $\mathbf{a_2-a_1}$ polarized lattice ($\alpha_0 = 1.9144$) with periodic boundary conditions given a sinusoidal wave perturbation on the top floppy edge with $ k_x = 0.0523$ (rad/unit cell) and $\varepsilon=1$ $\mu$rad. (b) 2D Fourier transform of (a). (c) Wave shapes of select rows (top, middle, and bottom) from the perturbed lattice for (a).}\label{a2a1_polarization_wave}
\end{figure}

\begin{figure}[htp!]
\centering
\subfloat[]{
\includegraphics[height=1.85in]{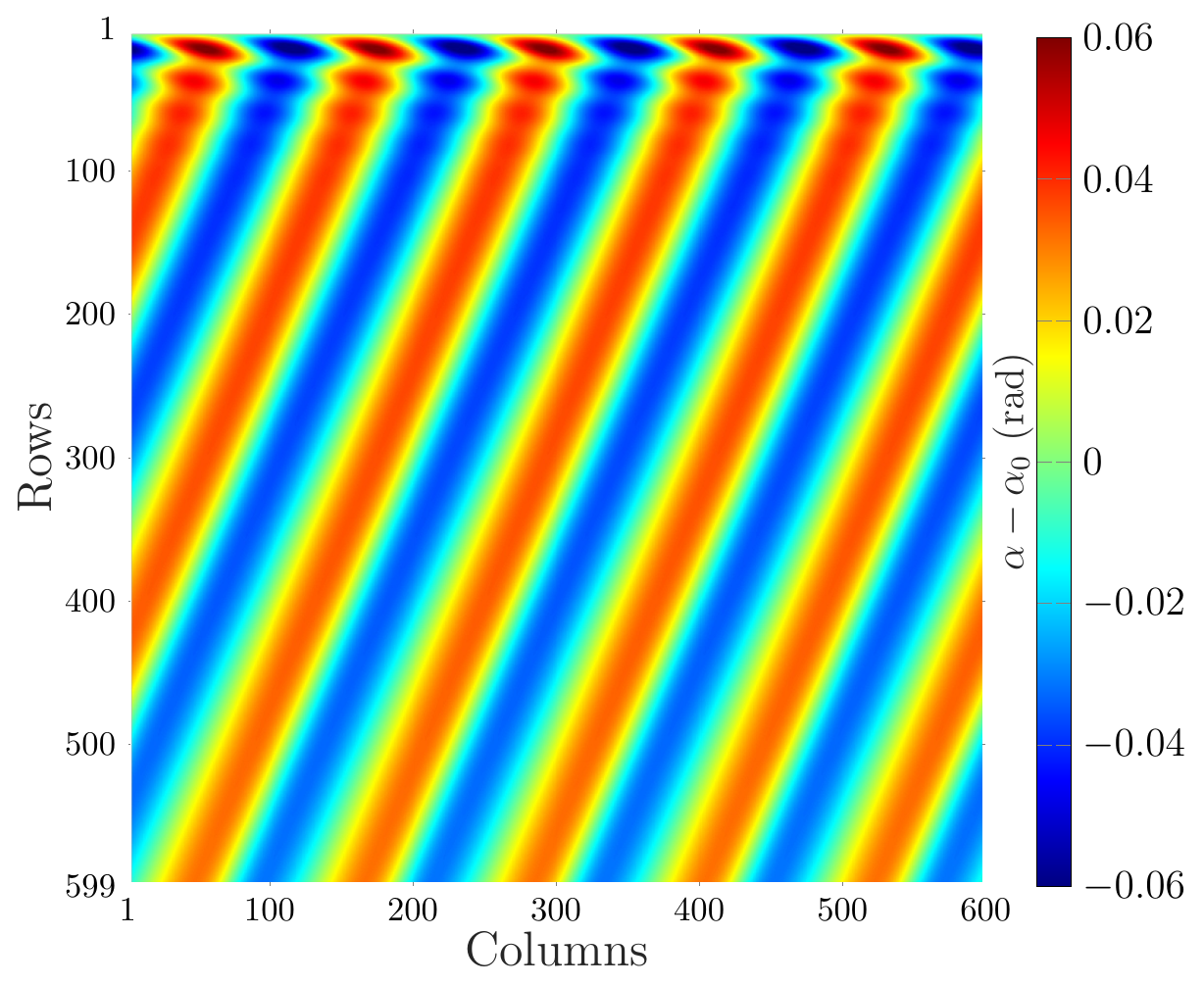}
}
\subfloat[]{
\includegraphics[height=1.45in]{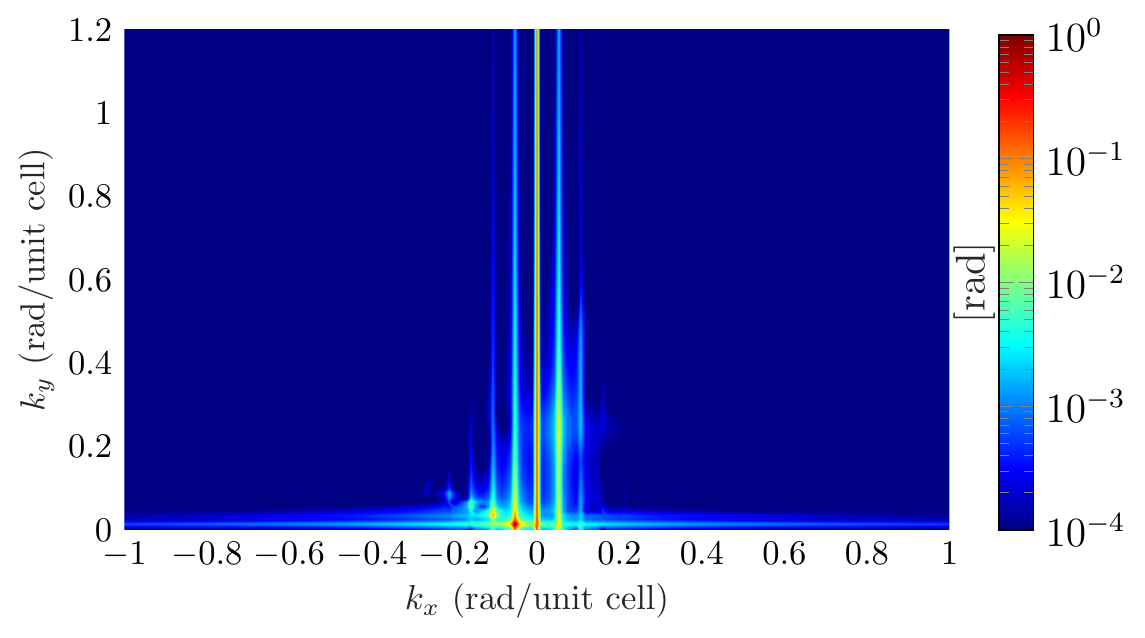}
}
\subfloat[]{
\includegraphics[height=1.8in]{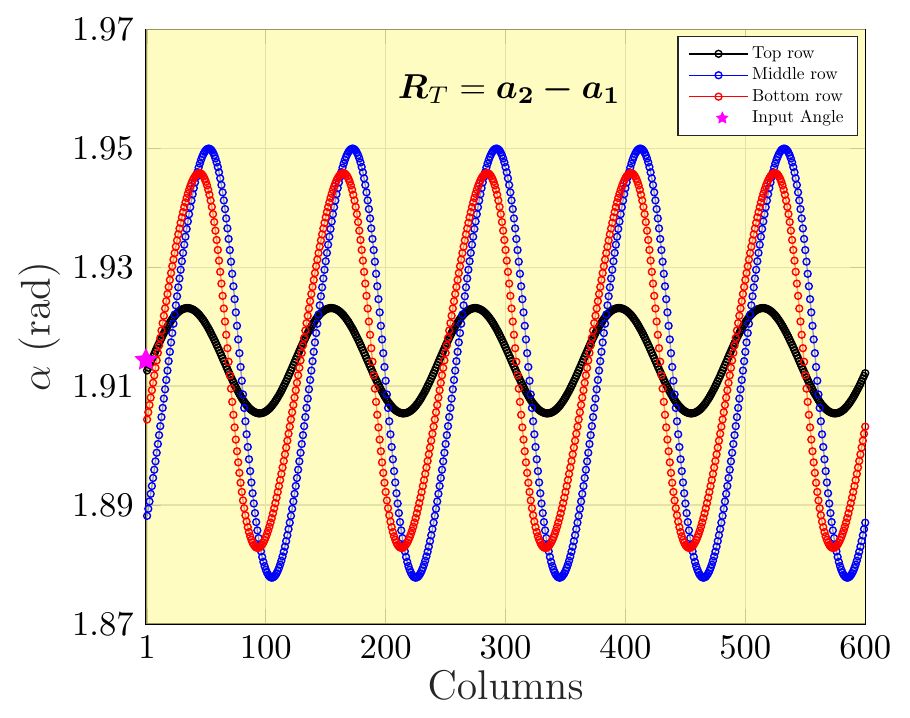}
}
\caption{(a) Perturbation from homogeneous configuration $\alpha-\alpha_0$ for a 600$\times$600 $\mathbf{a_2-a_1}$ polarized lattice ($\alpha_0 = 1.9144$) with periodic boundary conditions given a sinusoidal wave perturbation on the top floppy edge with $ k_x = 0.0523$ (rad/unit cell) and $\varepsilon=1$ mrad. (b) 2D Fourier transform of (a). (c) Wave shapes of select rows (top, middle, and bottom) from the perturbed lattice for (a).}\label{a2a1_polarization_wave_high}
\end{figure}

\noindent In addition to the simulations of the soft edge sinusoidal perturbation of $\mathbf{a_2}$ lattices in the main text (Figs. 1 and 2), here we show simulations with similar perturbation, but of $\mathbf{a_2}-\mathbf{a_1}$ polarized lattices. Figure~\ref{a2a1_polarization_wave} and Fig.~\ref{a2a1_polarization_wave_high} shows the simulated deformation field for the $\mathbf{a_2}-\mathbf{a_1}$ polarized lattices at low and high perturbation magnitudes, respectively. The key difference that can be observed between cases shown here and the $\mathbf{a_2}$ polarized lattices shown in the main text, is that the two excited modes propagate in opposite directions, rather than the same direction, as was the case for the $\mathbf{a_2}$ polarized lattices. 

%\begin{figure}[htp!]
%\centering
%\subfloat[]{%
%  \includegraphics[height=2in]{Rt0_highamp_3D_surface.pdf}%
%}
%\subfloat[]{%
%  \includegraphics[width=4.4in]{Rt0_real_space.pdf}%
%}
%\caption{(a) 3D visualization of the deformation field of a $\mathbf{R_T}=0$, $\alpha_0=1.114$ rad, $k_x=0.314$ rad/unit cell, and $\varepsilon=1e-2$ rad lattice that leads to breaking (pictured in (b)). The z-axis corresponds with perturbation magnitude $\alpha-\alpha_0$. (b) Real space deformation of (a) $\mathbf{R}_T=0$ polarization$\alpha_0=0.7144$ rad up until the point where the lattice breaks.}\label{real_space_Rt0}
%\end{figure}

\pagebreak

\section{Additional Wave Amplification Simulations}

\begin{figure}[htp!]
\centering
\subfloat[]{
\includegraphics[height=1.78in]{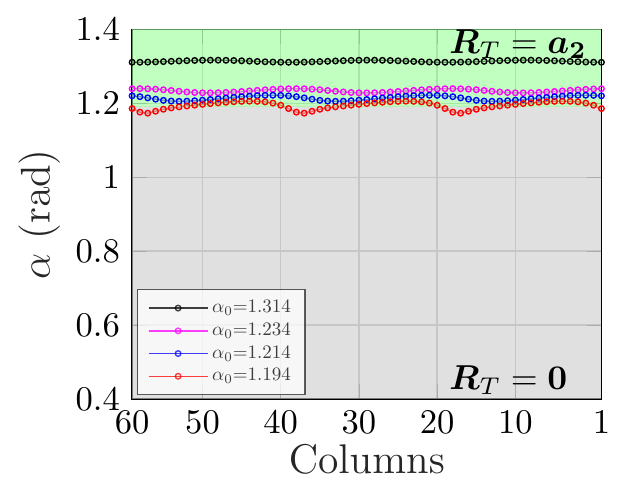}
}
\subfloat[]{
\includegraphics[height=1.78in]{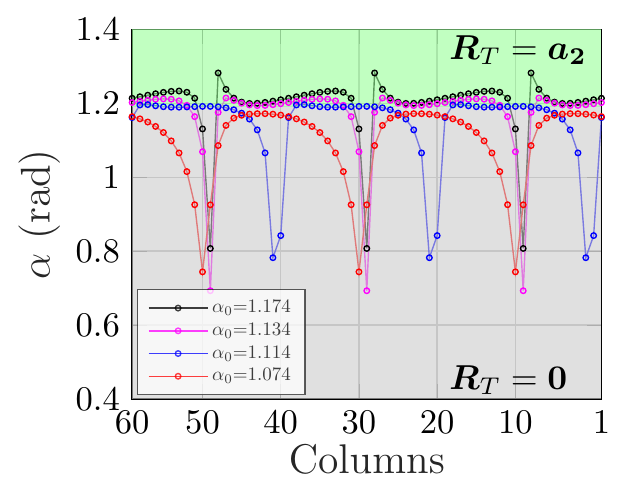}
}
\subfloat[]{
\includegraphics[height=1.78in]{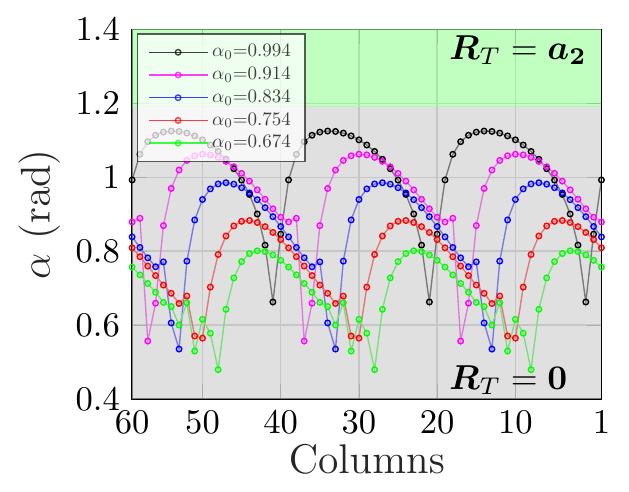}
}%

\subfloat[]{
\includegraphics[height=1.78in]{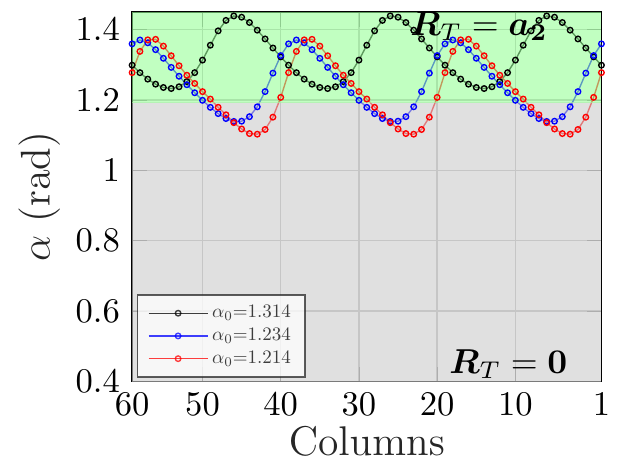}
}
\subfloat[]{
\includegraphics[height=1.78in]{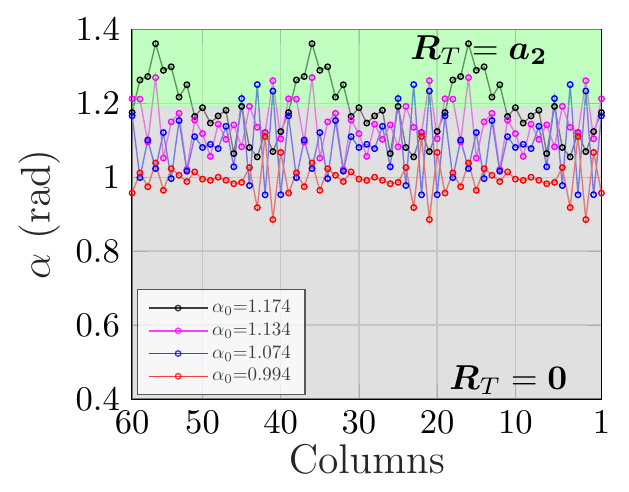}
}
\subfloat[]{
\includegraphics[height=1.78in]{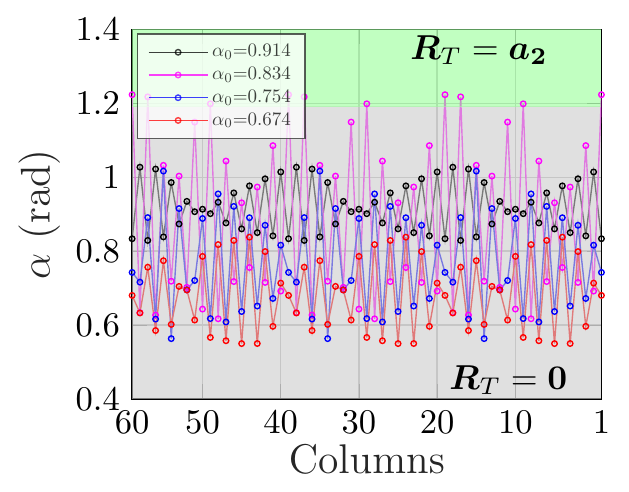}
} \caption{Computational results ($\alpha$) of wave amplification in $60$-column-wide lattices with periodic boundary conditions due to sinusoidal perturbation ($k_x=0.314$) for an array of $\alpha_0$ values. (a,d) $\alpha_0$ originally in $\mathbf{a_2}$ polarization, near the boundary with $\mathbf{R}_T=0$. (b,e) $\alpha_0$ in $\mathbf{R_T=0}$ polarization near the boundary with $\mathbf{a_2}$. (c,f) $\alpha_0$ further in $\mathbf{R}_T=0$ than before. (a-c) Perturbation amplitude $\varepsilon=1$ mrad. (a) Bottom edges after perturbing at the top (soft) edges. (b,c) Bottom edges after perturbing at the top edge of the unpolarized lattices. (d-f) Perturbation amplitude $\varepsilon=1$ $\mu$rad. (d) Top edges after perturbing at the bottom (hard) edges. (e,f) Top edges after perturbing the bottom of the unpolarized lattices. In (a), $60$ rows is chosen; In (b-f), the displayed row is the last row before the lattice breaks.}\label{a2Rt0_topbottomrow}
\end{figure}

\noindent In Fig.~\ref{a2Rt0_topbottomrow}, we show additional computational examples of wave amplification for 60-column-wide lattices with periodic boundary conditions over an array of $\alpha_0$ values. We observe polarization domain switching in Fig.~\ref{a2Rt0_topbottomrow}(a,b,d,e,f) and high frequencies generation (compared to the input wavelength) in Fig.~\ref{a2Rt0_topbottomrow}(e,f). 

\pagebreak

% \clearpage
\pagebreak

%%%%%%%%%%%%%%%%%%%%%%%%%%%%%
%%%%%%%%%%%%%%%%%%%%%%%%%%%%%

\section{Solitary-Wave-Like Behavior of the Deformed Kagome Lattice}

\begin{figure}[htp!]
\centering
\subfloat[]{
\includegraphics[height=3.3in]{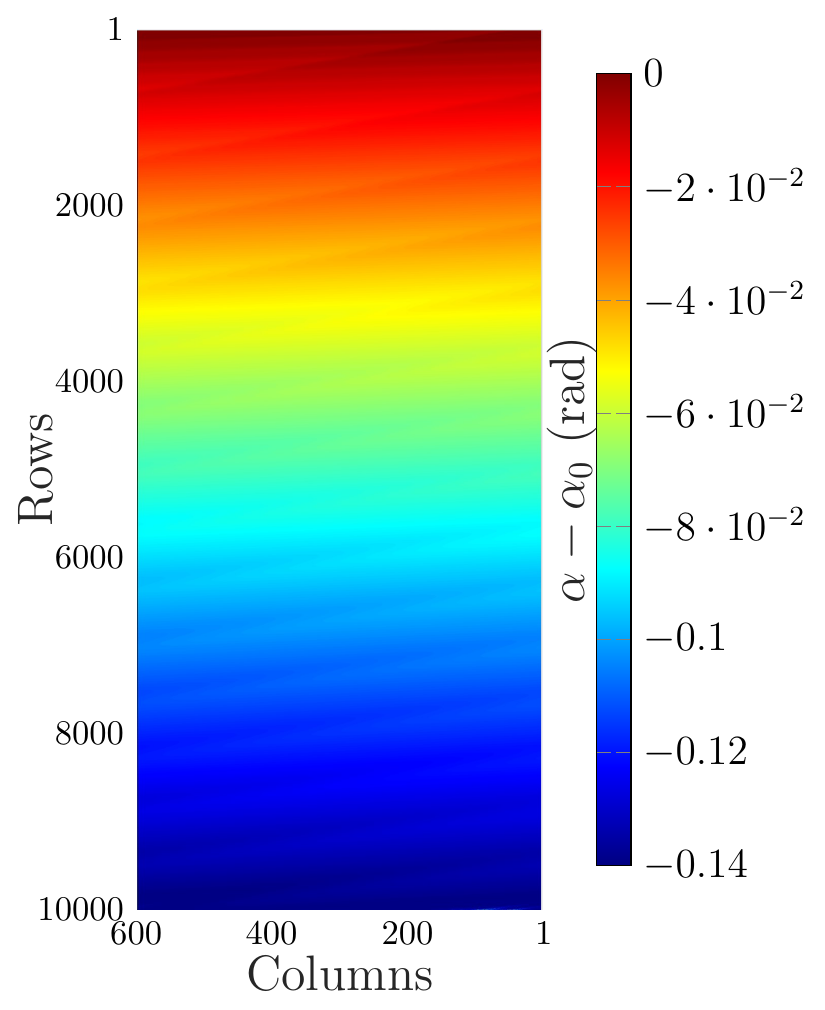}
}
\subfloat[]{
\includegraphics[height=3.3in]{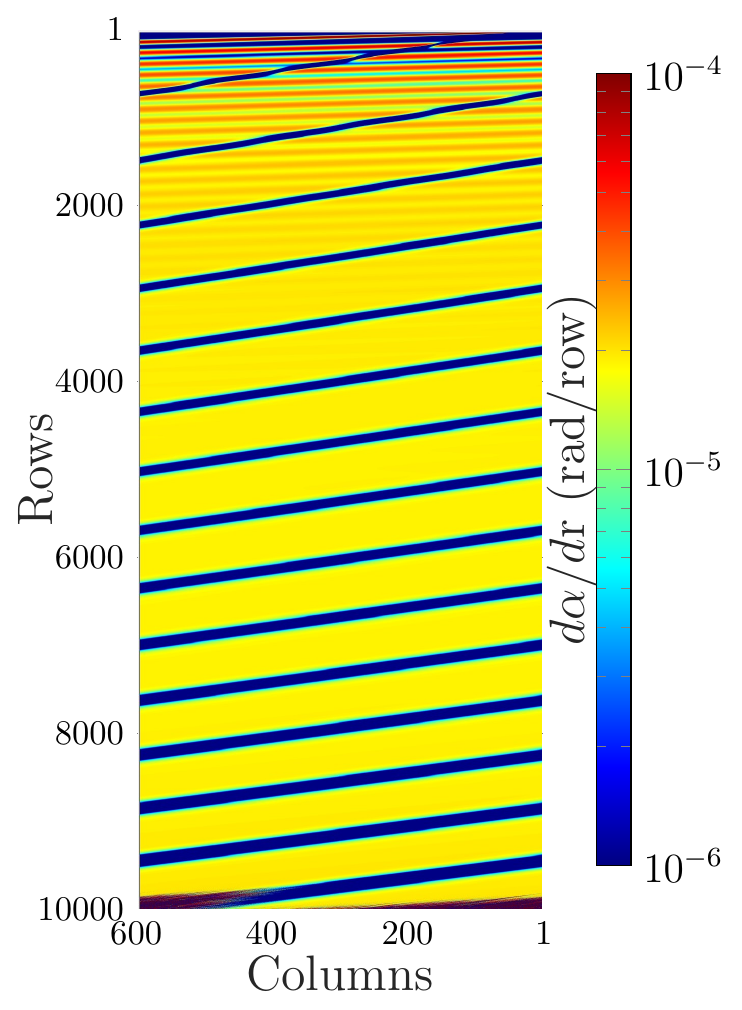}
}
\subfloat[]{
\includegraphics[height=3.3in]{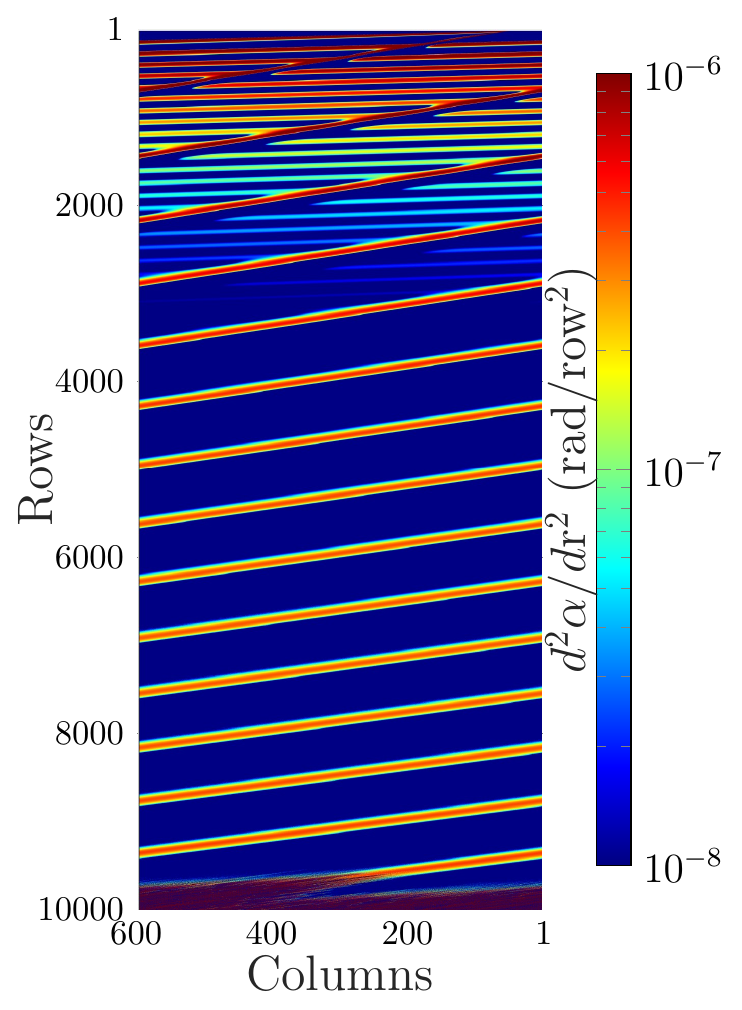}
}
\caption{Computational results of a 10000$\times$600 $\mathbf{a_2}$ polarized lattice with periodic left-right boundary conditions. The lattice has $\alpha_0=1.3144$ rad, and the point perturbation is applied on the top row at column 50. (a) $\alpha-\alpha_0$, (b) $d\alpha/dr$, and (c) $d^2\alpha/dr^2$. Logarithmic scale is used in the colorbars of (b) and (c). Note, (b) [(c)], $d\alpha/dr$ [$d^2\alpha/dr^2$] is saturated at the lower limit of $10^{-6}$ rad [$10^{-8}$ rad].}
\label{soliton_1}
\end{figure}

\begin{figure}[h!]
\centering
\subfloat[]{%
  \includegraphics[height=1.7in]{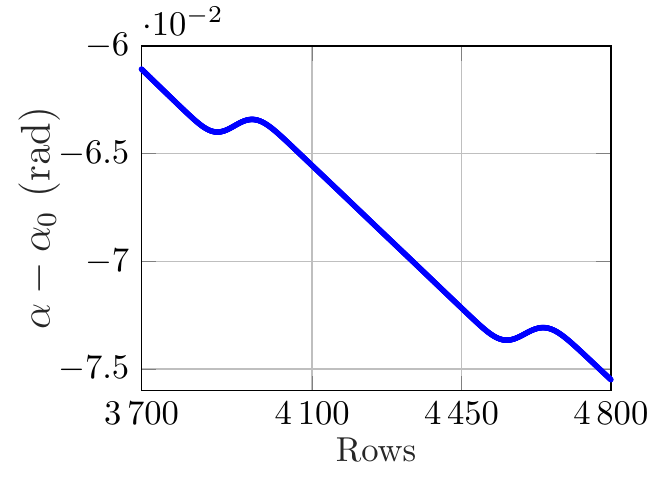}%
}
\subfloat[]{%
  \includegraphics[height=1.7in]{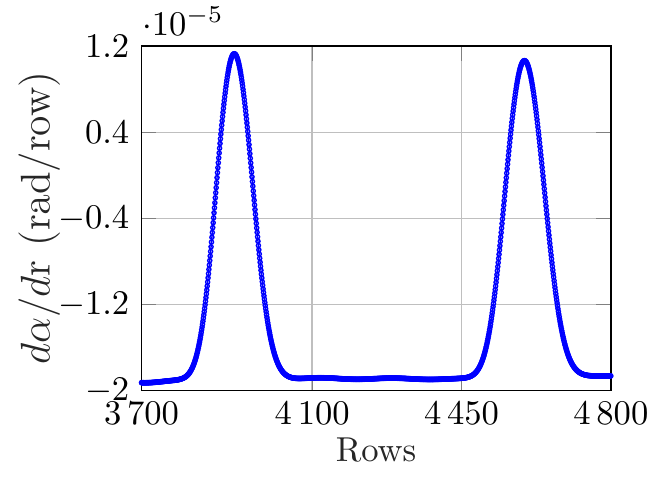}%
}
\subfloat[]{%
  \includegraphics[height=1.7in]{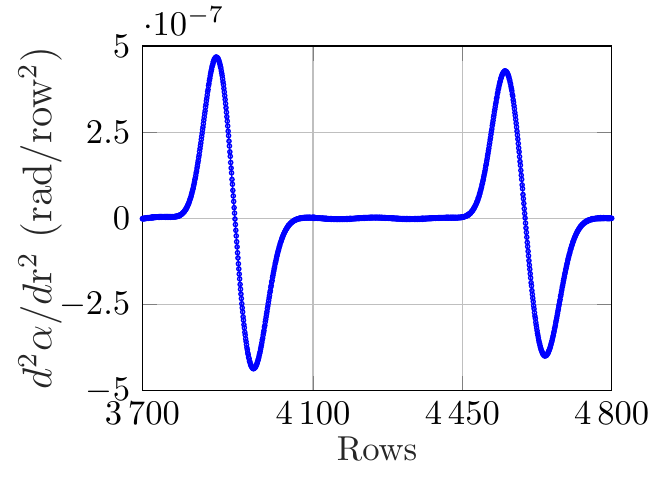}%
}

\subfloat[]{
\includegraphics[height=1.7in]{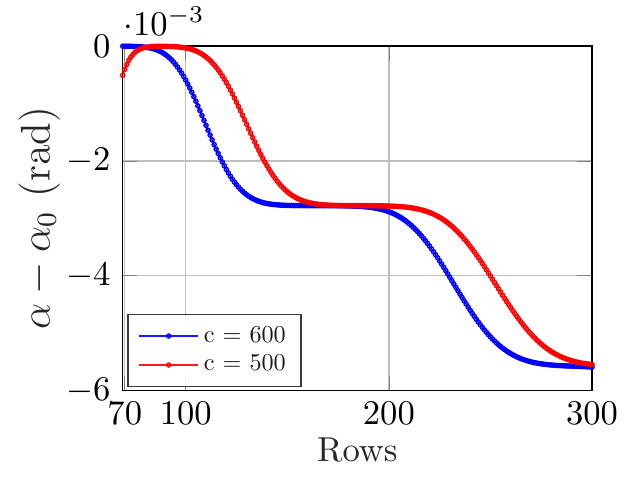}
}
\subfloat[]{
\includegraphics[height=1.7in]{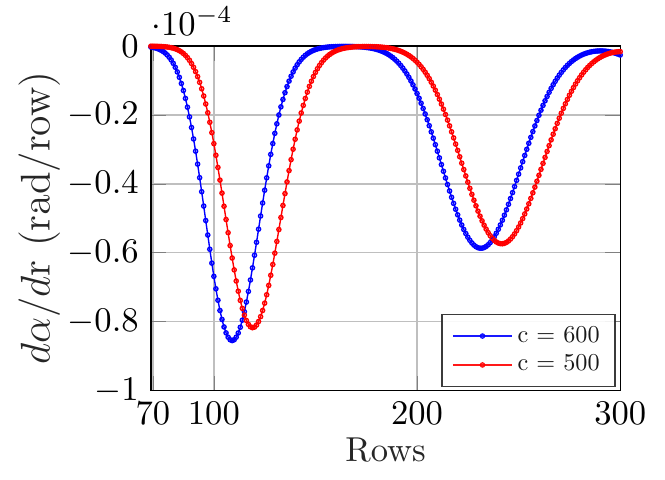}
}
\subfloat[]{
\includegraphics[height=1.7in]{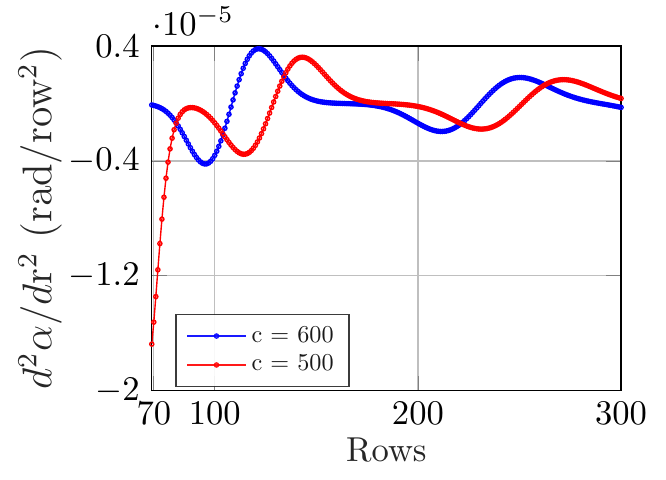}
}
\caption{Selected computational results for the two solitary waves observed in the lattice in Fig.~\ref{soliton_1}. (a-c) Slower moving solitary wave: (a) $\alpha-\alpha_0$, (b) $d\alpha/dr$, and (c) $d^2\alpha/dr^2$ at column 300 (middle column). Rows $3700-4800$ are chosen to minimize the interaction with the faster moving solitary wave. 
(d-f) Faster moving solitary wave: (d) $\alpha-\alpha_0$, (e) $d\alpha/dr$, and (f) $d^2\alpha/dr^2$ at columns 500 (red) and 600 (blue), respectively. Rows $70-300$ are chosen to avoid the effect of substantial amplitude decay. Note that we avoid very early rows $1-70$ to give the solitary waves adequate time to separate.}\label{fig:soliton2}
\end{figure}

\noindent In addition to the example of the $10000 \times 3000$ $\mathbf{a_2}$ polarized lattice in the main text, here we show a $10000 \times 600 $ $\mathbf{a_2}$ polarized lattice with a point perturbation applied at the top (floppy) edge. The computational results are shown in Figs.~\ref{soliton_1} and \ref{fig:soliton2}. 

% fig.s9

% fig.s10

%%%%%%%%%%%%%%%%%

% fig.s11
\begin{figure}[h!]
\centering
\subfloat[]{%
  \includegraphics[height=1.7in]{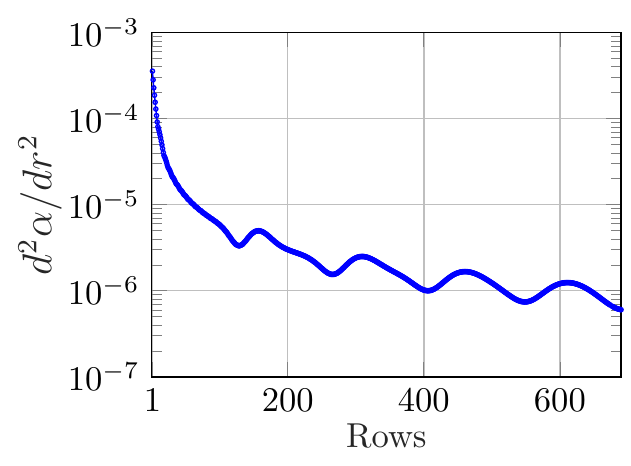}%
}
\subfloat[]{%
  \includegraphics[height=1.7in]{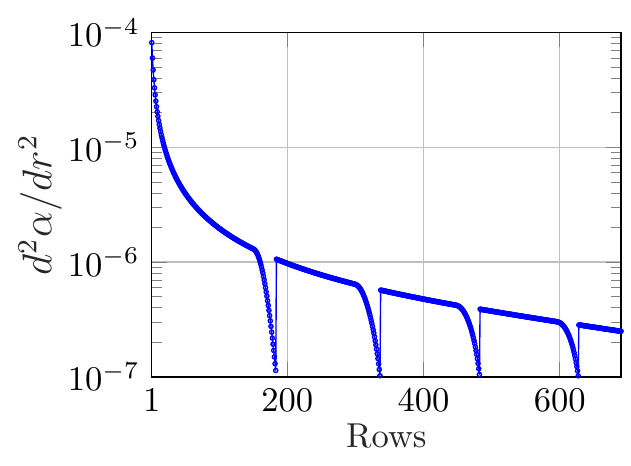}%
}
\subfloat[]{
\includegraphics[height=1.75in]{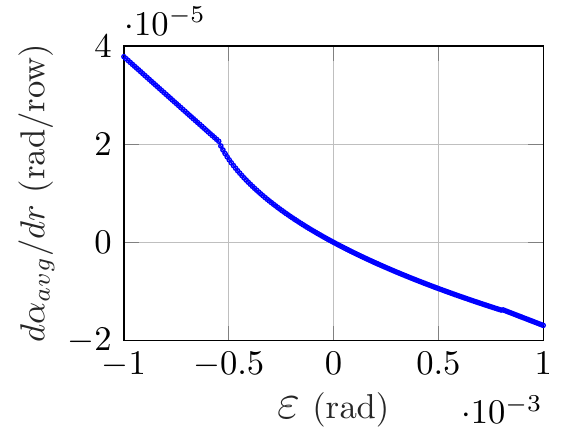}
}

\subfloat[]{%
  \includegraphics[height=1.65in]{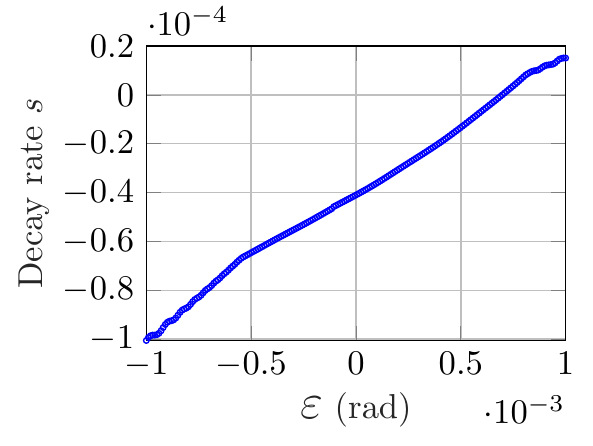}%
}
\subfloat[]{%
  \includegraphics[height=1.63in]{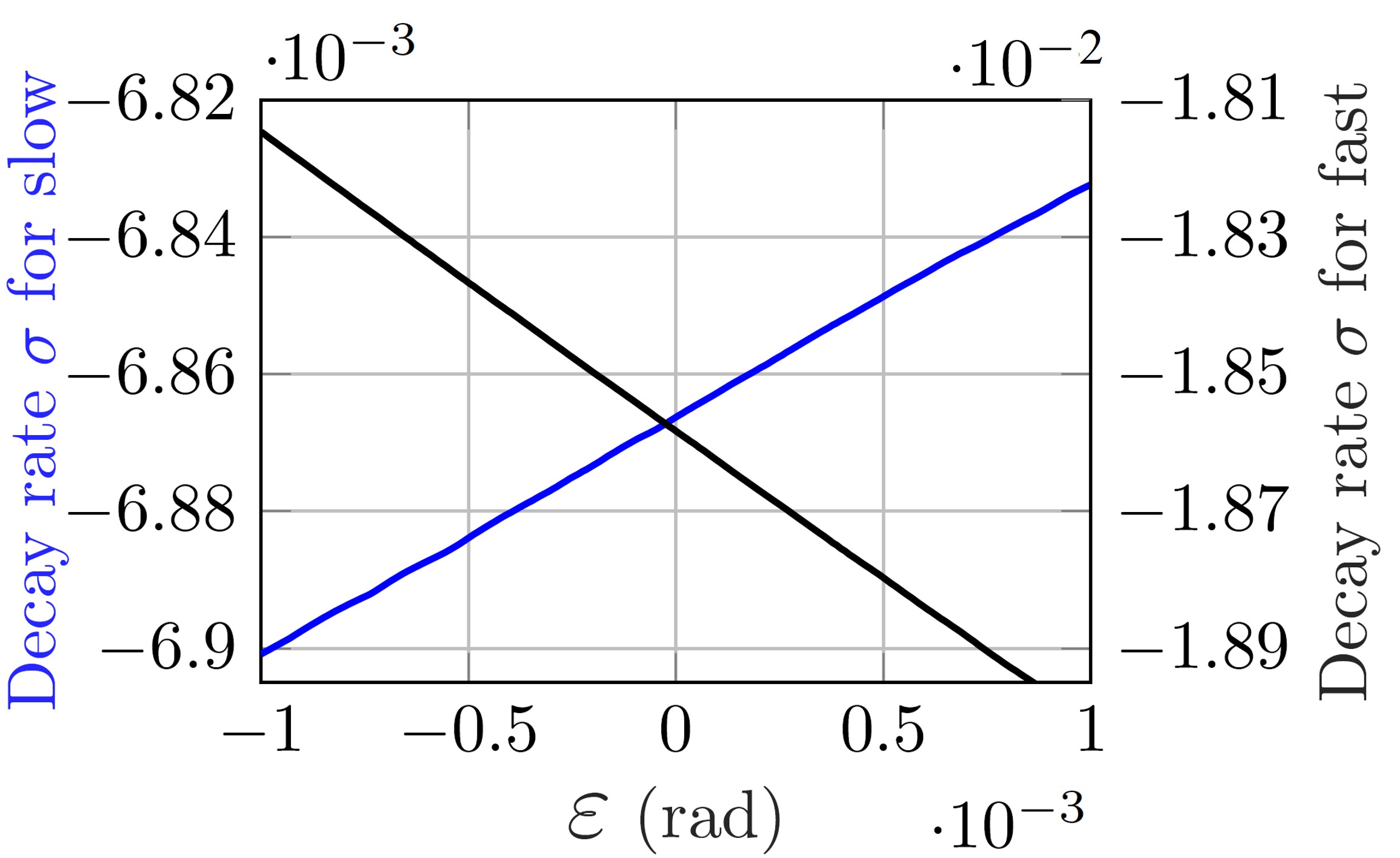}%
}
\subfloat[]{
\includegraphics[height=1.6in]{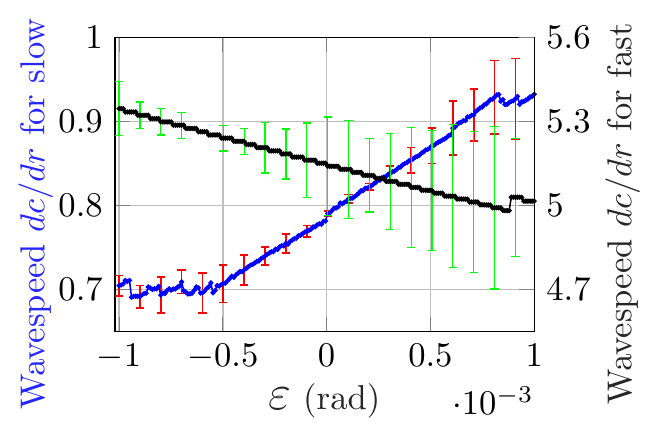}
}
\caption{(a,b) Peak magnitude of $d^2\alpha/dr^2$ as a function of rows for (a) slower and (b) faster moving solitary waves. The fluctuation of magnitude is due to the interaction of two solitary waves. As a function of point perturbation magnitude $\epsilon$ the: (c) rate of change of $\alpha_{avg}$ with respect to rows, (d) decay rate $s$ based on the amplitude $\alpha_{max}-\alpha_{min}$ at each row of the slower moving solitary wave, 
(e) decay rate $\sigma$ based on the peak-to-peak magnitude of $d^2\alpha/dr^2$ at each row for both solitary waves, and 
(f) wave-speeds of slower and faster moving solitary waves versus the point perturbation value $\epsilon$.}
\label{fast_soliton}
\end{figure}

% fig.s12

\begin{figure}[htp!]
\centering
\subfloat[]{%
  \includegraphics[height=1.8in]{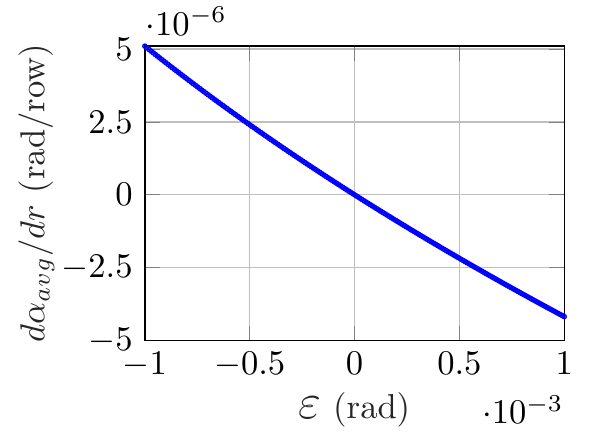}%
}
\subfloat[]{%
  \includegraphics[height=1.8in]{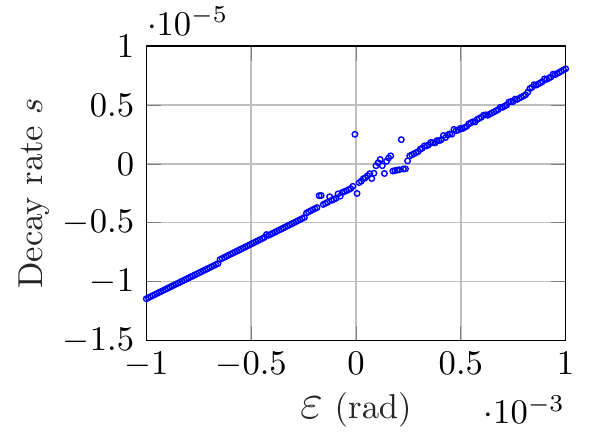}%
}
\caption{Additional computational results of a 10000$\times$3000 $\mathbf{a_2}$ polarized lattice ($\alpha_0 = 1.3144$ rad) with periodic left-right boundary conditions and
the point perturbation applied on the top row at column 50. As a function of point perturbation magnitude $\epsilon$, the: (a) rate of change of $\alpha_{avg}$ with respect to rows, and (b) decay rate $s$ based on the amplitude $\alpha_{max}-\alpha_{min}$ at each row of the slower moving solitary wave.}
\label{fast_soliton_as}
\end{figure}

In Fig.~\ref{fast_soliton}, we further analyze the simulated results from Fig.~\ref{soliton_1}. Fig.~\ref{fast_soliton}(a) [(b)] is obtained by tracking the peak magnitude of $d^2\alpha/dr^2$ (where derivatives with respect to $r$ are found via a finite difference formulation subtracting over columns) in each row for slower [faster] moving solitary wave using the \textit{findpeaks} command in MATLAB. Figure~\ref{fast_soliton}(c) shows the rate of change of average $\alpha$, where $\alpha_{avg}$ is the mean value in each row given by $\alpha_{avg} = \sum \alpha/c=\alpha/600$, with respect to rows, as a function of point perturbation value $\varepsilon$. This is tracked by taking $\alpha_{avg}$ at each row and performing a linear fitting $\alpha_{avg}=m r+b$ ranging the entire lattice (rows $1-10000$), using the \textit{polyfit} command in MATLAB. $d\alpha_{avg}/dr$ is thus given by the fitting parameter $m$. 

Figure~\ref{fast_soliton}(d) shows the fitted decay rate $s$ of the slower moving solitary wave, as a function of point perturbation value $\varepsilon$, obtained by finding $\alpha_{max}$ and $\alpha_{min}$ using the \textit{max} and \textit{min} commands in MATLAB, and then calculating the peak-to-peak amplitude $\alpha_{max}-\alpha_{min}$ at each row. The behavior of the peak to peak amplitude is oscillatory and decaying, in order to fit the decay rate, we first find the peak values of $\alpha_{max}-\alpha_{min}$ in the plot of $\alpha_{max}-\alpha_{min}$ versus $r$ using the \textit{findpeak} command in MATLAB, then fit these peak values to an exponential function $A e^{s r}$. Figure~\ref{fast_soliton}(e) shows the fitted decay rate $\sigma$ for both slower and faster moving solitary waves, as a function of point perturbation value $\varepsilon$, based on ${d^2\alpha/dr^2}_{max}-{d^2\alpha/dr^2}_{min}$ at each row, instead of $\alpha_{max}-\alpha_{min}$. Here, for each solitary wave, ${d^2\alpha/dr^2}_{max}$ and ${d^2\alpha/dr^2}_{min}$ represent the local maximum and minimum of ${d^2\alpha/dr^2}$ at each row. For the results shown in Fig.~\ref{fast_soliton}(e), rows $50-300$ are the chosen range for fitting ${d^2\alpha/dr^2}_{max}-{d^2\alpha/dr^2}_{min} = Ae^{\sigma r}$ since the solitary waves separate from each other and have not collided again. 

The ``speed'' of the solitary waves for the 600 column wide lattice is shown in Fig.~\ref{fast_soliton}(f). Differing algorithms were used to find the speed for the 3000 (main text) and 600 (here) column wide lattices, since the higher number of interactions in the 600 column wide lattice makes it difficult to locate the individual solitary wave peaks in each row. For the 3000-column-wide lattice shown in the main text, the speed $d(col)/d(row)\rightarrow dc/dr$ is obtained by tracking the location of peaks for $d^2\alpha/dr^2$ at each row, which can correspond with slower (bigger peak) and faster solitary waves (smaller peak). Again, the \textit{findpeaks} command in MATLAB is used here. During collision no data is collected. Once peaks of $d^2\alpha/dr^2$ are found at each row, a linear fitting is performed on rows $50-400$. The row range is chosen for the same reason as the previous fitting in Fig.~\ref{fast_soliton}(e). For each solitary wave, columns can be related to rows in the linear form $r=m'c+b'$ the slope $m'$ gives wave-speed $dc/dr$. For the 600 column wide lattice, the local maximum is found using the same peak detection algorithm, however this time, the data analyzed comes from one column for a prescribed set of rows, such as in Fig~\ref{fig:soliton2}(c). The peaks denote when the solitary wave passes the column of interest, the distance between peaks $\Delta r$ is calculated. The columns traversed is the width of the lattice, thus the velocity is given by $dc/dr=\Delta c/\Delta r = 600/\Delta r$. In order to avoid the interaction between the two solitary waves at the beginning rows of the lattice, the slower moving solitary wave is investigated in the range of $r/3-4r/5$. The final speed in main text Fig.~5 is calculated as the average of $600/\Delta r$ for all $\Delta r$ in the specified range. The error bars are calculated by taking the maximum and minimum solitary wave velocities for a particular $\varepsilon$ value. The amplitude of the faster moving solitary wave is found by looking at the far right column (column 600) and the first 500 rows of the acceleration term $d^2\alpha/dr^2$. The peaks and their corresponding row locations of the fast solitary wave are obtained using the \textit{findpeaks} command, which gives $\Delta r$, the faster solitary wave's velocity is then found in the same manner as before $dc/dr=600/(\Delta r)$. The error is found similar to the slower solitary wave case by taking the difference between the maximum and the minimum velocities.

Additional computational results of a $10000\times 3000$ $\mathbf{a_2}$ polarized lattice are shown in Fig.~\ref{fast_soliton_as}.

\pagebreak
%%%%%%%%%%%%%%%%%%%%%%%%%%%%%

%%%%%%%%%%%%%%%%%%%%%%%%%%%%%
%%%%%%%%%%%%%%%%%%%%%%%%%%%%%
\section{Experimental Setup and Image Processing}
\begin{figure}[htp!]
    \centering
    \includegraphics[width=5in]{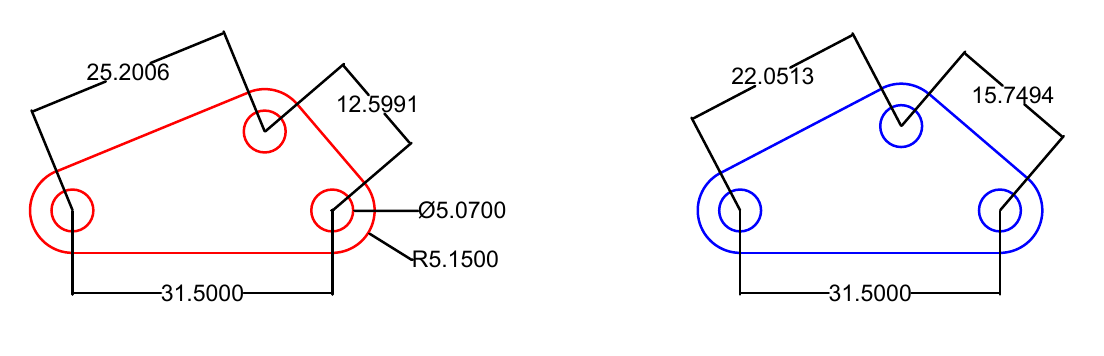}
    \caption{Dimensions (in the unit of mm) of laser cut acrylic triangles in the unit cell used in Fig.~7(a) of the main text.}
    \label{fig:triangle_dim_real}
\end{figure}

\begin{figure}[htp!]
    \centering
    \includegraphics[width=5in]{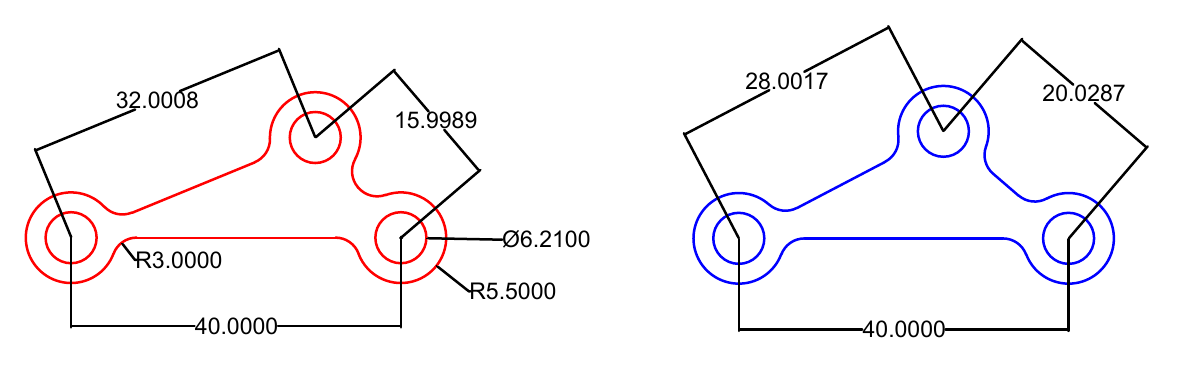}
    \caption{Dimensions (in the unit of mm) of laser cut acrylic triangles in the unit cell used in Fig.~7(d) of the main text.}
    \label{fig:triangle_dim_real2}
\end{figure}

\begin{figure}[htp!]
\centering
\subfloat[]{%
  \includegraphics[height=1.8in]{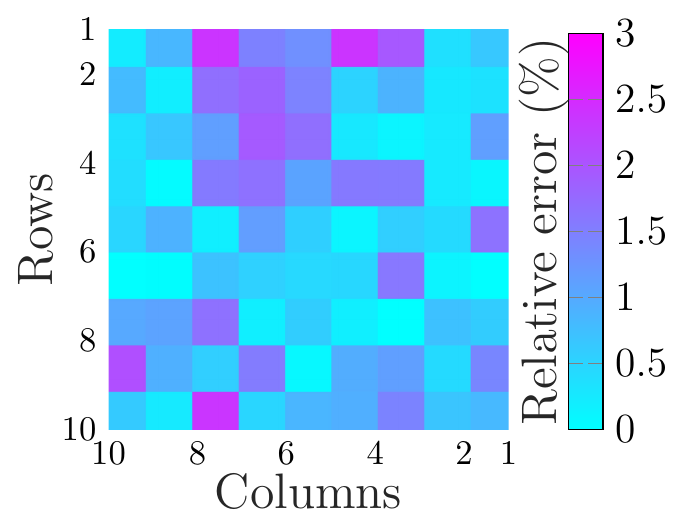}%
}
\subfloat[]{%
  \includegraphics[height=1.8in]{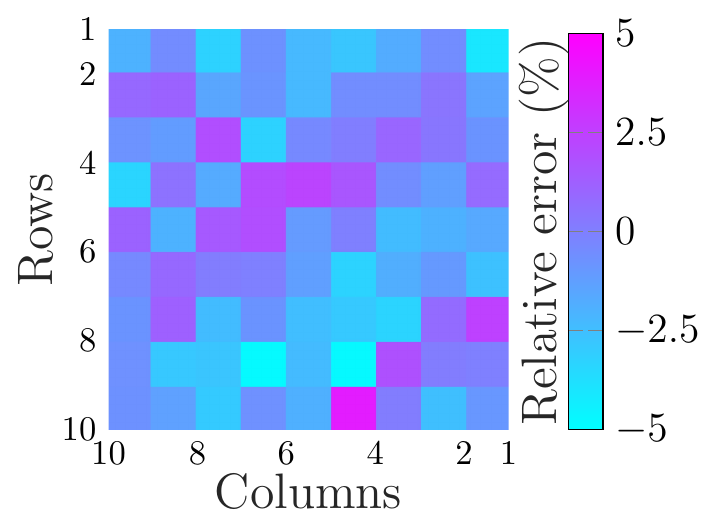}%
}
\caption{Difference between measured and simulated angles normalized by homogeneous angle $\alpha_0=1.3144$ for the two examples shown in Fig.~7 of the main text. (a) Sinusoidal perturbation for $\varepsilon=0.1$ rad, $k_x=0.6283$ rad/unit cell. (b) Point perturbation for $\varepsilon=45$ mrad applied at column $3$.}
\label{expeirmnet_error}
\end{figure}

\noindent Photographs of the assembled lattices were taken and MATLAB was used to post process the images. The hinge points were found using the \textit{imfindcircles} function, which allows the lattice to be reconstructed and analyzed. The lattices composed of laser cut black [red and blue] acrylic triangles in both the SI and the main text are shown in Fig.~\ref{fig:triangle_dim_real} [Fig.~\ref{fig:triangle_dim_real2}], respectively. Figure~\ref{expeirmnet_error} shows the error between the measured and simulated lattices shown in Fig.~7 of the main text. 

% Loading bibliography database
\setlength{\itemsep}{1mm}\footnotesize
\bibliography{Ref_SM}